\documentclass[11pt,a4paper]{article}

\RequirePackage{ifpdf} 
\usepackage{amsmath} 
\usepackage{mathtools}

\usepackage{jheppub}
\usepackage{pstricks}
\usepackage[final]{pdfpages} 
\usepackage{ifpdf} 
\usepackage{slashed}

\usepackage{hyperref}

\usepackage{color} 
\usepackage{graphics}

\usepackage{etoolbox} 
\usepackage{fixmath}

\usepackage{caption} 
\usepackage{subcaption} 
\usepackage{amsfonts}

\usepackage{multirow}
\usepackage{epstopdf}

\usepackage{relsize}
\usepackage{float}

\usepackage{tikz}
\usetikzlibrary{positioning,arrows}
\usetikzlibrary{decorations.pathmorphing}
\usetikzlibrary{decorations.markings}
\usetikzlibrary{shapes.geometric}
\tikzset{
    vector/.style={decorate, decoration={snake}, draw},
    provector/.style={decorate, decoration={snake,amplitude=2.5pt}, draw},
    antivector/.style={decorate, decoration={snake,amplitude=-2.5pt}, draw},
    fermion/.style={draw=black,
      postaction={decorate},decoration={markings,mark=at position .55
        with {\arrow[draw=black]{>}}}}, 
    fermionbar/.style={draw=black, postaction={decorate},
                       decoration={markings,mark=at position .55 with {\arrow[draw=black]{<}}}},
    fermionnoarrow/.style={draw=black},
    gluon/.style={decorate, draw=black,decoration={coil,amplitude=4pt, segment length=4pt}},
    scalar/.style={dashed,draw=black,
      postaction={decorate},decoration={markings,mark=at position .55
        with {\arrow[draw=black]{>}}}}, 
    scalarbar/.style={dashed,draw=black,
      postaction={decorate},decoration={markings,mark=at position .55
        with {\arrow[draw=black]{<}}}}, 
    scalarnoarrow/.style={dashed,draw=black},
    electron/.style={draw=black,
      postaction={decorate},decoration={markings,mark=at position .55
        with {\arrow[draw=black]{>}}}}, 
    bigvector/.style={decorate, decoration={snake,amplitude=4pt}, draw},
}

\title{NNLO QCD Corrections to the Drell-Yan Cross Section in Models of TeV-Scale Gravity}

\author[a,b]{Taushif Ahmed,} 
\author[a,b]{Pulak Banerjee,}
\author[a,b]{Prasanna K. Dhani,}
\author[c]{M.C. Kumar,}
\author[d]{Prakash Mathews,} 
\author[a,b]{Narayan Rana} 
\author[a]{and V. Ravindran}

\affiliation[a]{The Institute of Mathematical Sciences, IV Cross Road,
  CIT Campus, Chennai 600 113, Tamil Nadu, India}
\affiliation[b]{Homi Bhabha National Institute, Training School Complex, Anushakti Nagar, Mumbai 400085, India}
\affiliation[c]{Department of Physics, Indian Institute of Technology Guwahati, Guwahati 781039,
  India} 
\affiliation[d]{Saha Institute of Nuclear Physics, 1/AF Bidhan Nagar, Kolkata 700 064, West Bengal, India}

\emailAdd{taushif@imsc.res.in} \emailAdd{bpulak@imsc.res.in}
\emailAdd{prasannakd@imsc.res.in} \emailAdd{mckumar@iitg.ac.in}
\emailAdd{prakash.mathews@saha.ac.in} \emailAdd{rana@imsc.res.in}
\emailAdd{ravindra@imsc.res.in}

\abstract{ The first results on the complete next-to-next-to-leading order (NNLO) Quantum Chromodynamic (QCD) corrections to the production of
di-leptons at hadron colliders in large extra dimension models with spin-2 particles are reported in this article.
In particular, we have computed these corrections to the invariant mass distribution of the di-leptons taking into account all
the partonic subprocesses that contribute at NNLO.  In these models, spin-2 particles couple through the energy-momentum tensor of the Standard Model (SM) with the universal coupling strength.  The tensorial nature of the interaction 
and the presence of both quark annihilation and gluon fusion channels at the Born level make it challenging computationally and interesting phenomenologically.  We have demonstrated
numerically the importance of our results at the Large Hadron Collider (LHC) energies. The two loop corrections contribute an additional 10\% to the total cross section. We find that the QCD corrections are not only 
large but also important to make the predictions stable under renormalisation and 
factorisation scale variations 
providing an opportunity to stringently constrain the parameters of the models with a spin-2 particle.}

\preprint{} 

\begin{document}
\allowdisplaybreaks[4]
\unitlength1cm
\maketitle
\flushbottom


\def\D{{\cal D}}
\def\DD{\overline{\cal D}}
\def\g{\overline{\cal G}}
\def\gm{\gamma}
\def\M{{\cal M}}
\def\ep{\epsilon}
\def\epm1{\frac{1}{\epsilon}}
\def\epm2{\frac{1}{\epsilon^{2}}}
\def\epm3{\frac{1}{\epsilon^{3}}}
\def\epm4{\frac{1}{\epsilon^{4}}}
\def\unM{\hat{\cal M}}
\def\ashat{\hat{a}_{s}}
\def\asmur{a_{s}^{2}(\mu_{R}^{2})}
\def\sigbar{{{\overline {\sigma}}}\left(a_{s}(\mu_{R}^{2}), L\left(\mu_{R}^{2}, m_{H}^{2}\right)\right)}
\def\sigbarn{{{{\overline \sigma}}_{n}\left(a_{s}(\mu_{R}^{2}) L\left(\mu_{R}^{2}, m_{H}^{2}\right)\right)}}
\def\sigh{\hat{\sigma}}
\def\unas{ \left( \frac{\hat{a}_s}{\mu_0^{\epsilon}} S_{\epsilon} \right) }
\def\rnM{{\cal M}}
\def\bt{\beta}
\def\cD{{\cal D}}
\def\cC{{\cal C}}
\def\ca{\text{\tiny C}_\text{\tiny A}}
\def\cf{\text{\tiny C}_\text{\tiny F}}
\def\ct{{\red []}}
\def\sv{\text{SV}}
\def\murOmu{\left( \frac{\mu_{R}^{2}}{\mu^{2}} \right)}
\def\bb{b{\bar{b}}}
\def\bt0{\beta_{0}}
\def\bt1{\beta_{1}}
\def\bt2{\beta_{2}}
\def\bt3{\beta_{3}}
\def\gm0{\gamma_{0}}
\def\gm1{\gamma_{1}}
\def\gm2{\gamma_{2}}
\def\gm3{\gamma_{3}}
\def\l{\left}
\def\r{\right}

\newcommand{\dis}{}
\newcommand{\overbar}[1]{mkern-1.5mu\overline{\mkern-1.5mu#1\mkern-1.5mu}\mkern
1.5mu}

\newcommand{\nn}{\nonumber\\}
\newcommand{\be}{\begin{equation}}
\newcommand{\ee}{\end{equation}}
\newcommand{\bea}{\begin{eqnarray}}
\newcommand{\eea}{\end{eqnarray}}


\section{Introduction}
\label{sec:intro}

At hadron colliders, the production of a pair of leptons from the decay of electroweak gauge boson is not only a  clean process 
but also it is immensely important for physics studies at the LHC~\cite{CMS:2014jea, Aad:2016zzw}.  The
experimental signature involves two high $p_T$ leptons as a result of a
neutral gauge boson decay or a single high $p_T$ lepton and missing transverse
energy in the case of charged counterpart.  The parton model ideas
intended for the deep-inelastic scattering of lepton-proton were formally extended
to the proton-proton collisions to produce a pair of leptons (Drell-Yan (DY) process)~\cite{Drell:1970wh}.

The massive electroweak gauge bosons ($W^{\pm}$ and $Z$) were subsequently discovered using this
production process.  The high production rate and clean experimental
final state make the DY process a very important experimental tool and
can be used to determine electroweak model parameters.   For example, measurements
of the $W$ boson production at the Tevatron~\cite{Aaltonen:2013iut} lead to an accurate determination of the $W$ mass and width.
DY processes play an important role in constraining the parton distribution functions (PDF)~\cite{Harland-Lang:2014zoa,Ball:2014uwa,Dulat:2015mca} of the proton 
and also serve as luminosity
monitor of hadron collider.

While Run-I at the LHC culminated in the discovery of the Higgs boson~\cite{Aad:2012tfa, Chatrchyan:2012xdj},
Run-II is currently in operation and the SM is being scrutinised
at unprecedented levels
of precisions. To fully benefit from the
experimental program at the LHC, precise
theoretical predictions for both signals of new physics and SM background are very
essential. The leading order (LO) predictions are often very crude at the colliders due to missing higher order effects and the presence of unphysical scales resulting from ultraviolet renormalisation and mass factorisation. In addition, the choice of PDFs also influence the predictions. Hence, the predictions based on LO results are unreliable and they cannot constrain the model parameters stringently. We must go beyond LO.
The dominant next-to-leading order (NLO) corrections to the LO
DY result come from QCD and are large at LHC energies. In
addition, an estimate of the theoretical uncertainties due to truncation
of the perturbative expansion in the strong coupling constant, $a_s$, reduces
on the inclusion of the higher order terms in $a_s$.  For the
DY process, the next-to-next-to-leading order (NNLO) corrections in QCD
are available for inclusive cross section \cite{Hamberg:1990np}, rapidity
 distributions \cite{Anastasiou:2003yy, Anastasiou:2003ds}, 
fully exclusive distributions including $\gamma$-$Z$ interference, the leptonic
decay of gauge bosons and finite width effects are also included
\cite{Melnikov:2006di, Melnikov:2006kv,Catani:2009sm}.  The current accuracy 
of the DY process is next-to-next-to-next-to-leading order (N$^3$LO) corrections to the production cross section near
the partonic threshold \cite{Ahmed:2014cla,Catani:2014uta,Li:2014afw}.

Searches for physics beyond the SM involve looking for deviations from the SM predictions. 
The excess in the di-photon channel reported by the LHC collaborations~\cite{Aad:2012hf, ATLAS:2015-1410174, Chatrchyan:2012oaa, CMS:2015dxe} triggered enormous 
interest among theorists to interpret it in terms of a new resonance of mass 750 GeV.   
While several models with a new particle of mass 750 GeV explaining this excess have already 
been proposed, the conclusive and the most plausible interpretation
is possible only with more data.  Though the interpretation with a heavy spin-0 
particle could explain the excess in most of the scenarios, the data do not rule out 
the possibility of a spin-2 particle decaying into
a pair of photons.  
Massive spin-2 particles have been
phenomenologically well studied in the context of models with extra spatial dimensions
which could be flat as in the large extra dimension model, namely ADD~\cite{ArkaniHamed:1998rs, Antoniadis:1998ig, ArkaniHamed:1998nn}, or warped as in
the RS model~\cite{Randall:1999ee} or any other new physics scenario with spin-2.    
They couple to all the SM particles universally through energy-momentum tensor
of the SM.  There are also some studies with non-universal coupling 
of a spin-2 particle with the particles of the SM~\cite{Artoisenet:2013puc}.  
A generic spin-2 particle can also contribute to other production channels, namely di-lepton or di-vector boson
productions at the LHC.  
In this article, we will restrict ourselves to study the invariant mass of di-lepton pair 
in the ADD model with spin-2 particle.  The extension to other production channels is straightforward.

To match the theoretical accuracy of the SM DY process, the di-lepton
final states including a spin-2 intermediate state should also be calculated
to the same order of accuracy in QCD.  Presently for the ADD and RS
model, NLO QCD corrections are available for most of the di-final state
process with a trivial colour flow {\it viz.}: di-lepton \cite{Mathews:2004xp,
Mathews:2005zs, Kumar:2006id}, 
di-photon \cite{Kumar:2008pk, Kumar:2009nn}, 
$ZZ$ \cite{Agarwal:2009xr, Agarwal:2009zg} and 
$W^+ W^-$ \cite{Agarwal:2010sp, Agarwal:2010sn}.  In addition,
these processes have been extended to NLO+Parton Shower accuracy
\cite{Frederix:2012dp, Frederix:2013lga, Das:2014tva, Das:2016pbk}.  
These corrections are found to be large i.e K-factors are turned out to be order of 1.6. Needless to say,  
in going from 
LO to NLO the theoretical uncertainties gets reduced, but for most of
these processes the renormalisation scale $(\mu_R)$ dependence begins
at the NLO level and to compensate the $\mu_R$-dependence, going beyond
NLO is inevitable. Only at NNLO the renormalisation scale dependence starts getting compensated. Unlike the SM DY the gluon-gluon subprocess starts
at LO itself for spin-2 process and so the NLO corrections are large
where the spin-2 effects are dominant as compared to the SM.  

To go beyond NLO to NNLO, it is prudent to take incremental steps.
A general feature of the production of a large invariant mass system in
hadronic collisions is that the dominant contributions are often given
by the threshold approximation.  In \cite{deFlorian:2013wpa},
the relevant form factors such as the gluon-gluon
$\to$ spin-2 and quark-antiquark $\to$ spin-2 at  
two-loop level in QCD \cite{deFlorian:2013sza} were computed to
obtain threshold corrections at NNLO in QCD to the invariant mass distribution of di-leptons at hadron
colliders in ADD model and  
to a resonant production of a graviton in RS model.   
In \cite{Ahmed:2015qia}, three loop QCD corrections
were computed for these form factors in order to study the universal infra-red structure of
the QCD amplitudes involving spin-2 particle in the external states.  
In \cite{Ahmed:2014gla}, the
two-loop QCD corrections to the amplitudes of massive spin-2 resonance $\to$ 3 gluons
relevant for the production of a spin-2 particle plus jet were carried out.  

Going beyond threshold corrections is inevitable
in order to make accurate predictions. The contributions resulting from  
hard part of the cross section in 
quark annihilation and gluon fusion channels and those from other partonic channels 
may contribute significantly.  We will demonstrate in this article that this is indeed the
case for the invariant mass distribution of di-leptons by explicitly computing
the full NNLO QCD corrections.    
We also find that the contributions from quark-gluon initiated processes
both at NLO as well as NNLO levels are not only negative but also large, 
hence affects the threshold approximation.    
This is one of the main results of the present paper.

The paper is organised as follows: in Sec.~\ref{sec:Theory}, we introduce the effective action that describes
the interaction of the spin-2 particles with the SM fields, in particular, the part that is
relevant for our computation and then present the theoretical framework to compute the invariant mass
of the di-leptons at hadron colliders up to NNLO level in QCD.  Sec.~\ref{sec:3} is devoted to the methodology employed to 
compute all the partonic cross sections that contribute. 
In Sec.~\ref{sec:5}, we present the numerical impact of our new results.  Appendix~\ref{app:res} and \ref{app:ident} contain partonic 
coefficient functions resulting from all the channels up to NNLO level along with some useful identities
involving multiple polylogarithms.  

\section{Theoretical Framework}
\label{sec:Theory}

\subsection{The Effective Action}

In an effective theory, the spin-2 field, $h^{\mu\nu}$, couples to the
SM ones through the conserved SM energy momentum tensor,
$T^{\text{SM}}_{\mu\nu}$. The effective
action~\cite{ArkaniHamed:1998nn, ArkaniHamed:1998rs,
  Antoniadis:1998ig, Randall:1999ee} describing this 
interaction reads: 

\begin{equation}
\label{eq:1}
S= S_{\rm{SM}}+S_{h}- \frac{\kappa}{2}\int d^4x
~T_{\mu\nu}^{\rm{QCD}}(x)~h^{\mu\nu}(x) 
\end{equation}
where, $S_{\text{SM}}$ and $S_{h}$ represent the actions of the SM and
spin-2 fields, respectively. $\kappa$ is a dimensionful coupling
constant and $T_{\mu\nu}^{\rm{QCD}}$ is the conserved energy momentum
tensor of QCD which is given by  

\begin{align}
\label{eq:2}
T^{\rm{QCD}}_{\mu\nu} &= -g_{\mu\nu} {\cal L}_{\rm{QCD}} - F_{\mu\rho}^a F^{a\rho}_\nu
- \frac{1}{\xi} g_{\mu\nu} \partial^\rho(A_\rho^a\partial^\sigma A_\sigma^a)
+ \frac{1}{\xi}(A_\nu^a \partial_\mu(\partial^\sigma A_\sigma^a) + A_\mu^a\partial_\nu
(\partial^\sigma A_\sigma^a))
\nonumber\\
&+\frac{i}{4} \Big[ \overline \psi \gamma_\mu (\overrightarrow{\partial}_\nu -i g_s T^a A^a_\nu)\psi
-\overline \psi (\overleftarrow{\partial}_\nu + i g_s T^a A^a_\nu) \gamma_\mu \psi
+\overline \psi \gamma_\nu (\overrightarrow{\partial}_\mu -i g_s T^a A^a_\mu)\psi
\nonumber\\
&-\overline \psi (\overleftarrow{\partial}_\mu + i g_s T^a A^a_\mu) \gamma_\nu \psi\Big]
+\partial_\mu \overline \omega^a (\partial_\nu \omega^a - g_s f^{abc} A_\nu^c \omega^b)
\nonumber\\
&+\partial_\nu \overline \omega^a (\partial_\mu \omega^a- g_s f^{abc} A_\mu^c \omega^b).
\end{align}
$g_{s}$ is the strong coupling constant and $\xi$ is the gauge fixing
parameter which is set to 1 for working in Feynman gauge. $T^{a}$ and
$f^{abc}$ are the Gell-Mann matrices and structure constants of SU(N)
gauge theory, respectively. Presence of the ghost fields, $\omega^{a}$,
in the interaction part of the action is a reflection of the fact
that spin-2 fields couple to everything democratically. In the action,
Eq.~(\ref{eq:1}), we have presented only the QCD interaction term as we
are interested only in this regime.


\subsection{Invariant Lepton Pair Mass Distribution $d\sigma/dQ^2$}
\label{ss:inv}

We consider the production of a leptonic pair, $l^{+}$ and $l^{-}$,
through the scattering of two hadrons, represented by $H_{1}$ and
$H_{2}$:
\begin{equation}
\label{eq:3}
H_{1}(P_{1})+ H_{2}(P_{2})  \rightarrow l^{+}(l_{1}) + l^{-}(l_{2}) + X(P_{X})\,.
\end{equation}
$X$ denotes the final inclusive state. The terms inside the
parentheses represent the 4-momenta of the corresponding particles. In the
QCD improved parton model, the hadronic cross section is related to
the partonic one through 
\begin{align}
\label{eq:4}
2S \frac{d\sigma^{H_1 H_2}}{dQ^2} \big(\tau, Q^2\big)  = \sum_{ab =
  q,\bar{q},g} \int_{0}^{1}dx_1  
\int_{0}^{1}dx_2  f_{a}^{H_1}(x_1)  f_{b}^{H_2} (x_2)  
\nonumber\\
\times \int_{0}^{1}dz \, 2 \hat{s}\frac{d\hat{\sigma}^{ab}}{dQ^2}\big(z,
  Q^2\big)\delta(\tau - zx_1x_2)\,. 
\end{align}
In the above expression, $S$ is the square of the hadronic center of
mass energy which is related to the partonic one, ${\hat s}$, through
${\hat s}=x_{1}x_{2}S$. The invariant mass square of the final state
leptonic pair, $m^{2}_{l^{+}l^{-}}$ is represented through
$Q^{2}$. $f_{a}$ and $f_{b}$ are the partonic distribution functions of the
initial state partons $a$ and $b$, respectively. The other parameters
are defined as
\begin{align}
\label{eq:5}
\tau \equiv \frac{Q^2}{S},   \quad   z \equiv \frac{Q^2}{\hat{s}}
  \quad\text{and}\quad  \tau = x_1x_2z\,.
\end{align}   
The underlying partonic process corresponding to the hadronic
one~(\ref{eq:3}) is $$a(p_{1})+b(p_{2})
\rightarrow j(q) + \sum\limits_{i=1}^{m} X_{i}(q_{i}) \rightarrow
l^{+}(l_{1}) + l^{-}(l_{2}) + \sum\limits_{i=1}^{m} X_{i}(q_{i})$$
where, $j$ can be photon ($\gamma^{*}$), 
Z-boson ($Z$) or spin-2 particle. $X_{i}$ stands for the real QCD hard radiations
from the initial state partons $a$ and $b$. In perturbative quantum
field theory (pQFT), the cross section for the Drell-Yan
process can be factored out into partonic ($ab\rightarrow j$) and
leptonic ($j \rightarrow l^{+}l^{-}$) parts:
\begin{align}
\label{eq:7}
2 \hat{s}\frac{d{\hat \sigma}^{ab}}{dQ^2} =& 
 \frac{1}{2\pi} \sum\limits_{j,j'=\gamma^{*},Z,h} \int dPS_{m+1} |{\cal
  M}^{ab \rightarrow jj'}|^{2} \cdot P_{j}(q) \cdot P_{j'}^{*}(q)
  \,\cdot{\cal L}^{jj' \rightarrow l^{+}l^{-}}(q) 
\end{align}
where, the $m+1$ body phase space in $n$-dimensions is defined as 
\begin{align}
\label{eq:7a}
\int dPS_{m+1} =&  \int
  \prod_{i=1}^{m} \left( \frac{d^{n}q_{i}}{(2\pi)^{n}}
  2\pi\delta^{+}(q_{i}^{2})\right) 
\times 
\frac{d^{n}q}{(2\pi)^{n}}2\pi\delta^{+}(q^{2}-Q^{2})
\nonumber\\
&
\times(2\pi)^{n}
  \delta^{(n)}\left(p_{1}+p_{2}-q-\sum\limits_{i=1}^{m}q_{i}\right)\,.
\end{align}
and the quantity ${\cal L}^{jj'\rightarrow l^{+}l^{-}}$ is given by
\begin{align}
\label{eq:8}
{\cal L}^{jj'\rightarrow l^{+}l^{-}}(q) = \prod_{i=1}^{2} \left( \frac{d^{n}l_{i}}{(2\pi)^{n}}
  2\pi\delta^{+}(l_{i}^{2})\right) \times (2\pi)^{n}
  \delta^{n}(q-l_{1}-l_{2}) |{\cal M}^{jj'\rightarrow l^{+}l^{-}}|^{2}\,.
\end{align}
${\cal
  M}^{ab \rightarrow jj'}$ and ${\cal
  M}^{jj' \rightarrow l^{+}l^{-}}$ are the partonic and leptonic part
of the matrix elements, respectively. $j\neq j'$ reflects the
interference terms between the channels $j$ and $j'$. In the above
Eq.~(\ref{eq:7}), the sum over Lorentz indices 
between matrix element squared and the propagators is implicit through
a symbol `dot product'. The propagators are
\begin{align}
\label{eq:9}
&P_{\gamma,\mu\nu} (q) = -\frac{i}{Q^2} \eta_{\mu\nu} \equiv  \eta_{\mu\nu}
                 \tilde{P_{\gamma}} ( Q^2), 
\nonumber\\
&P_{Z,\mu\nu} (q) = - \frac{i}{( Q^2 - M_{Z}^2 - i M_{Z}
                      \Gamma_{Z} ) } \eta_{\mu\nu} \equiv \eta_{\mu\nu}
                      \tilde{P_{Z}} ( Q^2), 
\nonumber\\
&P_{h,\mu\nu\rho\sigma}(q) =  {\cal D}(Q^2) B_{\mu\nu\rho\sigma}(q) \equiv
           B_{\mu\nu\rho\sigma}(q) \tilde{P_{h}} ( Q^2) 
\end{align} 
where 
\begin{align}
\label{eq:10}
B_{\mu\nu\rho\sigma}(q) =& \left( \eta_{\mu\rho} -
  \frac{q_{\mu}q_{\rho}}{q.q} \right) \left( \eta_{\nu\sigma} -
  \frac{q_{\nu}q_{\sigma}}{q.q} \right) +
\left( \eta_{\mu\sigma} -
  \frac{q_{\mu}q_{\sigma}}{q.q} \right) \left( \eta_{\nu\rho} -
  \frac{q_{\nu}q_{\rho}}{q.q} \right)
\nonumber\\
&- \frac{2}{n -1}  \left( \eta_{\mu\nu} -
  \frac{q_{\mu}q_{\nu}}{q.q} \right) \left( \eta_{\rho\sigma} -
  \frac{q_{\rho}q_{\sigma}}{q.q} \right)\,,
\end{align}
$\eta_{\mu\nu}=\text{diag}[1,-1,-1,-1,\cdots]$ and ${\cal D}(Q^{2})$, the summation over the virtual Kaluza-Klein
(KK) modes in the time like propagators~\cite{Han:1998sg} in $(4+d)$-dimensions, is
\begin{align}
\label{eq:6}
{\cal D}(Q^2) = 16 \pi \l(\frac{Q^{d-2}}{\kappa^2 M_s^{d+2}}\r) I \l( \frac{M_s}{Q}\r).
\end{align}
The integral $I$ is regulated presumably by a cutoff of the order of
$M_{S}$ in ultraviolet (UV) region~\cite{Han:1998sg}. This cutoff sets the limit on the
applicability of the effective theory. For the DY process, this implies
$Q<M_{S}$. The quantity $q^{2}=Q^{2}=m_{l^{+}l^{-}}^{2}$.

Hence, the computation of the partonic level cross section boils down
to the evaluation of partonic and leptonic parts. The leptonic part
comes out to be
\begin{align}
\label{eq:11}
{\cal L}^{ j j' \rightarrow l^+ l^- }(q)  &=  g_{\mu\nu} (q)
                                            L_{jj'}(Q^2) , \qquad
                                            jj' = \{\gamma\gamma, ZZ,
                                            \gamma Z\}\,,
\nonumber\\
{\cal L}^{ hh \rightarrow l^+ l^- }(q)  &= B_{\mu\nu\rho\sigma}(q)
                                           {L}_{hh}(Q^2)\,,
\end{align}
where 
\begin{align}
&L_{hh} (Q^2)   =   Q^4 \frac{\kappa^2}{640 \pi},   \qquad\qquad
  L_{ZZ}(Q^2)  =  Q^2 \frac{2\alpha }{3 c_w^2 s_w^2 } 
\left( (g_{e}^{V})^2 + (g_{e}^{A})^2\right)\,,
\nonumber\\
&L_{\gamma Z} (Q^2)  =  -Q^2 \frac{2\alpha g_{e}^{V}}{3 c_w s_w },
  \qquad~ L_{\gamma\gamma}(Q^2) =  Q^2 \frac{2\alpha}{3}\,,
\nonumber\\
\text{and}~~&g_{\mu\nu}(q)\equiv\eta_{\mu\nu}-\frac{q_{\mu}q_{\nu}}{q.q}\,.
\end{align}
In the above equation, $\alpha$ is the fine structure constant,
$c_{W}\equiv\cos\theta_{W}, s_{W}\equiv\sin\theta_{W}$ and $\theta_{w}$
is the weak mixing angle. 
$ g_{f}^{V}$ and $g_{f}^{A}$ can be expressed in terms of charge $Q_{f}$ of the
fermions ($f$) i.e. quarks, leptons and weak isospin $ T_f^3$:
\begin{equation}
\label{eq:12}
g_{f}^{V}  =  \frac{1}{2} T_f^3 - s_w^2 Q_f,  \quad   g_{f}^{A}  = - \frac{1}{2} T_f^3\,.
\end{equation}
Hence, the hadronic cross section~(\ref{eq:4}) can be rewritten as
\begin{align}
\label{eq:13}
2S\frac{d\sigma^{H_1 H_2}}{dQ^2} = \frac{1}{2\pi} \sum_{j,j' =\{
  \gamma^*,Z,h\}} \tilde{P_j}(Q^2) \,  \tilde{P}_{j'}^*(Q^2) L _{ j j'}
  (Q^2) W_{jj'}^{H_1 H_2}( \tau, Q^2) 
\end{align}
where, the hadronic structure function $W$ is 
\begin{align}
\label{eq:14}
W_{j j'}^{H_1 H_2}( \tau, Q^2)   =&  \sum_{a,b,j,j'} \int dx_1 \int
                                   dx_2 {f_a}^{H_1}(x_1)
                                   {f_b}^{H_2}(x_2) \int dz \delta (\tau - zx_1x_2)
\nonumber\\
& \times \int dPS_{m+1} |{\cal M}^{ab\rightarrow jj'}|^{2} T_{jj'}(q)
\end{align}
with
\begin{align}
\label{eq:15}
T _ {j j'}(q) = 
\begin{cases}
g_{\mu\nu}(q), \qquad \quad\, j j' = \gamma\gamma, \gamma\,
  Z, ZZ
\\
B_{\mu\nu\rho\sigma} (q), \qquad j j' = hh\,.
\end{cases}
\end{align}
As a consequence, the computation of the $Q^{2}$ distribution of the di-lepton pairs
requires the evaluation of the integrals in a suitable frame over $dPS_{m+1}$
and $dz$ after substituting the matrix element squared $|{\cal
  M}^{ab\rightarrow jj'}|^{2} T_{jj'}(q)$ in
Eq.~(\ref{eq:14}). We define the bare partonic coefficient function
${\hat \Delta}^{jj'}_{ab} \left( z, Q^2, 1/\epsilon \right)$ through
\begin{align}
\label{eq:33}
{\hat \Delta}^{jj'}_{ab} \left( z, Q^2, 1/\epsilon \right)
&= C_{jj'} \int dPS_{m+1} |{\cal M}^{ab \rightarrow jj'}|^2 T_{jj'}(q)
\end{align}
where
\begin{align}
\label{eq:34}
C_{jj'} = 
\begin{cases}
\frac{1}{e^2} \qquad\quad jj' = \gamma\gamma, ZZ, \gamma Z\,,
\\
\frac{1}{Q^2 \kappa^2} \qquad jj'= hh\,.
\end{cases}
\end{align}
The partonic cross section receives contributions from two different
classes of processes: first one happens through a virtual photon or a
$Z$-boson whereas the second one contains a spin-2 particle in the
intermediate state. Interestingly, on performing the phase space integration, the interference
term between the two classes of diagrams up to NNLO identically vanishes,
this was earlier noted to NLO~\cite{Mathews:2004xp}. The underlying reason behind the
vanishing of this interference term is also explained in that article~\cite{Mathews:2004xp}.
Hence, our result does not receive any contribution
from the interference terms. 

\begin{figure}[h]
\qquad\quad
  \begin{minipage}{3in}
\begin{tikzpicture}[line width=0.6 pt, scale=0.7]
 \draw[fermion] (-2,1.5) -- (0,0);
 \draw[fermionbar] (-2,-1.5) -- (0,0);
 \draw[scalarnoarrow] (0,0) -- (2.5,0); 
 \node at (3.8,0) {$\gamma^{*}/Z/h$};
\end{tikzpicture}
  \end{minipage}
  \qquad
  \begin{minipage}{3in}
\begin{tikzpicture}[line width=0.6 pt, scale=0.7]
 \draw[gluon] (-2,1.5) -- (0,0);
 \draw[gluon] (-2,-1.5) -- (0,0);
 \draw[scalarnoarrow] (0,0) -- (2.5,0);
 \node at (3,0) {$h$}; 
\end{tikzpicture}
  \end{minipage}
\caption{Leading order processes for the DY}
  \label{fig:1}
\end{figure}
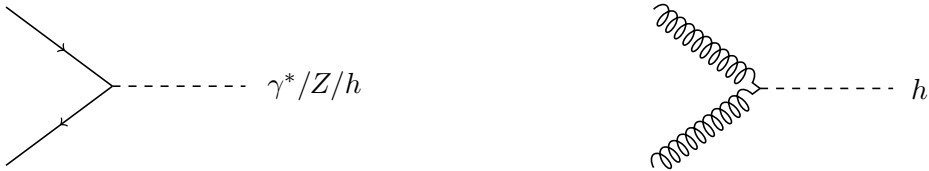

In case of spin-2 appearing as an intermediate state, at
leading order (LO) we can have
gluon initiated process as well, in addition to the quark initiated
one. Upon inclusion of the spin-2 contribution, at the 
LO we have (See Fig.~\ref{fig:1})  
\begin{align}
\label{eq:16}
q+{\bar q} \rightarrow \gamma^{*}/Z/h\,, \quad g+g \rightarrow h\,.
\end{align}
Beyond LO, the contributions arise from virtual as well as real
emission diagrams. At next-to-leading order (NLO), we get contributions
from pure virtual and pure real emission diagrams. On the other hand, in
addition to these two types of contributions, virtual-real diagrams also
contribute at next-to-next-to-leading order (NNLO)

The evaluation of the virtual as well as real emission diagrams
exhibits divergences of three kinds: ultraviolet (UV), soft and
collinear (IR). In particular, diagrams involving virtual particles give
rise to all the above-mentioned divergences whereas the real emissions
cause only soft and collinear ones. We regulate the UV as well as IR
divergences using dimensional regularisation where the space-time
dimensions $n$ is chosen to be equal to $4+\epsilon$. All the
divergences manifest themselves as the poles in dimensional
regularisation parameter $\epsilon$: $1/\epsilon^{\alpha}$ with
$\alpha \in [1,4]$. In ${\overline {\text{MS}}}$, the UV poles are
removed through strong coupling 
constant renormalisation using
\begin{align}
\label{eq:21}
{\hat a}_{s} S_{\epsilon} = \left( \frac{\mu^{2}}{\mu_{R}^{2}}  \right)^{\epsilon/2}
  Z_{a_{s}} a_{s}
\end{align}
where,
\begin{align}
\label{eq:22}
&S_{\epsilon} = {\rm exp} \left[ (\gamma_{E} - \ln 4\pi)\epsilon/2
\right]\,, \qquad \gamma_{E} = 0.5772\ldots\,,
\nonumber\\
&Z_{a_{s}} = 1+ a_s\left[\frac{2}{\epsilon} \beta_0\right]
             + a_s^2 \left[\frac{4}{\epsilon^2 } \beta_0^2
             + \frac{1}{\epsilon}  \beta_1 \right]
             + a_s^3 \left[\frac{8}{ \epsilon^3} \beta_0^3
             +\frac{14}{3 \epsilon^2}  \beta_0 \beta_1 +  \frac{2}{3
             \epsilon}   \beta_2 \right]\,,
\nonumber\\
&a_{s} \equiv a_{s}(\mu_{R}^{2}) \equiv \frac{g_{s}^{2}}{16\pi^{2}}
\end{align}  
and $\mu$ is the scale introduced to keep the unrenormalised strong
coupling constant ${\hat a}_{s}$ dimensionless in $n$-dimensions. The
corresponding renormalisation scale is denoted by
$\mu_{R}$. $\beta_{i}$'s are the coefficients of QCD
$\beta$-function~\cite{Gross:1973id, Politzer:1973fx, Caswell:1974gg, Tarasov:2013zv, Larin:1993tp}. Since, the spin-2 particles couple to the SM ones
through the conserved energy momentum
tensor, the universal gravitational coupling constant $\kappa$ is protected from any ultraviolet (UV)
renormalisation. Hence, there is no additional UV renormalisation
required other than the strong coupling constant renormalisation. The
soft divergences arising from virtual diagrams cancel exactly against the same
coming from real emission ones, thanks to the Kinoshita-Lee-Nauenberg (KLN) theorem~\cite{Kinoshita:1962ur, Lee:1964is}. The
collinear divergences are removed through mass factorisation,
performed at the factorisation scale $\mu_{F}$:
\begin{align}
\label{eq:23}
{\hat \Delta}^i_{ab}(z,Q^2,1/\epsilon) =
  \sum_{c,d=q,{\bar q}, g}\Gamma_{ca}(z,\mu_F^2,1/\epsilon) \otimes
  \Gamma_{db}(z,\mu_F^2,1/\epsilon)\otimes 
  \Delta_{cd}^i(z,Q^2,\mu_{F}^2)\,.
\end{align}
In the above expression, ${\hat \Delta} \equiv {\hat \sigma}/z$ is the
bare partonic coefficient function and the corresponding one after performing the mass
factorisation is denoted by $\Delta$. Further we have dropped the
double index $jj'$ from the partonic coefficient function (see
Eq.~(\ref{eq:33})) because of the vanishing interference terms between
the two classes of diagrams and
replace it by the single index $i$ instead. The mass factorisation kernel in the modified minimal subtraction
$(\overline{\text{MS}})$ scheme is given by
\begin{align}
\label{eq:24}
&\Gamma_{ab}(z, \mu_{F}^{2},1/\epsilon) = \sum\limits_{k=0}^{\infty}
  a_{s}^{k}(\mu_{F}^{2})\Gamma^{(k)}_{ab}(z, \mu_{F}^{2},1/\epsilon) \nonumber
\intertext{with}
&\Gamma_{ab}^{(0)} = \delta_{ab} \delta(1-z)\,,
\nonumber\\
&\Gamma_{ab}^{(1)} = \frac{1}{\epsilon}  P^{(0)}_{ab}(z)\,,
\nonumber\\
&\Gamma_{ab}^{(2)} = \frac{1}{\epsilon^{2}} \Bigg( \frac{1}{2} P^{(0)}_{ac} \otimes
  P^{(0)}_{cb} + \beta_{0} P^{(0)}_{ab}\Bigg) + \frac{1}{\epsilon}
  \Bigg( \frac{1}{2} P^{(1)}_{ab}\Bigg)\,. 
\end{align}
$P^{(i)}_{ab}$ are the Altarelli-Parisi splitting
functions~\cite{Altarelli:1977zs, Floratos:1980hm, Floratos:1980hk, Curci:1980uw, Vogt:2004gi}. The symbol $\otimes$ stands for the
convolution:
\begin{align}
\label{eq:25}
\left( f\otimes g \right)(z) \equiv \int\limits_z^1 \frac{dx}{x} f(x)
  g\left( \frac{z}{x} \right)\,.
\end{align}
Expanding the unrenormalised coefficient function in Eq.~(\ref{eq:33})
and the mass factorised one in Eq.~(\ref{eq:23}) in powers of strong
coupling constant as
\begin{align}
\label{eq:35}
{\hat \Delta}^{i}_{ab} &= \sum\limits_{k=0}^{\infty} {\hat a}_s^k
                         S_{\epsilon}^k \l(\frac{Q^2}{\mu^2}\r)^{k
                         \frac{\epsilon}{2}} {\hat
                         \Delta}^{i,(k)}_{ab}\,,
\nonumber\\
\Delta^{i}_{ab} &= \sum\limits_{k=0}^{\infty} a_{s}^{k}(\mu_{F}^{2})
  \Delta^{i, (k)}_{ab}
\end{align}
and using the Eq.~(\ref{eq:24}), we can get all the contributions to
NNLO arising from all the subprocesses $ \Delta^{i, (k)}_{ab}$.
From the results of the bare coefficient functions and the known
splitting functions, we can obtain the finite $\Delta^{i}_{ab}$. 
This in turn gives us the $Q^{2}$
distribution of the leptonic pair in the DY process:
\begin{align}
2 S{d \sigma^{H_1H_2} \over dQ^2}(\tau,Q^2)&=
2 S{d \sigma^{H_1H_2}_{\rm SM} \over dQ^2}(\tau,Q^2)
\nonumber\\
&+\sum{\cal F}_{h} \int_0^1 {d x_1 } \int_0^1 
{dx_2} \int_0^1 dz \delta(\tau-z x_1 x_2)
\nonumber\\&
\times \Bigg[ 
H_{q{\bar q}} 
             \sum\limits_{k=0}^{2} a_{s}^{k} \Delta^{h, (k)}_{q{\bar q}} 
+
H_{g g} \sum\limits_{k=0}^{2} a_{s}^{k} \Delta^{h, (k)}_{gg} 
+ \Big( H_{gq} + H_{qg}  \Big)
             \sum\limits_{k=1}^{2} a_{s}^{k} \Delta^{h, (k)}_{gq}
\nonumber\\&+
H_{q q} \sum\limits_{k=2}^{2}
             a_{s}^{k} \Delta^{h, (k)}_{qq}  
+ H_{q_{1} q_{2}}  \sum\limits_{k=2}^{2}
             a_{s}^{k} \Delta^{h, (k)}_{q_{1}q_{2}}  
\Bigg]\,.
\end{align}
In the above expression
\begin{align}
\label{eq:31}
{\cal F}_{h}=\;&{\kappa^4 Q^6 \over 320 \pi^2 }|{\cal D}(Q^2)|^2\,,
\nonumber\\
\Delta^{i, (k)}_{ab} ~&= \Delta^{i, (k)}_{ab}(z, \mu_{F}^{2})
\end{align}
and the renormalised partonic distributions are
\begin{align}
\label{eq:32}
H_{q \bar q}(x_1,x_2,\mu_F^2)&=
f_q^{H_1}(x_1,\mu_F^2) 
f_{\bar q}^{H_2}(x_2,\mu_F^2)
+f_{\bar q}^{H_1}(x_1,\mu_F^2)~ 
f_q^{H_2}(x_2,\mu_F^2)\,,
\nonumber\\
H_{q q}(x_1,x_2,\mu_F^2)&=
f_q^{H_1}(x_1,\mu_F^2) 
f_{q}^{H_2}(x_2,\mu_F^2)
+f_{\bar q}^{H_1}(x_1,\mu_F^2)~ 
f_{\bar q}^{H_2}(x_2,\mu_F^2)\,,
\nonumber\\
H_{q_1 q_2}(x_1,x_2,\mu_F^2)&=
f_{q_1}^{H_1}(x_1,\mu_F^2) 
\Big( f_{q_2}^{H_2}(x_2,\mu_F^2) + f_{\bar q_2}^{H_2}(x_2,\mu_F^2) \Big)
\nonumber\\&+f_{\bar q_1}^{H_1}(x_1,\mu_F^2)~ 
\Big( f_{q_2}^{H_2}(x_2,\mu_F^2) + f_{\bar q_2}^{H_2}(x_2,\mu_F^2) \Big)\,,
\nonumber\\
H_{g q}(x_1,x_2,\mu_F^2)&=
f_g^{H_1}(x_1,\mu_F^2) 
\Big(f_q^{H_2}(x_2,\mu_F^2)
+f_{\bar q}^{H_2}(x_2,\mu_F^2)\Big)\,,
\nonumber\\
H_{q g}(x_1,x_2,\mu_F^2)&=
H_{g q}(x_2,x_1,\mu_F^2)\,,
\nonumber\\
H_{g g}(x_1,x_2,\mu_F^2)&=
f_g^{H_1}(x_1,\mu_F^2)~ 
f_g^{H_2}(x_2,\mu_F^2)\,.
\end{align}
In this article, we extend this
distribution of the DY pair to NNLO QCD from the existing NLO 
result~\cite{Mathews:2004xp} in models of TeV scale gravity. The
contributions arising from solely SM
are already available in the
literature~\cite{Altarelli:1978id, Matsuura:1987wt, Matsuura:1988sm,
  Hamberg:1990np}. The missing parts, namely, the contributions coming from
the presence of the spin-2 particles, $\Delta^{h,(2)}_{ab}$ are computed in this article. In the next Sec.~\ref{sec:3}, we discuss the methodology of this
computation in great details.


\section{Methodology}
\label{sec:3}

The computation of the partonic cross section beyond leading order
consists of the evaluation of the loop integrals arising from the
virtual diagrams and the phase space integrals. The developments of
the techniques to evaluate the former one takes place quite rapidly
compared to the latter one. In the very first 
computation of the NNLO QCD correction to the DY pair production
in~\cite{Hamberg:1990np}, the phase space integrals were performed
through evaluation of the two parametric and two angular integrations
in three different frames. Later, to calculate the inclusive production
cross section of the Higgs boson three different techniques were
employed. In~\cite{Harlander:2002wh}, the partonic cross section was
obtained by performing an 
expansion around the soft limit. In the meantime a completely new and
elegant formalism was developed in~\cite{Anastasiou:2002yz} by
Anastasiou and Melnikov to get 
the same result. The phase space integrals were converted to loop
integrals by using the idea of reverse unitarity. So, the evaluation of the phase space integrals boils down to
the evaluation of the loop integrals. Hundreds of different loop
integrals were reduced to only a few number of 
master integrals (MIs) by making use of the integration-by-parts
(IBP)~\cite{Tkachov:1981wb, Chetyrkin:1981qh} and
Lorentz invariance (LI)~\cite{Gehrmann:1999as} identities. The
resultant MIs were computed using the 
techniques of differential equations to arrive at the final
result. The same result was again reproduced
in~\cite{Ravindran:2003um} using the conventional 
method of evaluating loop and phase space integrals. The method of
reverse unitarity was latter employed to obtain the 
  state-of-the-art result, namely N$^3$LO QCD corrections to the
  inclusive Higgs boson
  production~\cite{Anastasiou:2014vaa, Anastasiou:2015ema, Anastasiou:2016cez}. In
  this article, we use the formalism developed 
in~\cite{Anastasiou:2002yz} to calculate the partonic cross section of
the DY pair production 
through intermediate spin-2 particle at NNLO QCD. In this section, we
demonstrate this methodology in brief.

At this order we need to calculate three different contributions which
are mentioned in the last section:
\begin{itemize}
\item \textbf{\textit{double-real:}} the self-interference of the tree level
  amplitudes for the processes contributing to pure double-real emissions.
  For example, for the process $q+{\bar q} \rightarrow h+q+{\bar q}$, presented through the Fig.~\ref{fig:2}, we have
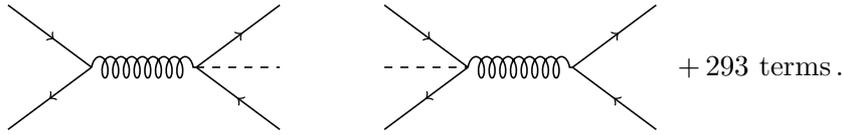
\begin{figure}[h]
\begin{center}
\begin{tikzpicture}[line width=0.6 pt, scale=0.55]
 \draw[fermion] (-2,1.5) -- (0,0);
 \draw[fermionbar] (-2,-1.5) -- (0,0);
 \draw[gluon] (0,0) -- (2.5,0);
 \draw[fermion] (2.5,0) -- (4.5,1.5);
 \draw[fermionbar] (2.5,0) -- (4.5, -1.5);
 \draw[scalarnoarrow] (2.5,0) -- (4.5,0);
 \draw[fermion] (7,1.5) -- (9,0);
 \draw[fermionbar] (7,-1.5) -- (9,0);
 \draw[gluon] (9,0) -- (11.5,0);
 \draw[fermion] (11.5,0) -- (13.5,1.5);
 \draw[fermionbar] (11.5,0) -- (13.5, -1.5);
 \draw[scalarnoarrow] (7,0) -- (9,0);
 \node at (16,0) {$+\, 293$ terms\,.};
\end{tikzpicture}
\caption{Self-interference of double-real emissions}
\label{fig:2}
\end{center}
\end{figure}
\item \textbf{\textit{real-virtual:}} the interference of the one-loop and the
  tree level amplitudes.
  For example, for
  the process $q+{\bar q} \rightarrow h+g+1$-loop, drawn in the Fig.~\ref{fig:3}, we have
\begin{figure}[H]
\begin{center}
\begin{tikzpicture}[line width=0.6 pt, scale=0.55]
 \draw[fermion] (-2,1.5) -- (0,0);
 \draw[fermionbar] (-2,-1.5) -- (0,0);
 \draw[gluon] (0,0) -- (1,0);
 \draw[gluon] (2,0) circle (1);
 \draw[gluon] (3,0) -- (4.5,1.5);
 \draw[scalarnoarrow] (3,0) -- (4.5,-1.5);
 \draw[gluon] (7,1.5) -- (9,0);
 \draw[scalarnoarrow] (7,-1.5) -- (9,0);
 \draw[gluon] (9,0) -- (11,0);
 \draw[fermion] (11,0) -- (13,1.5);
 \draw[fermionbar] (11,0) -- (13,-1.5);
 \node at (15.5,0) {$+\, 171$ terms\,.};
\end{tikzpicture}
\caption{Interference of real-virtual with single real emission}
\label{fig:3}
\end{center}
\end{figure}
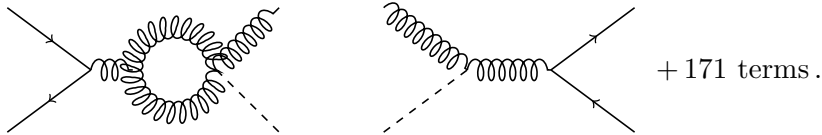
\item \textbf{\textit{double-virtual:}} the interference of the two loop and
  the tree level amplitudes.
  For example,
  for the process $q+{\bar q} \rightarrow h+2$-loop, represented through the Fig.~\ref{fig:4}, we have
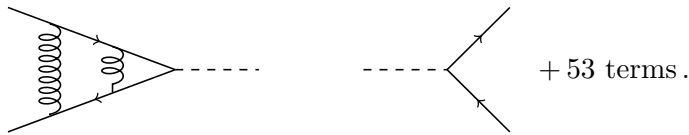
\begin{figure}[H]
\begin{center}
\begin{tikzpicture}[line width=0.6 pt, scale=0.55]
 \draw[fermion] (-4,1.5) -- (0,0);
 \draw[fermionbar] (-4,-1.5) -- (0,0);
 \draw[gluon] (-3,1.1) -- (-3,-1.1);
 \draw[gluon] (-1.5,0.55) -- (-1.5,-0.55);
 \draw[scalarnoarrow] (0,0) -- (2,0);
 \draw[scalarnoarrow] (4.5,0) -- (6.5,0);
 \draw[fermion] (6.5,0) -- (8,1.5);
 \draw[fermionbar] (6.5,0) -- (8, -1.5);
 \node at (10.5,0) {$+\, 53$ terms\,.};
\end{tikzpicture}
\caption{Interference of two loop with Born}
\label{fig:4}
\end{center}
\end{figure}
\end{itemize}
In addition, this contribution also arises from the square of one loop diagrams.

All the required Feynman diagrams are generated symbolically using
computer package 
QGRAF~\cite{Nogueira:1991ex}. The raw output is converted to a
suitable format using in-house code written in
FORM~\cite{Vermaseren:2000nd, Tentyukov:2007mu} for our further
computation. Below, we describe the methodology to evaluate the 
above three categories. However, since all of them follow very
similar techniques, we discuss only the evaluation of double-real
diagram in brief. 
%

\begin{figure}[h]
\begin{center}
\begin{tikzpicture}[line width=0.6 pt, scale=0.55]
 \draw[fermion] (-2,1.5) -- (0,0);
 \draw[fermionbar] (-2,-1.5) -- (0,0);
 \draw[gluon] (0,0) -- (2.5,0);
 \draw[scalarnoarrow] (2.5,0) -- (5.3,0);
 \draw[fermion] (2.5,0) arc (180:0:1.4);
 \draw[fermionbar] (2.5,0) arc (-180:0:1.4);
 \draw[gluon] (5.3,0) -- (7.8,0);
 \draw[fermion] (7.8,0) -- (9.8,1.5);
 \draw[fermionbar] (7.8,0) -- (9.8,-1.5);
 \node at (-2.5,1.5) {$p_{1}$};
 \node at (-2.5,-1.5) {$p_{2}$};
 \node at (10.3,1.5) {$p_{1}$};
 \node at (10.3,-1.5) {$p_{2}$};
 \draw[scalarnoarrow, line width=0.5mm, blue] (3.9,2.5) -- (3.9, -2.5); 
 \end{tikzpicture}
\caption{Effective two loop diagram with three cut propagators}
\label{fig:6}
\end{center}
\end{figure}
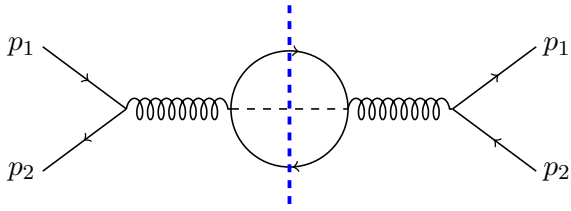

We take a sample double real emission diagram for illustrating the
methodology~\cite{Anastasiou:2002yz, Baikov:2000jg} to handle phase
space integrals: $q+{\bar q} \rightarrow h + q + {\bar q}$
\begin{figure}[H]
\begin{center}
\begin{tikzpicture}[line width=0.6 pt, scale=0.55]
 \draw[fermion] (-2,1.5) -- (0,0);
 \draw[fermionbar] (-2,-1.5) -- (0,0);
 \draw[gluon] (0,0) -- (2.5,0);
 \draw[fermion] (2.5,0) -- (4.5,1.5);
 \draw[fermionbar] (2.5,0) -- (4.5, -1.5);
 \draw[scalarnoarrow] (2.5,0) -- (4.5,0);
 \node at (-2.5,1.5) {$p_{1}$};
 \node at (-2.5,-1.5) {$p_{2}$};
 \node at (5,1.5) {$q_{1}$};
 \node at (5,-1.5) {$q_{2}$};
 \node at (5,0) {$q$};
 \draw (-3,1.7) -- (-3,-1.7);
 \draw (5.5,1.7) -- (5.5,-1.7);
 \node at (5.8,1.7) {$2$};
 \node at (14, 0) {$\propto\;\; {\mathlarger{\int}} \frac{d^{n}q_{1}}{(2\pi)^{n-1}}
   \frac{d^{n}q_{2}}{(2\pi)^{n-1}}  
     \delta_{+}(q_{1}^{2}) \delta_{+}(q_{2}^{2}) \delta_{+}(q^{2}-m_{h}^{2}) [\cdots]$};
\end{tikzpicture}
\caption{Self-interference of double-real emissions}
\label{fig:5}
\end{center}
\end{figure}
\noindent
where, $\delta_{+}(q^{2}-m^{2}) \equiv \delta(q^{2}-m^{2})
\theta(q^{0})$. According to Cutkosky rules~\cite{Cutkosky:1960sp},
the $\delta_{+}$ functions can be replaced by the difference between
two propagators with opposite prescriptions for their imaginary parts:
\begin{align}
\label{eq:28}
\delta_{+}(q^{2}-m^{2}) \sim \frac{1}{q^{2}-m^{2}+i\varepsilon} -
  \frac{1}{q^{2}-m^{2}-i\varepsilon} 
\end{align}
with $\varepsilon \rightarrow 0$. Upon this substitution, the 
square of the diagram, depicted through Fig.~\ref{fig:5} becomes equivalent to the forward scattering
amplitude, presented in Fig.~\ref{fig:6}, where, the blue dotted line denotes the cut propagators which should
be replaced by the RHS of Eq.~(\ref{eq:28}). 

We begin our computation
by evaluating the normal Born square of the above diagram (5-external
onshell legs) where the sum over colors and spins are performed. With the
final answer, we multiply the phase space factor which contains the
three $\delta_{+}$ functions corresponding to the final state
particles. Moreover, to convert it
into a cut two-loop Feynman diagram through the application of reverse
unitarity, we replace the $\delta_{+}$ functions by the difference of
the 
corresponding propagators using Eq.~(\ref{eq:28}). As a consequence,
the phase space integral can now be handled in the same way as the multiloop
integrals. We make use of the IBP and LI identities to reduce this two
loop diagram into a set of MIs. Since, the sign of the imaginary parts
of the cut propagators are irrelevant for the above identities,
the two terms of those propagators which are differed by the different
prescriptions of the 
imaginary parts give rise to same IBP relations. Each of these two
terms have the same form of the IBP relations as the original two-loop
integral without the cut. Hence, instead of considering the two terms,
we can take only one term. This is equivalent to substituting the
$\delta_{+}$ functions by its first
propagator from the RHS of Eq.~(\ref{eq:28}). Once the reduction is
done, we must put those MIs to zero which do
not contain any of the three cut propagators. In other words, the MIs
which contain all the three cut propagators are the only ones to
contribute to the original phase space integrals owing to the
Eq.~(\ref{eq:28}). While performing the reduction using the
Mathematica based 
package LiteRed~\cite{Lee:2012cn, Lee:2013mka}, we make sure not to
apply any transformation on the momenta of the cut propagators which
essentially helps to keep intact the cut propagators in its original
form even in the MIs. At the end, the $\delta_{+}$ functions need to be reinstated in
place of all the cut propagators which leads us to the final set of
phase space MIs. These integrals are identified with the ones
appearing as phase space MIs for the evaluation of the NNLO QCD
correction to the inclusive production cross section of the Higgs
boson which are obtained in the article~\cite{Anastasiou:2012kq}. Same
set of MIs were also evaluated in~\cite{Pak:2011hs}.

The evaluation of the processes under real-virtual and virtual follow exactly
the similar method.  The 
polarisation sum of the external gluons is carried out in axial gauge
to ensure the exclusion of the unphysical degrees of freedom. We
include the ghost loops to cancel the unphysical 
degrees of freedom of the internal gluons present in the
virtual loops.

Considering all the subprocesses, we have 2979 number of
double real, 948 real-virtual and 207 double virtual Feynman
diagrams. In this present article, the computations of the double real and real-virtual
contributions are performed mostly using our in-house codes written in
FORM~\cite{Vermaseren:2000nd, Tentyukov:2007mu} and Mathematica. The
color simplification is done in general SU(N) gauge theory. The Dirac and
Lorentz algebra are carried out in $n$-dimensions
($n=4+\epsilon$). After performing the IBP reduction of the phase
space integrals to reduce these to a smaller set of MIs following the
techniques described above, we borrow the
analytical results of these MIs from~\cite{Anastasiou:2012kq} to get
the final answer in powers of $\epsilon$. However, instead if directly
using the results of the MIs presented in~\cite{Anastasiou:2012kq}, we
make use of some identities to convert the expressions into a form
which is manifestly real. The results of the two loop virtual diagrams are available
from~\cite{deFlorian:2013sza} which were computed by some of us. Using
the results of all the subprocesses belonging to the above discussed three categories
and performing the appropriate mass factorisation using
Eq.~(\ref{eq:23}), we get the completely finite partonic cross sections
or partonic coefficient functions at NNLO QCD. All the final results
of the partonic coefficient functions involving spin-2 particle,
Eq.~(\ref{resultspartonic}), are presented in the Appendix~\ref{app:res}. 
We have also provided these results as an ancillary file in Mathematica format.

\section{Numerical Implications}
\label{sec:5}

In this section, we present the numerical impact of two-loop QCD corrections on the
di-lepton production in ADD model at the LHC.  The LO, NLO and NNLO corrected hadronic cross sections 
are obtained by convoluting the partonic coefficient functions order-by-order with 
the corresponding parton distribution functions (PDFs) taken from {\tt
  lhapdf}~\cite{Whalley:2005nh}.   
We have used the strong coupling constant $a_s$ supplied by the corresponding PDF set. 
The fine structure constant $\alpha_{em} = 1/128$ and the weak mixing 
angle $\text{sin}^2\theta_W=0.227$. The results are presented for $n_f=5$ flavours 
and in the massless limit of quarks. Unless mentioned otherwise, our default choice of the
PDF set is {\tt MSTW2008lo/nlo/nnlo}. Except for studying the scale variations, the factorisation
and the renormalisation scales are set equal to the invariant mass of the di-lepton, i.e.,
$\mu_F = \mu_R = Q$. 
{Before proceeding further, we note that in the past there have been a series of experimental 
searches for large extra dimensions using di-lepton events at both Tevatron and the LHC.
Consequently, stringent bounds have been obtained on the
scale $M_{s}$ of the ADD model as a function of the number of extra dimensions $d$.
For instance, the lower limits on the scale $M_{s}$ obtained from both ATLAS and CMS 
collaborations using 7 TeV data are $M_{s}= 2.4 (3.9)$ TeV corresponding to $d=7(3)$~\cite{Aad:2012bsa, Chatrchyan:2012kc}.
With the availability of 8 TeV data~\cite{Aad:2014wca, Khachatryan:2014fba}, the lower limits on these parameters are further pushed to about $M_{s}=3.3 (4.9)$ TeV corresponding to $d=7(3)$.
There have already been some preliminary results on
search for narrow resonances in di-lepton final state using 13 TeV data~\cite{Khachatryan:2016zqb}.

It is worth mentioning that both ATLAS and CMS detectors have recorded di-lepton events with
invariant mass as large as 1800 GeV using 8TeV LHC data corresponding to a luminosity of about $20 fb^{-1}$~\cite{Aad:2014wca, Khachatryan:2014fba}.
With 13 TeV data the experimental sensitivity will further improve to measure events with larger di-lepton 
invariant masses. }
%
%
For the illustration of
the impact of QCD corrections, we choose the model parameters to be 
$M_s=4$ TeV and $d=3$.  
 
\begin{figure}[htb]
\centerline{
\epsfig{file=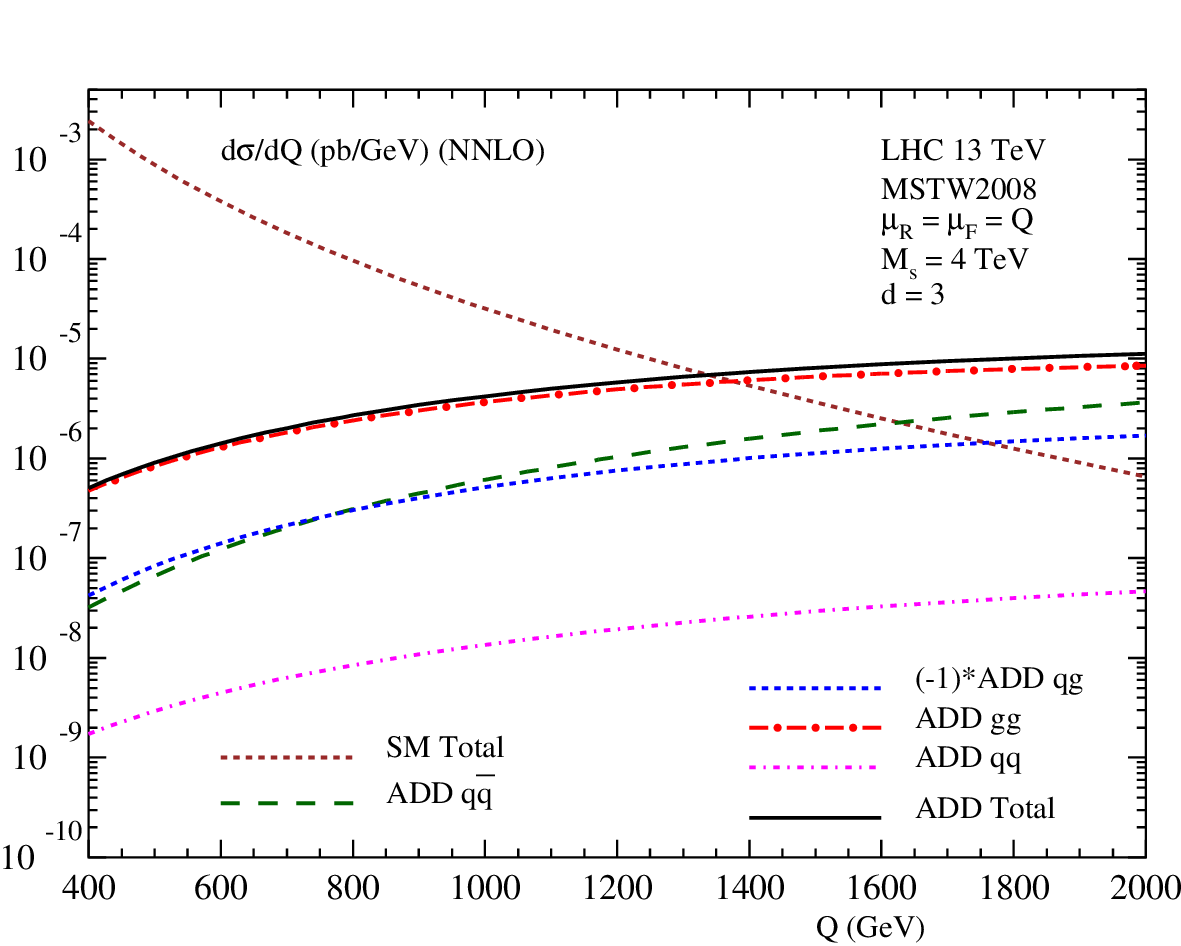,width=8cm,height=7.0cm,angle=0}
}
\caption{\sf Various sub-process contributions to the di-lepton production
computed at ${\cal O}(a_s^2)$ QCD in ADD model. The SM background contains
the full $a_s^2$ correction.}
\label{sub}
\end{figure}

Let us begin by discussing the relative contributions of various partonic 
channels that contribute to the hadronic cross section at NNLO level.  
The contributions from individual channels are not physical while their 
sum is. The bare partonic cross sections are ill defined due to the presence of infra-red
divergences and are removed by mass factorisation
in a scheme dependent way.  
Hence, the resulting channel-wise contributions
depend on the scheme, which in our case is 
 ${\overline {\text{MS}}}$.  In the Fig.~\ref{sub},  we present the $Q$ distributions
for various subprocesses at NNLO in the ADD model along with
the contribution from SM at NNLO~\cite{Matsuura:1987wt,Matsuura:1988sm,Hamberg:1990np}. 

At LO, the quark anti-quark initiated sub-process ($q\bar{q}$) contributes both in the SM and in the ADD model.
However, the gluon fusion sub-process ($gg$) starts contributing at the LO in the ADD model unlike in the 
SM where its contribution begins at NNLO.  We note that the contributions arising from the $gg$ sub-process in the ADD model dominates over the rest, because of 
the large gluon flux at the LHC.
Recall that the production cross section for the Higgs boson at the LHC is 
also dominated by gluon fusion sub-process.  The crucial difference between these two production
channels is the presence of strong coupling constant $a_s(\mu_R)$ at the leading order 
for the Higgs boson production cross section.  The other interesting aspect that one can not 
ignore is the numerical impact of quark-gluon ($qg$) sub-process beyond LO. 
The major difference between them 
is that in the ADD model at NLO, through mass factorisation, it receives collinear subtraction terms 
due to the presence of $q\bar{q}$ and $gg$ born sub-processes, whereas in the SM  
it is due to only $q\bar{q}$ Born sub-process. Irrespective of this difference, 
the $qg$ sub-process contribution both in the SM and in the ADD model is found to be negative but 
significantly large in magnitude. The same trend continues even at NNLO.
Particularly, we notice that the NNLO QCD corrections from $qg$ sub-process are considerably larger
in magnitude than the sum of all the quark initiated sub-processes 
($q\bar{q}, qq,q_{1}q_2, q_1\bar{q_2}$).
The other channels,
as can be seen from the Fig.~\ref{sub}, contribute very little to the
total inclusive cross section but they are important to stabilise the cross section under 
renormalisation and factorisation scale variations through renormalisation group equations. A generic
pattern in all of these sub-processes is that their contributions increase with $Q$, simply 
because of the increase in the number of accessible KK-modes with $Q$.

\begin{figure}[htb]
\centerline{
\epsfig{file=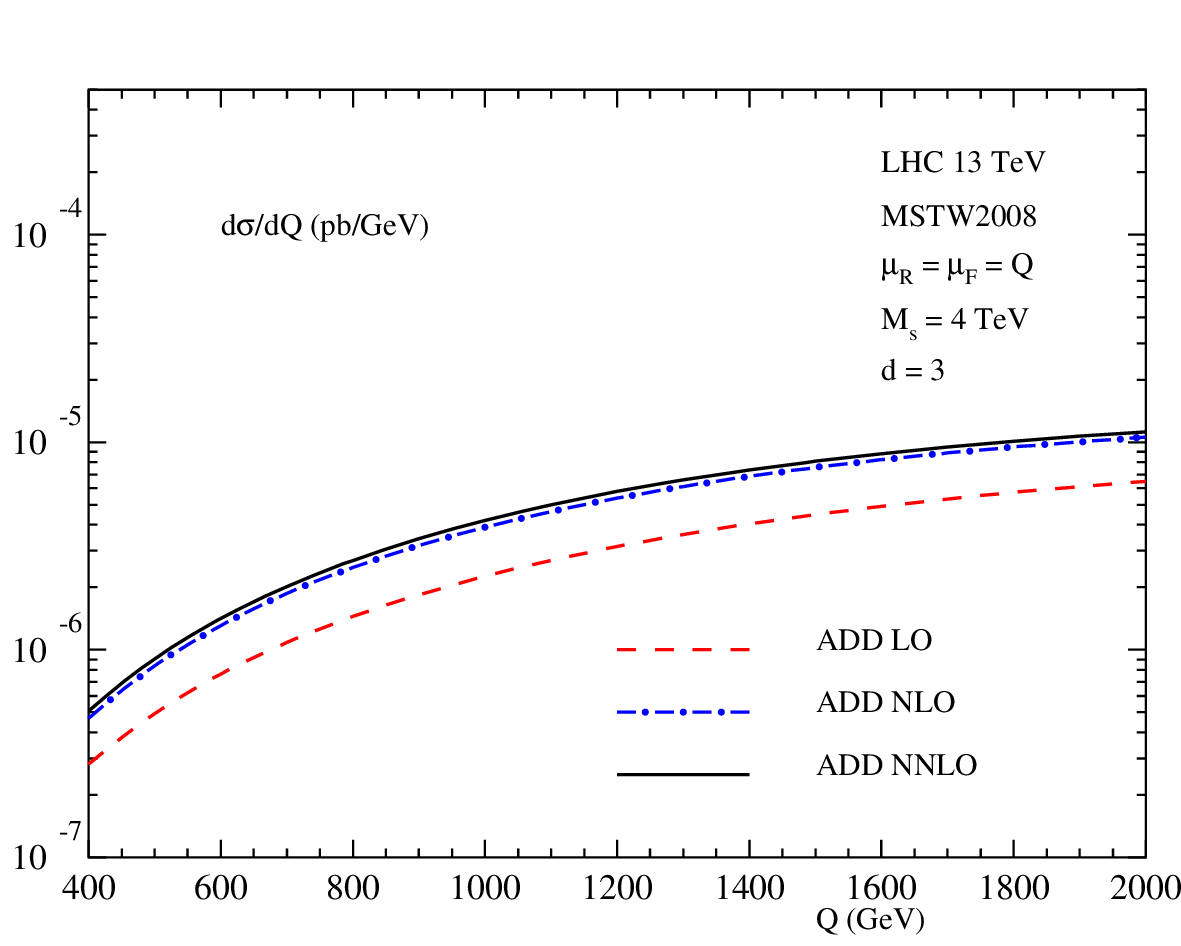,width=7.5cm,height=6.5cm,angle=0}
\epsfig{file=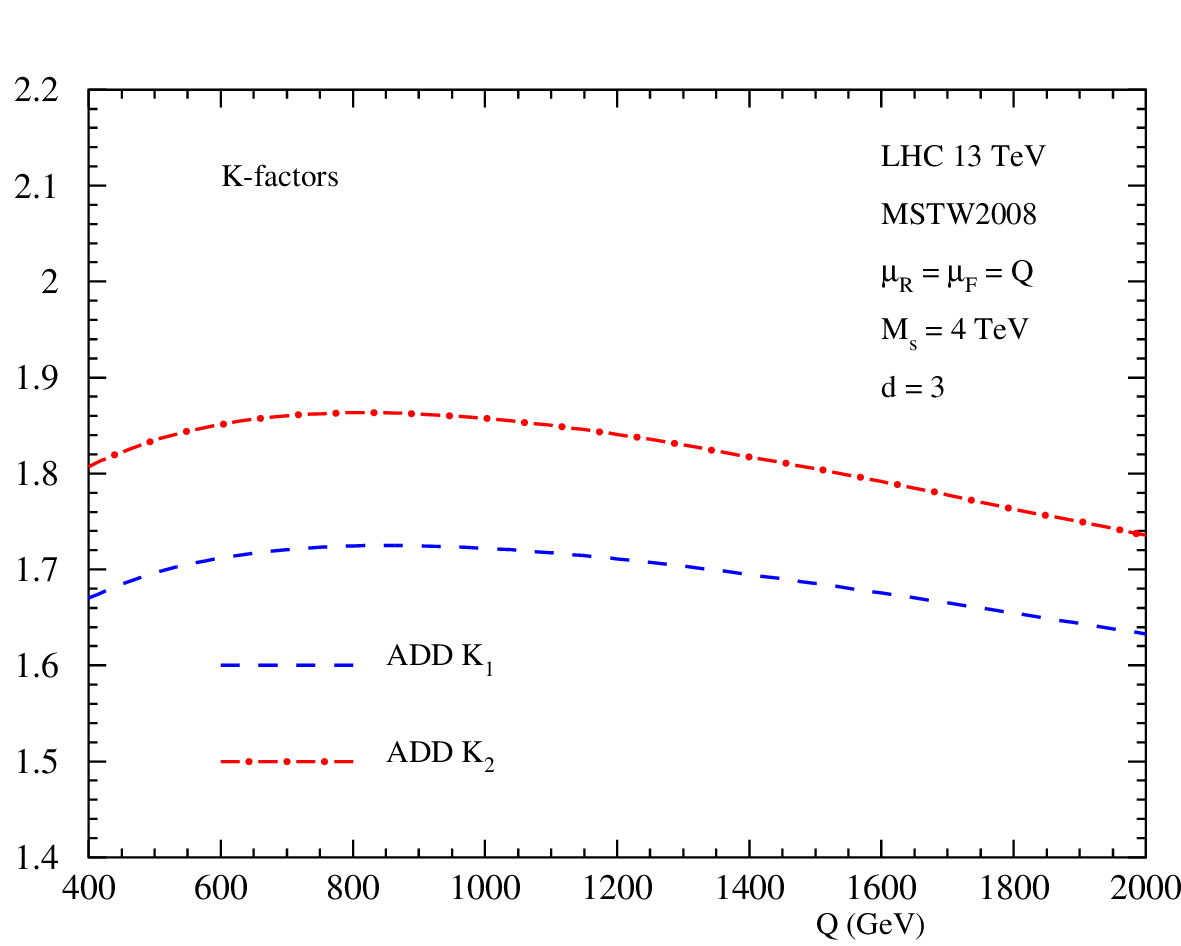,width=7.5cm,height=6.5cm,angle=0}
}
\caption{\sf Pure graviton contribution to the Drell-Yan production cross 
section (left panel) up to NNLO QCD in the ADD model for LHC13 and the
corresponding K-factors (right panel).}
\label{gravity}
\end{figure}
We next move on to the Fig.~\ref{gravity} where in the left panel we present
$d\sigma/dQ$ as a function of invariant mass Q at LO, NLO and NNLO for ADD model
(i.e. setting the SM contributions to zero).  We find that the contribution from the
interference terms between the SM and spin-2 is zero. 
It is also observed that the contributions arising from the ${\cal O}(a_s^2)$ increase the NLO cross section moderately.  In the  
right panel, we have plotted the K-factors that are defined as 
\begin{eqnarray}
\text{K}_{i}=\frac{d\sigma^i}{d\sigma^{\text{LO}}}, \quad \quad \quad i = 1\text{(NLO)}, 2\text{(NNLO)}
\end{eqnarray}
The NLO QCD corrections here increase the LO cross sections by about 68\% 
for $Q=1.5$ TeV, while the NNLO corrections that are still reasonably large 
contribute an additional 12\% (K$_1 = 1.68$ and K$_2 = 1.80$). With this considerably large contributions,
the reliability of perturbative QCD calls for the computations beyond NNLO.
The K-factors depend on the invariant mass through the logarithm corrections
both in partonic cross sections as well as in the evolution of PDFs.  Hence 
one is discouraged to use the constant K-factor for constraining the model parameters. 
Finally, we would like to make a remark that the conservative estimate of the K-factor for the 
Drell-Yan production in ADD model resembles closely to that of
the Higgs boson production. However, because of the large negative contribution from the $qg$ 
sub-process, the exact values of the $K$-factors differ in these two cases. In any case, we note 
that $K_2$ in ADD model alone is bigger than the corresponding one for the SM  
simply because of the dominance of $gg$ sub-process over others.

\begin{figure}[htb]
\centerline{
\epsfig{file=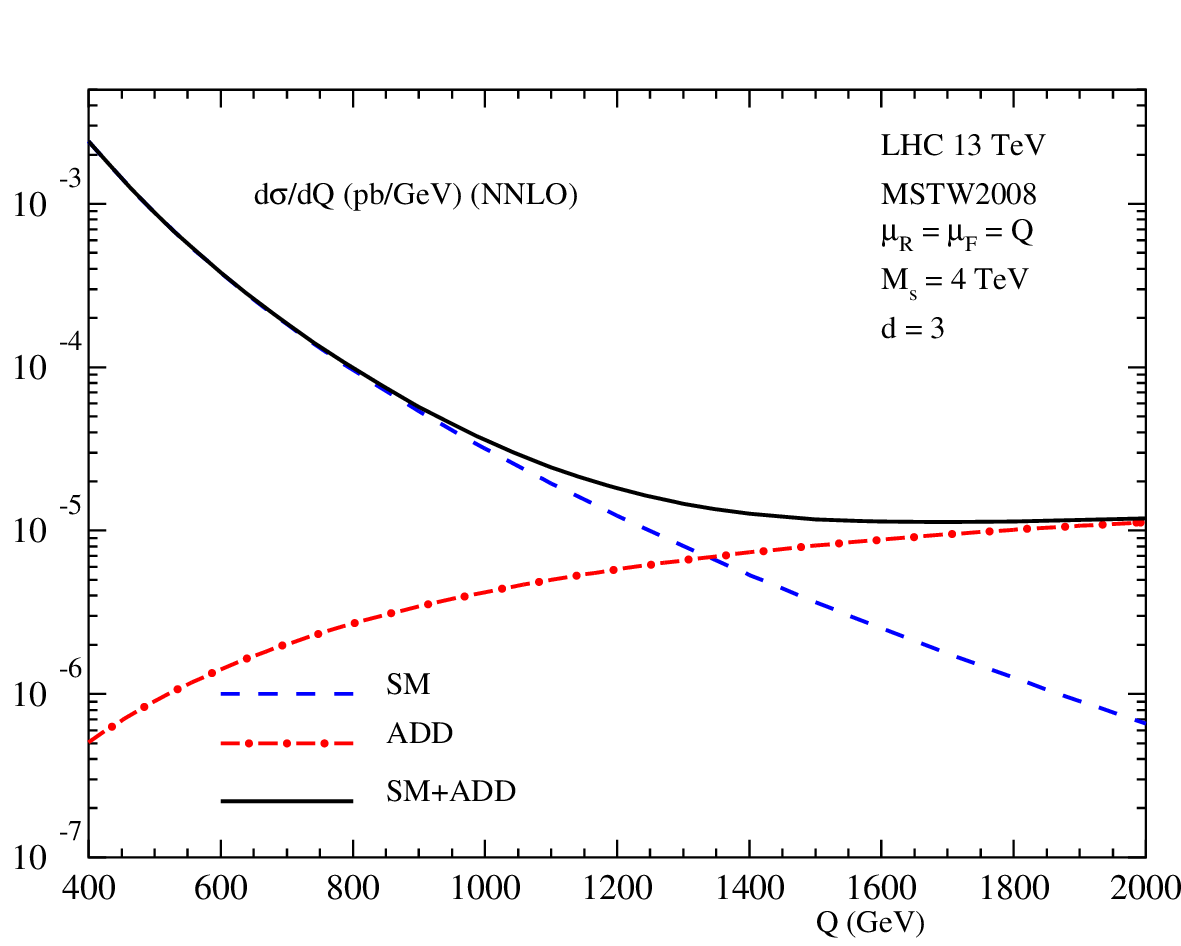,width=7.5cm,height=6.5cm,angle=0}
\epsfig{file=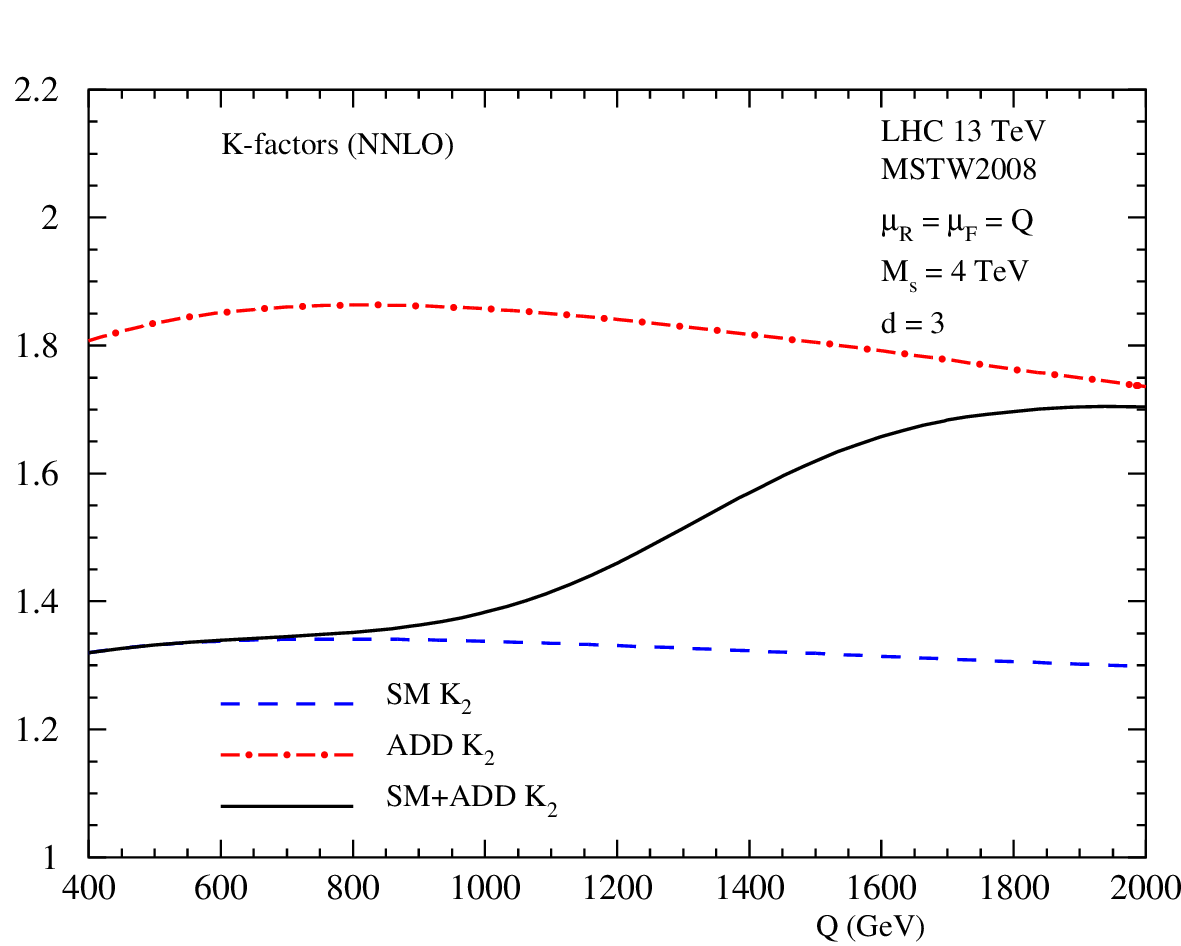,width=7.5cm,height=6.5cm,angle=0}
}
\caption{\sf Drell-Yan production cross section (left panel) for SM, GR and
the signal in the ADD model for LHC13 along with the corresponding K-factors
(right panel). Here, $M_s = 4$ TeV and $d=3$.}
\label{all-cs}
\end{figure}
%
\begin{figure}[t]
\centerline{
\epsfig{file=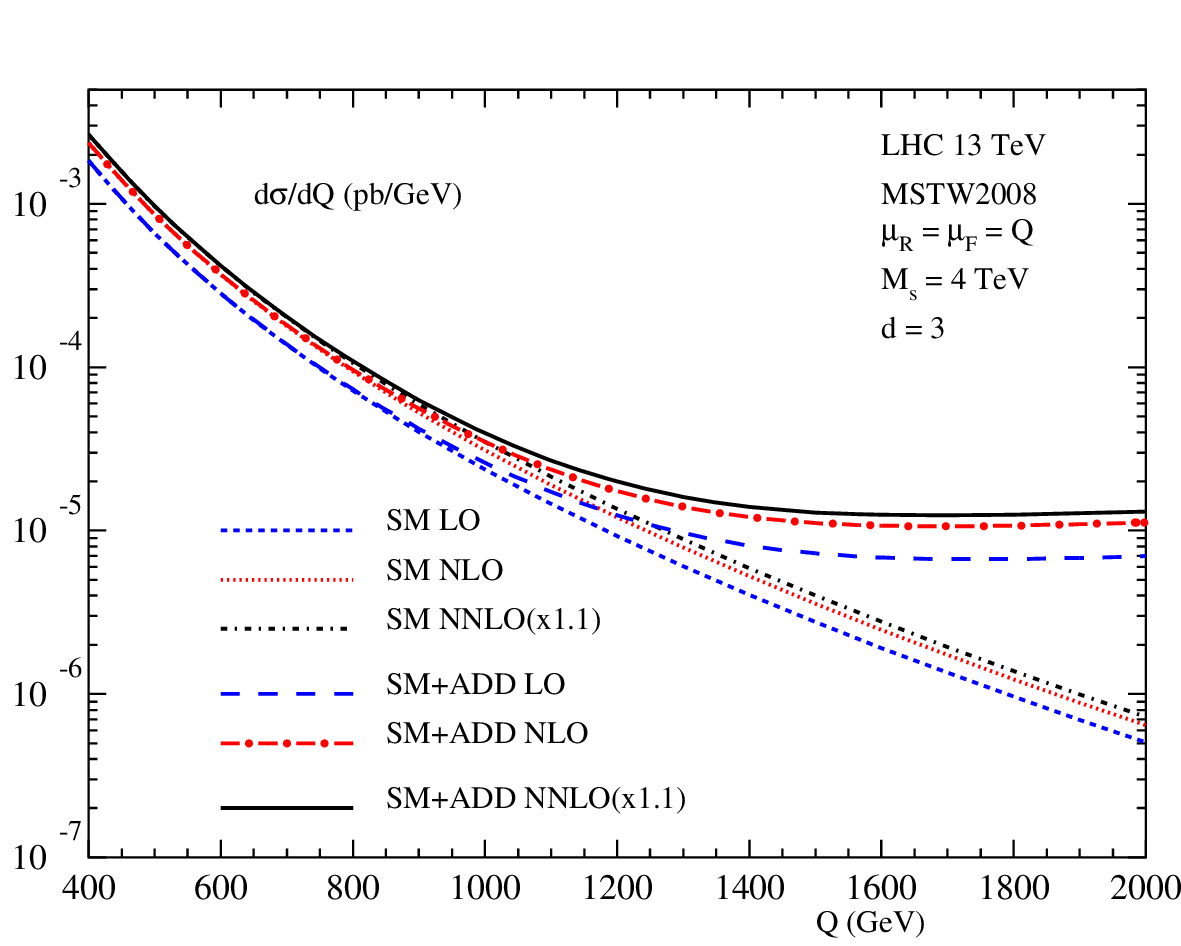,width=7.5cm,height=6.5cm,angle=0}
\epsfig{file=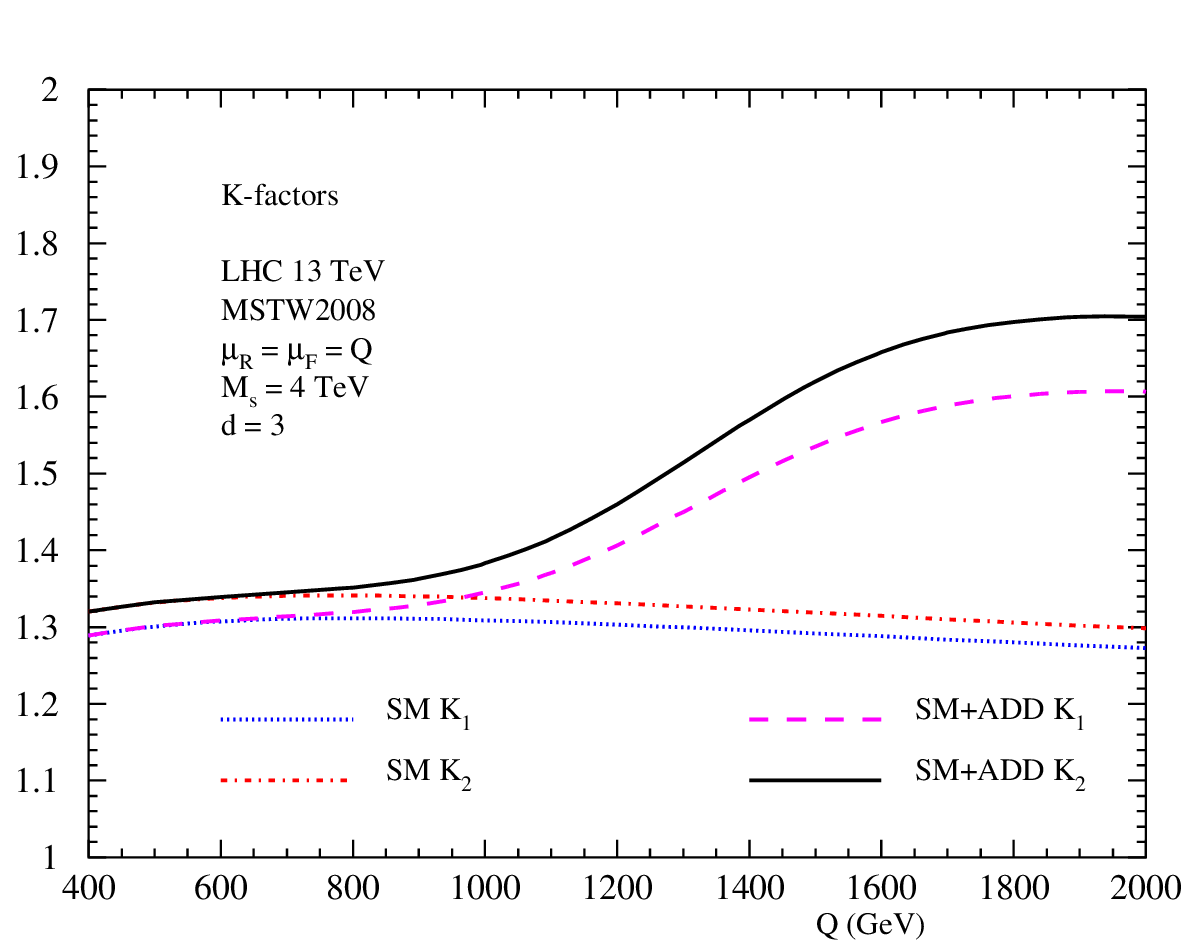,width=7.5cm,height=6.5cm,angle=0}
}
\caption{\sf Drell-Yan production cross section (left panel) for SM as well
as the signal in the ADD model for LHC13 along with the corresponding K-factors
(right panel).}
\label{sm-signal}
\end{figure}

In the left panel of Fig.~\ref{all-cs}, we present the NNLO cross sections for the SM, spin-2 (GR)
and the signal (SM+GR) together with the corresponding NNLO K-factor i.e. $K_2$ 
in the right panel.  
%
%
{The ADD model is an effective theory valid below the cut-off scale
$M_{s}$. Since the number of accessible KK modes will increase with Q 
as can be seen from Eq.~(2.11), the cross sections in the pure ADD model 
will increase with Q. Beyond the cut-off scale i.e. $Q > M_{s}$, the effective theory 
formalism ceases to be valid. Hence, in the kinematic regime $Q<M_{s}$, the spin-2 should 
give reliable predictions for the LHC. Because the spin-2 contributions increase with Q 
in the pure ADD model, they can dominate the SM contribution at some invariant mass $Q_{0} (< M_{s})$, 
the precise value of which depends on the choice of model parameters.
This simply leaves us with a phenomenologically interesting kinematic regime $Q_{0}<Q<M_{s}$ where the 
the spin-2 signals can give significant deviations from the SM predictions without breaking 
the effective theory formalism.}
%
For our default choice of model parameters, $Q_0$ is about $1.4$ TeV.
This implies that the signal is dominated by SM contributions well below $Q_0$ and by
ADD model contributions well above this $Q_0$. In the region closer
to this $Q_0$, which itself depends on the choice of the model parameters, the signal K-factor 
receives contributions both from SM and ADD model. 
%

\begin{figure}[htb]
\centerline{
\epsfig{file=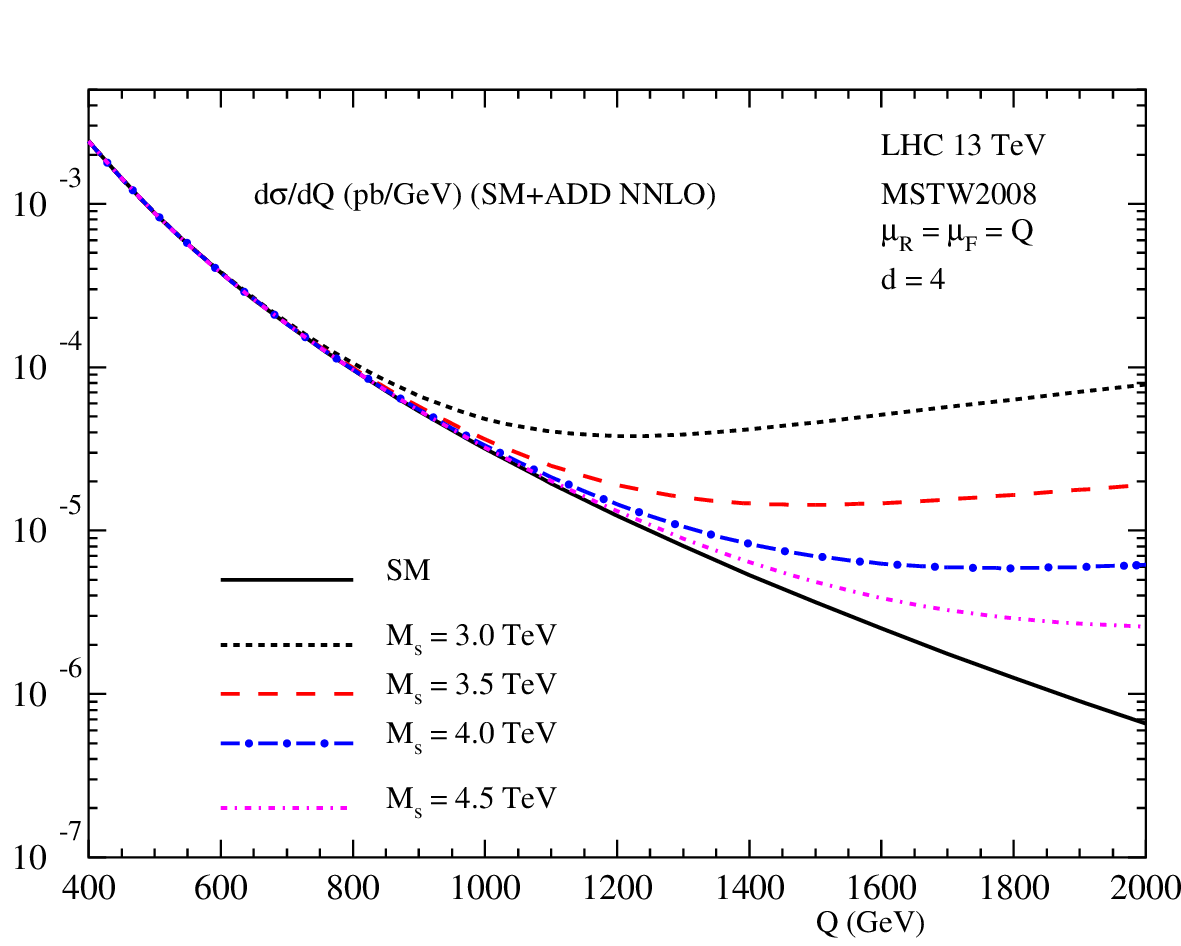,width=7.5cm,height=6.5cm,angle=0}
\epsfig{file=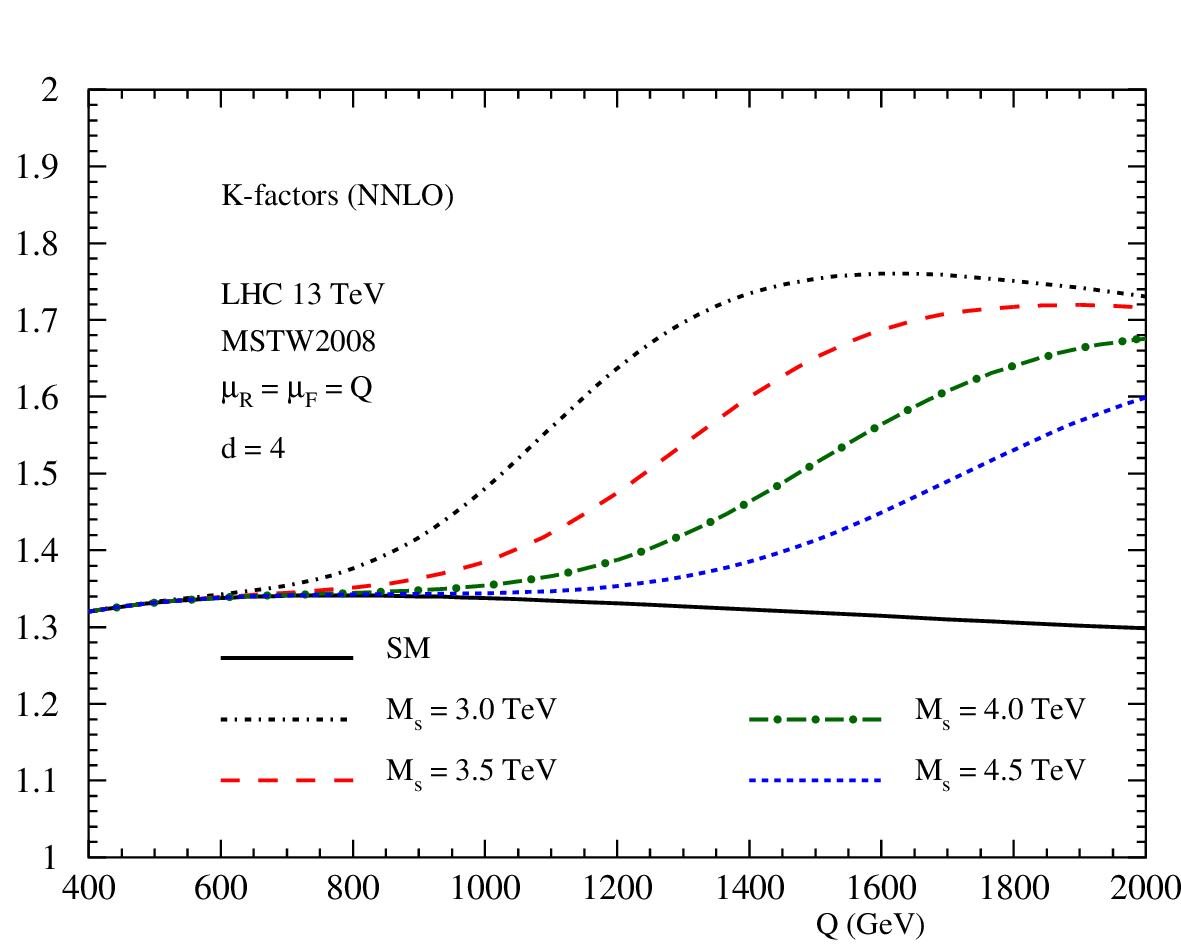,width=7.5cm,height=6.5cm,angle=0}
}
\caption{\sf Dependence of the signal production cross sections
at NNLO on the the scale of the ADD model $M_s$ (left panel) and the corresponding
signal K-factors(right panel).}
\label{ms-var}
\end{figure}
%
\begin{figure}[htb]
\centerline{
\epsfig{file=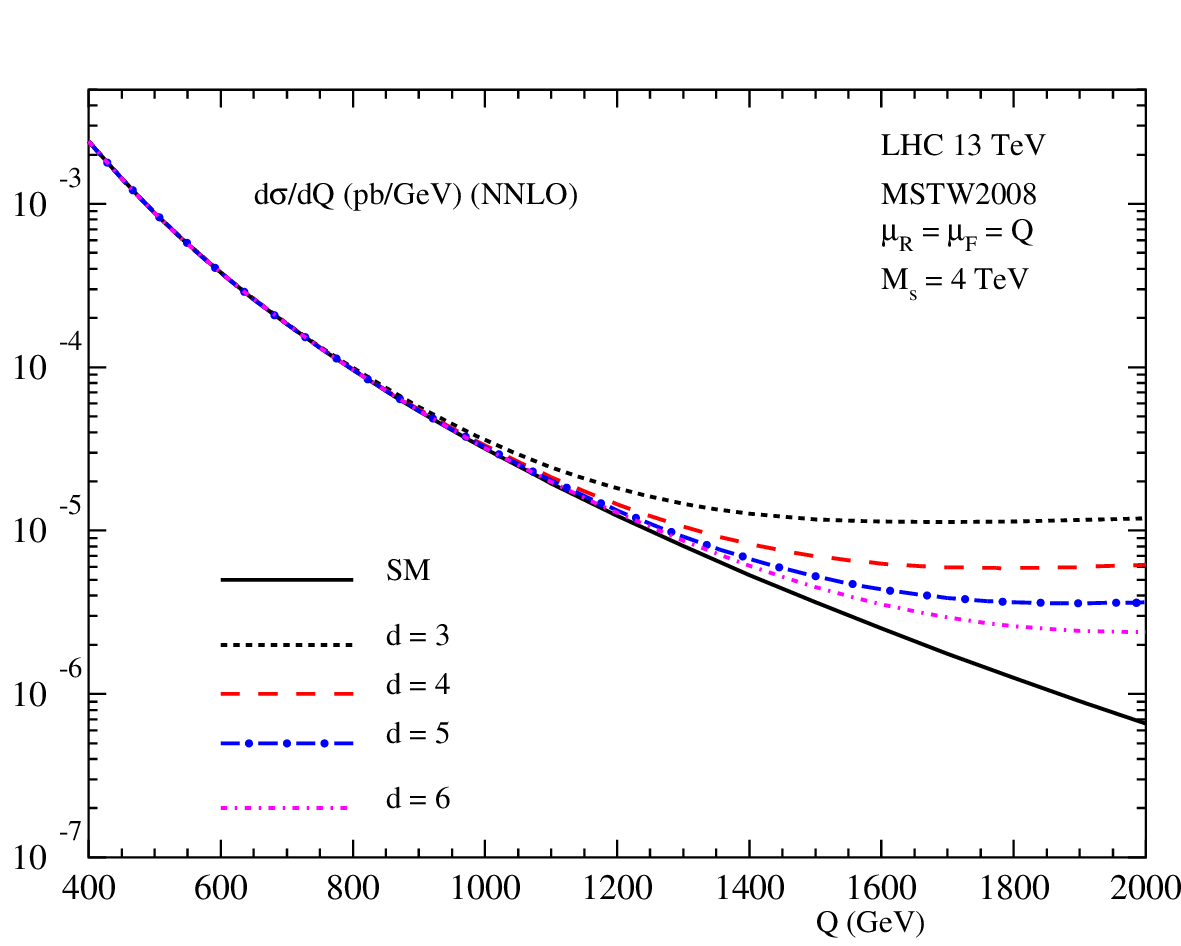,width=7.5cm,height=6.5cm,angle=0}
\epsfig{file=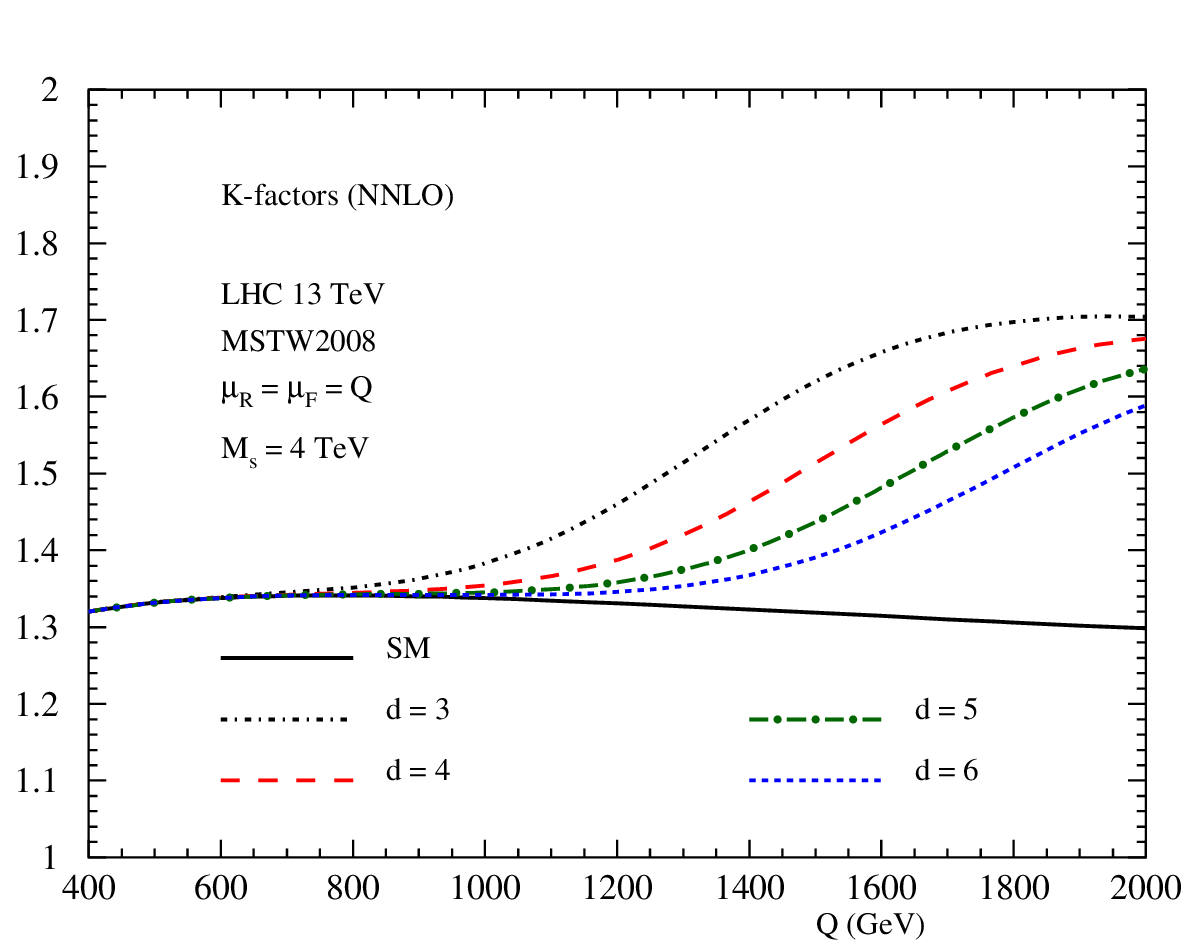,width=7.5cm,height=6.5cm,angle=0}
}
\caption{\sf Dependence of the signal production cross sections at NNLO
on the number of extra dimensions $d$ (left panel) and the corresponding 
K-factors (right panel).}
\label{d-var}
\end{figure}
From now on, we will focus on the signal contributions together with the corresponding SM 
background owing to their importance in the experimental searches for 
extra dimensions. In Fig.~\ref{sm-signal},
we present the results for the Q-distributions in the SM and ADD model order by order in the perturbation theory
in the left panel and the corresponding K-factors in the right panel.

So far, we have studied the Q-distribution by keeping the model parameters i.e. the scale of
extra dimensions ($M_s$) and the number of extra dimensions ($d$) fixed at some 
values that are consistent with the experimental bounds.  In Fig.~\ref{ms-var}, we demonstrate
it at NNLO level as a function of $M_s$.  As we decrease $M_s$, the
value of $Q_0$ also goes down as can be seen in the left panel of Fig.~\ref{ms-var}. 
However, 
for far beyond this $Q_0$, the SM contribution can be neglected altogether and hence
the SM+ADD K-factor assumes just the pure ADD K-factor that is insensitive to the
choice of the the model parameters. Hence far beyond $Q_0$, the SM+ADD K-factors tend
to converge to each other as can be seen in the right panel.

We also study the dependence on the number of extra dimensions, see Fig.~\ref{d-var}.
A similar explanation can be given as for the $M_s$ variation except noting that
the cross sections in SM+ADD decrease with increase in the number of extra dimensions $d$.
The leading order prediction is only a crude estimate of the true cross section.  In our case,
the LO prediction depends strongly on the factorisation scale $\mu_F$ through the parton distribution
functions.  It is often mild for the quark initiated processes while it is strong 
for the gluon initiated process.  The dependence on the $\mu_F$ scale 
starts getting reduced at higher orders leaving a residual scale dependence that is proportional
to $a_s^n ,n >1 $.   
%
\begin{figure}[htb]
\centerline{
\epsfig{file=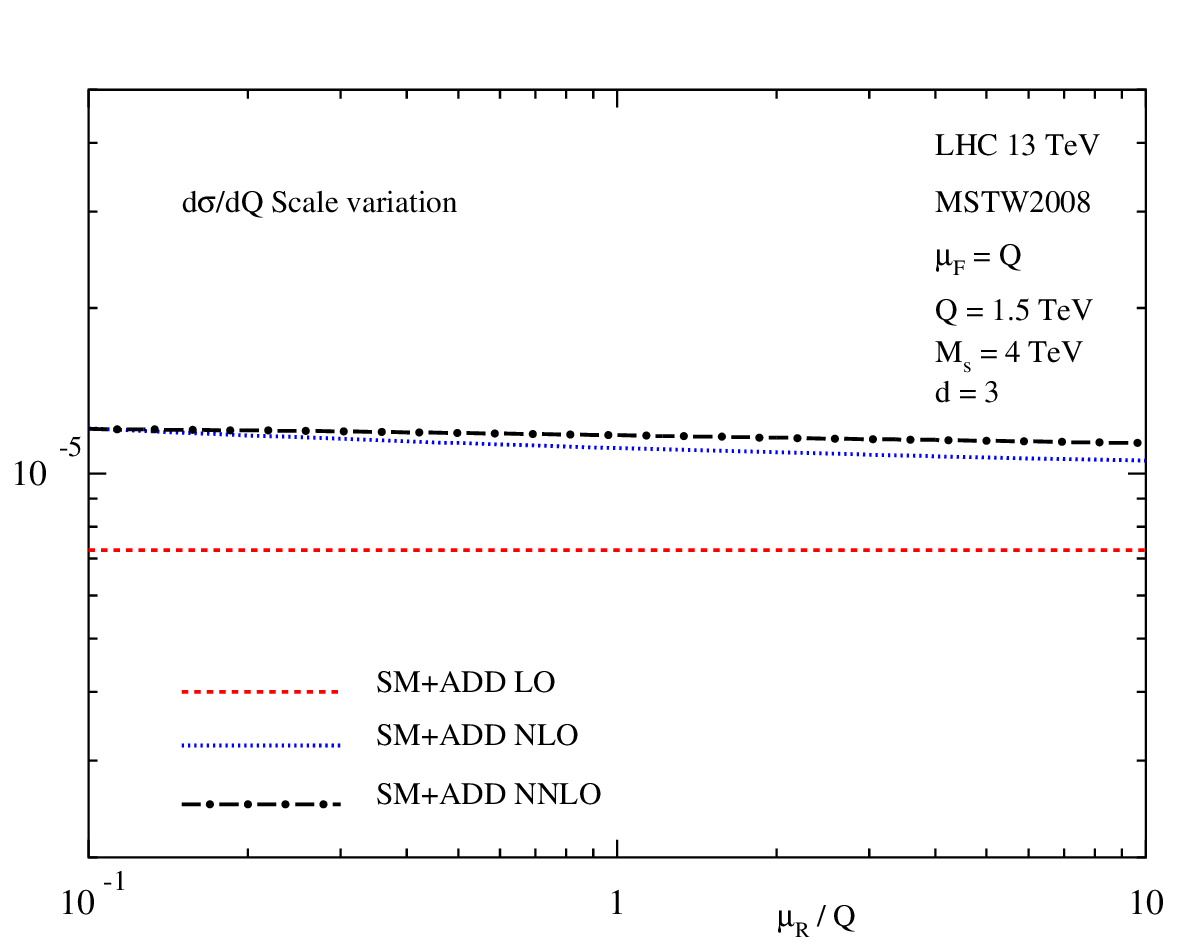, width=7.5cm,height=6.5cm,angle=0}
\epsfig{file=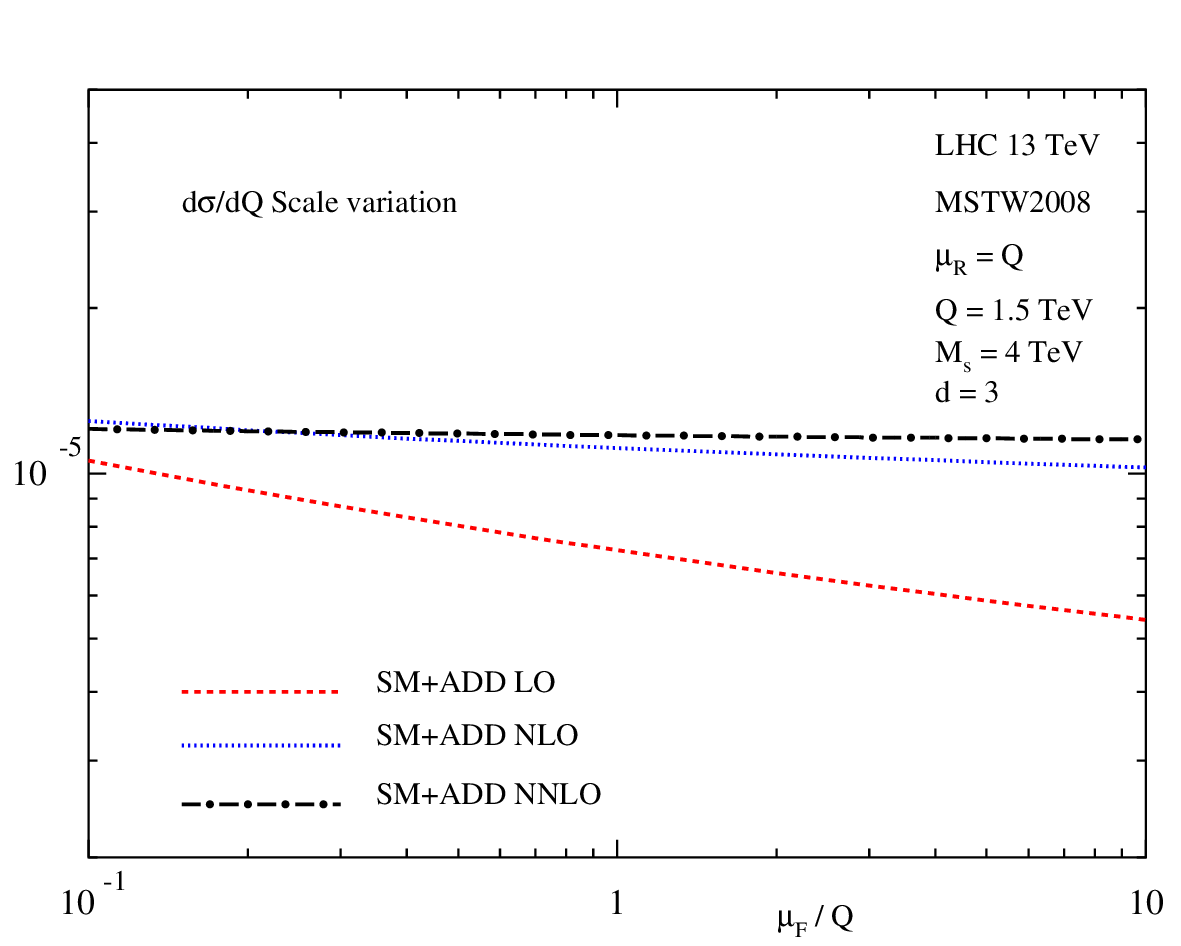, width=7.5cm,height=6.5cm,angle=0}
}
\caption{\sf Uncertainties in the signal production cross section due to the
choice of renormalisation scale $\mu_R$ (left panel) and 
factorisatioin scale $\mu_F$ (right panel).}
\label{scale}
\end{figure}
%
%
\begin{figure}[htb]
\centerline{
\epsfig{file=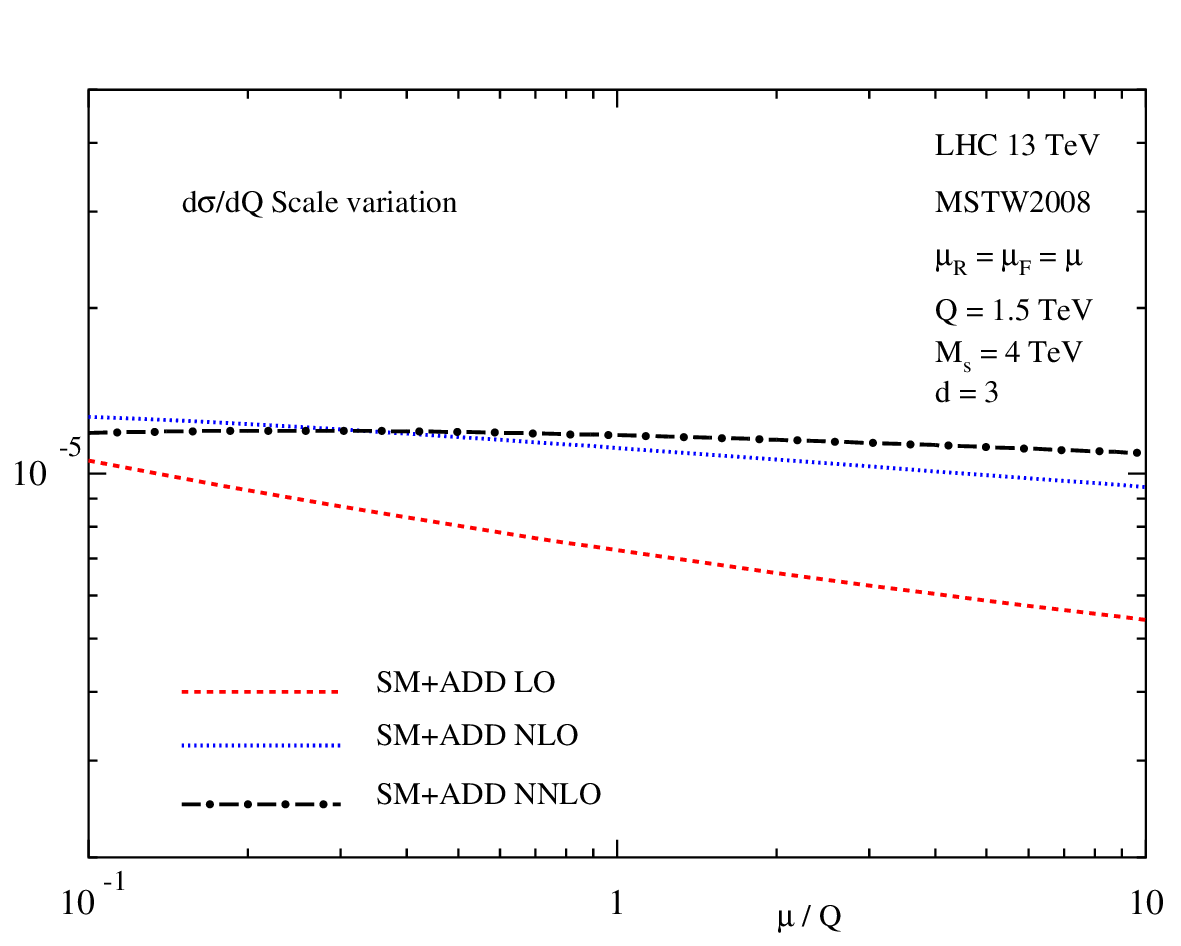,width=8cm,height=6.5cm,angle=0}
}
\caption{\sf Uncertainties in the signal production cross section due to the
choice of the scale $\mu=\mu_R=\mu_F$.}
\label{murf}
\end{figure}
%
\begin{figure}[htb]
\centerline{
\epsfig{file=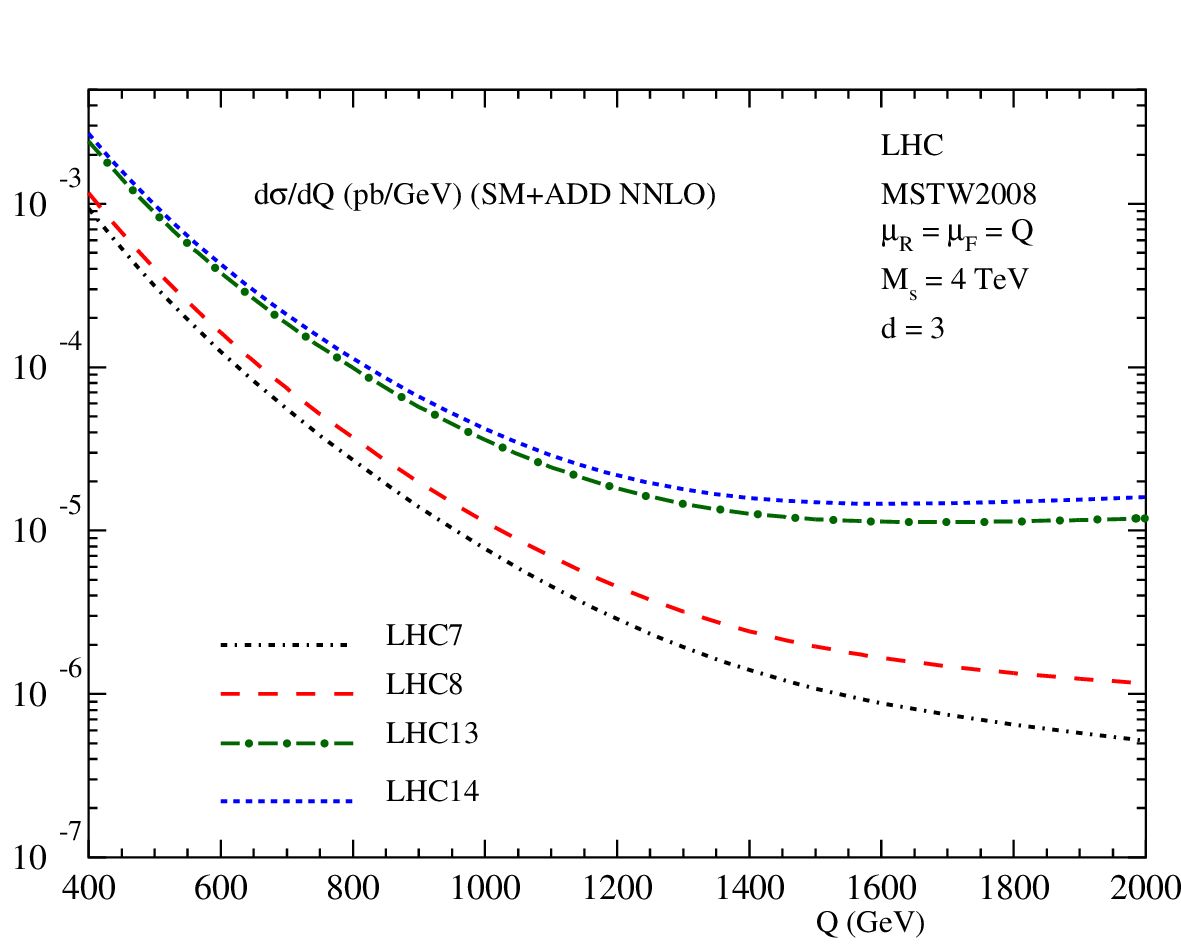,width=7.5cm,height=6.5cm,angle=0}
\epsfig{file=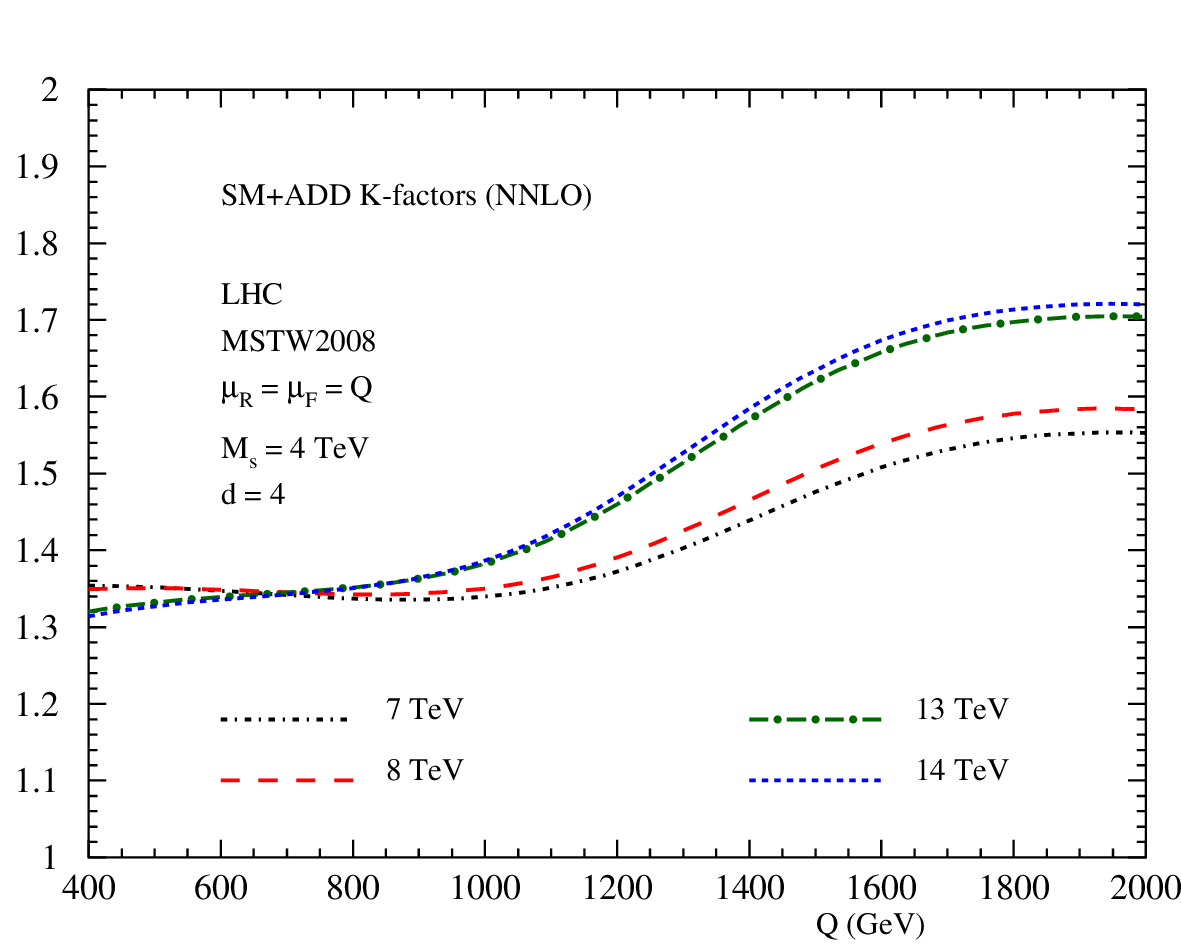,width=7.5cm,height=6.5cm,angle=0}
}
\caption{\sf Dependence of the signal production cross sections at NNLO
on the center of mass energy at LHC (left panel) and the corresponding
K-factors (right panel).}
\label{ecm-var}
\end{figure}
%
%
\begin{figure}[htb]
\centerline{
\epsfig{file=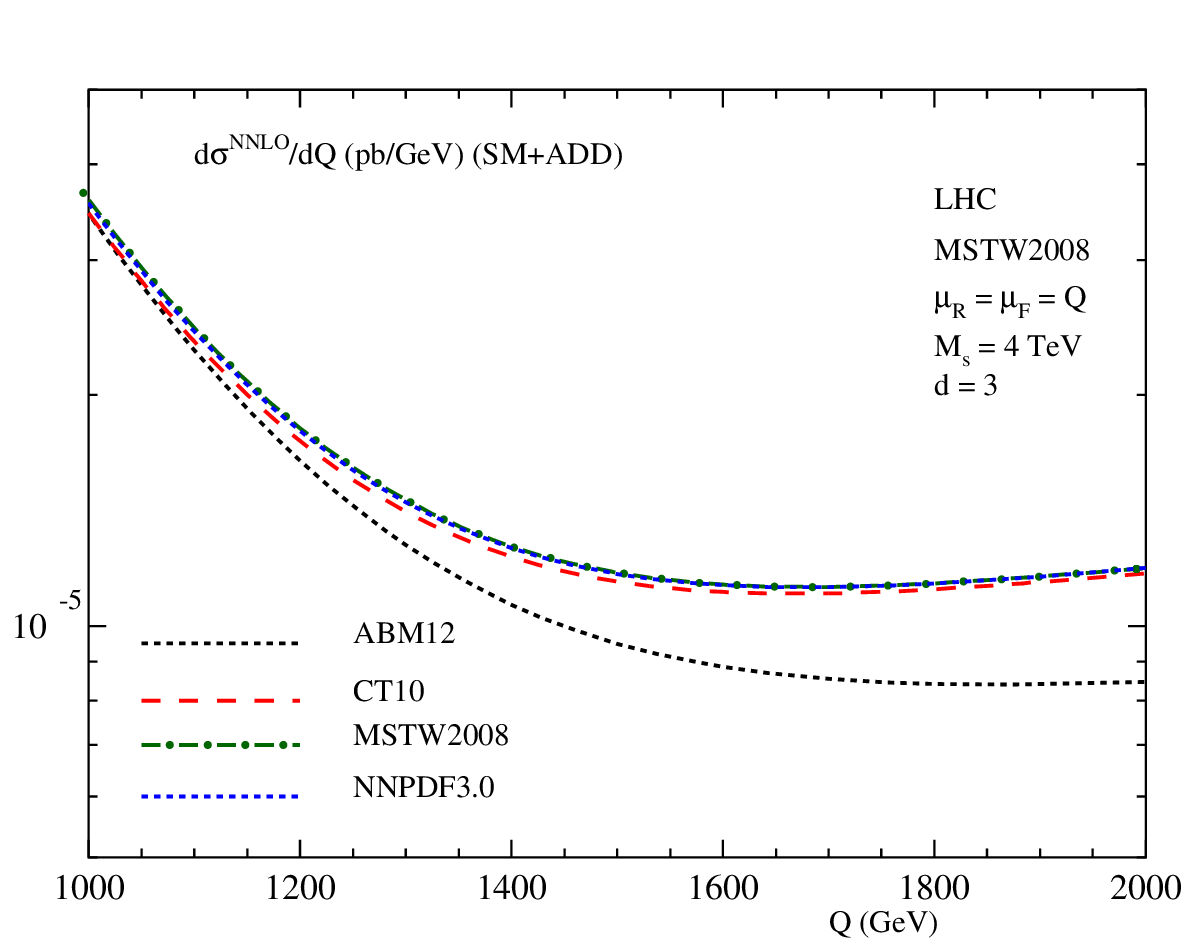,width=7.5cm,height=6.5cm,angle=0}
\epsfig{file=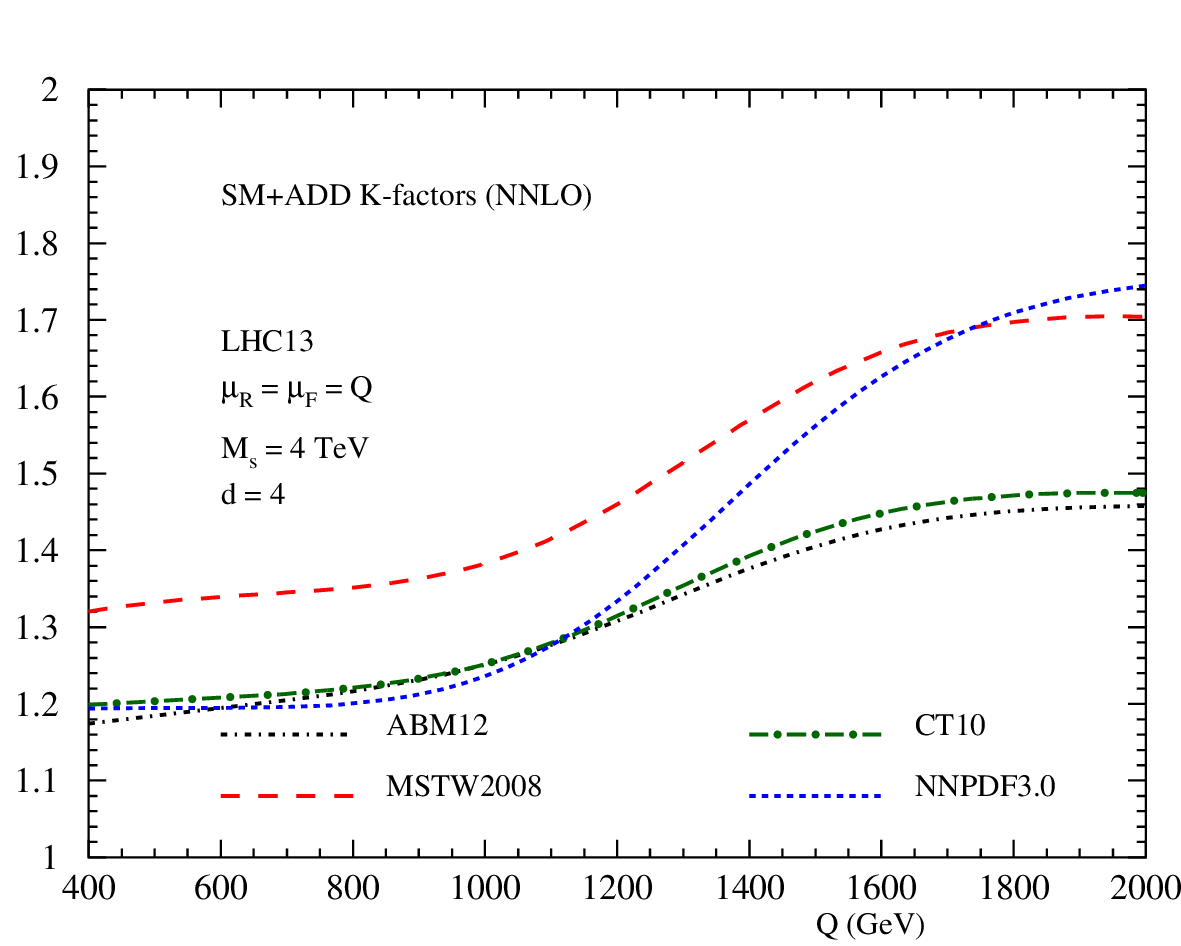,width=7.5cm,height=6.5cm,angle=0}
}
\caption{\sf Dependence of the signal production cross sections at NNLO on the choice of PDFs (left panel). Signal K­factors at NNLO for different PDFs (right panel).}
\label{pdf-var}
\end{figure}
At NLO level, for the first time the strong coupling constant $a_s(\mu_R)$ enters
our calculation. Since it depends on the renormalisation scale $\mu_R$, 
the result up to NLO level will now become sensitive to choice of $\mu_R$.  Hence, at NLO, while the factorisation 
scale dependence gets reduced, the renormalisation scale dependence crops up.  Renormalisation group equation
ensures that the inclusion of more and more higher order terms in the perturbation theory will
reduce its dependence and it will eventually go away if we know the result to
all orders in perturbation theory.  A similar statement can be made for the factorisation scale dependence
as well thanks to the fact that the factorised hadronic cross section is independent of 
$\mu_F$.  In order to demonstrate the reduction in the scale dependence, we have plotted the $d\sigma/dQ$ in the Fig.~\ref{scale}
at a fixed value of $Q=1.5$ TeV, the choice where the new physics dominates, as function of 
$\mu_R$ (left panel), $\mu_F$ (right panel) and then $\mu=\mu_F=\mu_R$ (see Fig.~\ref{murf}.
in the range between $Q/10$ to $10 Q$, for wider scale variations.
As expected, we find that inclusion of higher terms in the perturbation theory 
indeed reduce the dependence on these
unphysical scales.

\begin{figure}[htb]
\centerline{
\epsfig{file=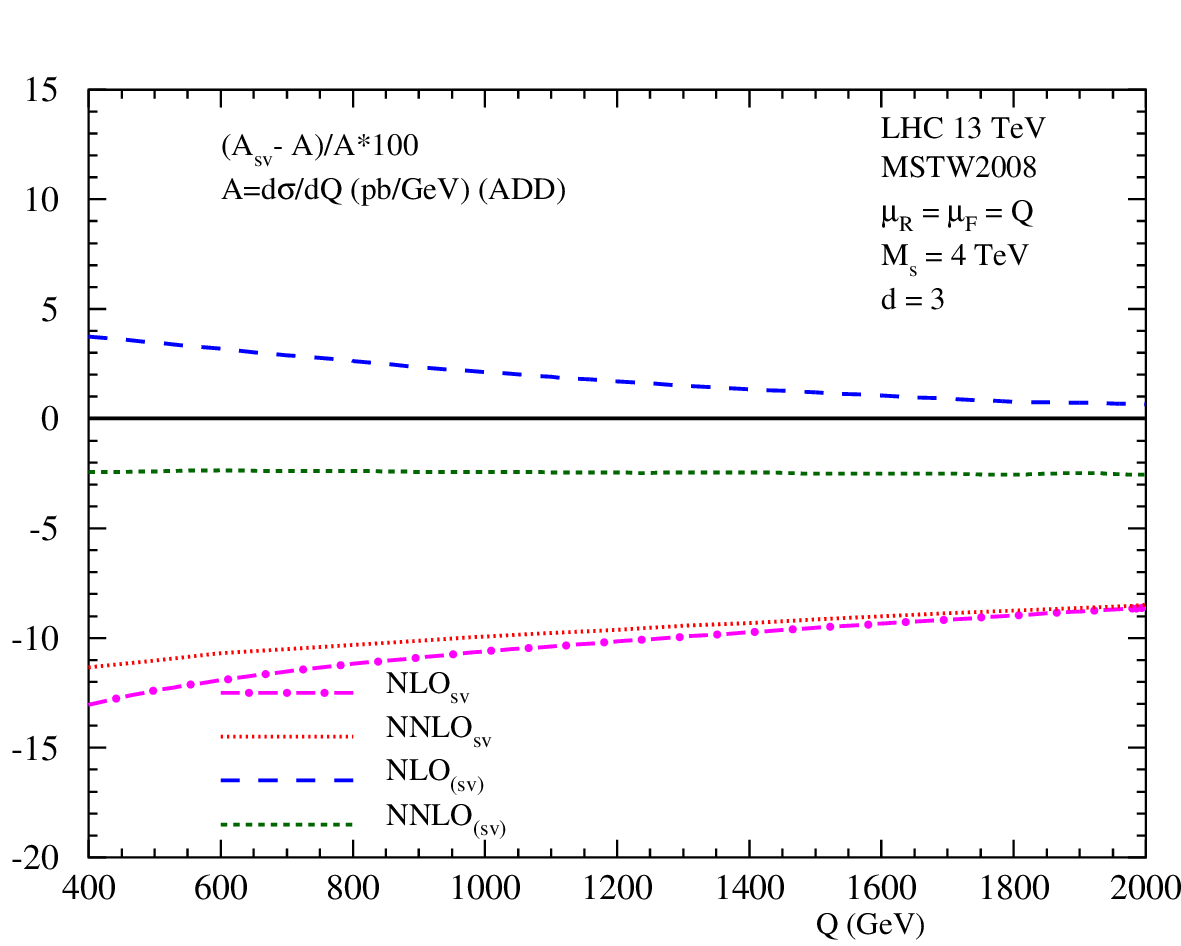,width=8cm,height=7.0cm,angle=0}
}
\caption{\sf NLO and NNLO predictions obtained from modified SV approximation
for the signal only with the $gg$ subprocess contribution.}
\label{sv}
\end{figure}
In Fig.~\ref{ecm-var}, we have presented the predictions for the invariant mass distribution for 
various center of mass energies, namely $7,8,13 ~\text{and}~ 14$ TeV at the LHC.  As the energy increases,
the parton fluxes particularly the gluon flux will increase and hence the sensitivity to 
the ADD model will also go up. Consequently, both the NNLO SM+ADD cross sections (left panel)
and the corresponding signal K-factors (right panel) will increase with the center of mass energy.

In addition to the choice of scale, the choice of PDFs do affect the predictions significantly.
The precise value of the strong coupling constant consistent with a given PDF set also 
influences the prediction.  
In order to study these effects, we have plotted the cross sections, in the left panel of Fig.~\ref{pdf-var}, using various PDF sets such as MSTW2008, ABM12, CT10, NNPDF3.0. In the right panel of Fig.~\ref{pdf-var}, we present the corresponding K­-factors. We note that for this PDF uncertainties, we have convoluted the partonic cross sections computed at a particular order in $\alpha_s$ with the PDFs extracted to the same order in $\alpha_s$ for all the PDF sets considered here except for ABM12 for which we have used  only the available NNLO PDFs for computing all the LO, NLO and NNLO hadron level cross sections. This approximately gives the sensitivity to the choice of PDF sets and $a_s$ ($\mu_R$), as well as the estimate of the error on the predictions. It is also worth noting here that although the difference in the cross sections is directly related to the difference in the parton fluxes from different PDF sets, the K-­factors may not show a similar pattern as that of the cross sections simply because PDFs of different orders enter in the ratio of K-­factors, as can be seen in the right panel of Fig.~\ref{pdf-var}.

Finally, we address the impact of soft-plus corrections on our fixed order predictions. Note that for ADD,
the numerical impact of soft-plus-virtual (SV) were already reported
in \cite{deFlorian:2013wpa}.  Now that we have
a complete result at NNLO level, it is important to study the validity of SV approximation. 
As mentioned before that the $gg$ initiated sub-process in the pure spin-2 case is 
similar to the SM Higgs production in gluon fusion channel. For the latter case,
the SV corrections (or rather with the modified parton fluxes) are found to be a very good approximate 
for the fixed order results. This indeed is the case even for our ADD model predictions provided 
we just take only the $gg$ initiated subprocesses. 
In addition, if we use the modified SV approximation as described
in~\cite{Ahmed:2015qda}, we find that it is closer to the exact result, resulting from $gg$ subprocesses alone (see Fig.~\ref{sv}).
Inclusion of $qg$ initiated sub processes spoil this approximation
as their contribution is negative and significantly large.  Hence, the SV approximation at $a_s^2$ does not seem
to be working very well unlike in the Higgs production in gluon fusion.

\section{Conclusions}
\label{sec:6}

In summary, we have performed the very first calculation involving a massive spin-2 particle at NNLO level in QCD for the production of a pair of leptons at hadron colliders. We have included all the relevant sub-processes that can contribute to the invariant mass distribution of the di-leptons. The methodology of reverse unitarity and IBP identities are systematically employed to achieve it. Unlike the DY process within the SM, the spin-2 mediated processes are dominated by the gluon initiated ones due to the large gluon flux at the LHC. In addition, the quark-gluon initiated sub-processes have negative but significantly large contributions at NNLO. The corrections at various orders are quantified through their respective K-factors (1.54 at NLO and 1.62 at NNLO). We find, that the corrections are not only large but also important to stabilise the predictions with respect to the unphysical renormalisation and factorisation scales. The scale uncertainties get reduced to 29\% at NLO from 71\% at LO, which further gets reduced to about 8\% at NNLO. 
The extensions to the scenarios where a spin-2 particle couples differently to various SM particles are straightforward. 

\section*{Acknowledgments}

We would like to thank C. Duhr, T. Gehrmann, R. Lee and F. Maltoni for useful discussions and timely
help. We also thank M. Mahakhud and M. K. Mandal for useful discussions. PB and PKD also like to thank their parents, siblings and friends for their wonderful love and continuous support. 

\appendix

\section{Results of the Partonic Cross Sections}
\label{app:res}

In this appendix, we present the renormalised and finite partonic coefficient functions
involving spin-2 particles, $\Delta^{h,(k)}_{ab} \l(z, Q^2, \mu_F^2\r)$ in
Eq.~(\ref{eq:35}), up to 
NNLO QCD ($k=0,1,2$). The results at NLO are in agreement with the existing
ones~\cite{Mathews:2004xp}. The soft-virtual
corrections i.e. the contributions arising
from the soft gluon emissions at NNLO were computed
in~\cite{deFlorian:2013wpa}. Our results are also consistent with these
ones. Below we present all of our findings after normalising the
components of the coefficient functions by the leading
order results:
\begin{align}
\label{eq:36}
&{\Delta}^{h,(k)}_{gg} \equiv \frac{\pi}{2(N^2-1)}
  {\bar{\Delta}}^{h,(k)}_{gg}\,,
\nonumber\\
&{\Delta}^{h,(k)}_{ab} \equiv \frac{\pi}{8N} {\bar{\Delta}}^{h,(k)}_{ab}
  \qquad\quad {\rm for}\quad ab \neq gg
\end{align}
and all the $\l(\frac{\log^{i}(1-z)}{1-z}\r)$ terms should be understood as distributions, ${\cal D}_{i}$ with
\begin{align}
\label{eq:dist}
{\cal D}_{i} \equiv \l[ \frac{\log^{i}(1-z)}{1-z} \r]_{+}\,, \qquad\qquad i = 0, 1, 2, \cdots\,\,.
\end{align}
The results are obtained as
\begin{align}
\label{resultspartonic}
   {\bar{\Delta}}_{gg}^{h,(0)} &=\delta(1-z)  \,,
 \nonumber\\
   {\bar{\Delta}}_{gg}^{h,(1)} &=  \mathbf{n_f}\Bigg\{ \delta(1-z)   \Bigg( \frac{35}{9} \Bigg)
+ \log \left(\frac{Q^2}{\mu_F^2}\right) \delta(1-z)   \Bigg(  - \frac{4}{3} \Bigg)\Bigg\}
\nonumber\\&+ \mathbf{C_A}\Bigg\{   \Bigg(  - 2 - \frac{22}{3} \frac{1}{z} + 2 z + \frac{22}{3} z^2 \Bigg)
\nonumber\\&+  \log(1-z)   \Bigg(  - 32 + 16 \frac{1}{z} + 16 \frac{1}{1-z} + 16 z - 16 z^2 \Bigg)
\nonumber\\&+  \log(z)   \Bigg( 16 - 8 \frac{1}{z} - 8 \frac{1}{1-z} - 8 z + 8 z^2 \Bigg)
+  \delta(1-z)   \Bigg(  - \frac{203}{9} \Bigg)
\nonumber\\&+  \log \left(\frac{Q^2}{\mu_F^2}\right)  \Bigg[ \Bigg(  - 16 + 8 \frac{1}{z} + 8 \frac{1}{1-z} + 8 z - 8 z^2 \Bigg)
 +\delta(1-z)   \Bigg( \frac{22}{3} \Bigg)\Bigg]
\nonumber\\&+ \zeta_2 \delta(1-z)   \Bigg( 8 \Bigg)\Bigg\}\,,
  \nonumber\\
     {\bar{\Delta}}_{gg}^{h,(2)} &= \mathbf{n_f^2}\Bigg\{ \delta(1-z)   \Bigg( \frac{1225}{324} \Bigg)
+  \log \left(\frac{Q^2}{\mu_F^2}\right) \delta(1-z)   \Bigg(  - \frac{70}{27} \Bigg)
\nonumber\\&+  \log^2 \left(\frac{Q^2}{\mu_F^2}\right) \delta(1-z)   \Bigg( \frac{4}{9} \Bigg)
+  \zeta_2 \delta(1-z)   \Bigg( \frac{8}{3} \Bigg)\Bigg\}
\nonumber\\&+ \mathbf{C_F n_f} \Bigg\{  \Bigg(  - \frac{664}{9} + \frac{1748}{27} \frac{1}{z} - \frac{389}{9} z + \frac{1411}{27} z^2 \Bigg)
\nonumber\\&+  S_{1,2}(1-z)   \Bigg( 60 - 16 \frac{1}{z} - 24 z - 28 z^2 \Bigg) +  S_{1,2}(-z)   \Bigg(  - 72 - 32 \frac{1}{z} - 48 z - 8 z^2 \Bigg)
\nonumber\\&+  {\rm Li}_{3}(1-z)   \Bigg(  - 72 + 8 \frac{1}{z} - 16 z + 32 z^2 \Bigg) +  {\rm Li}_{3}(-z)   \Bigg(  - 36 - 80 \frac{1}{z} - 56 z + 4 z^2 \Bigg)
\nonumber\\&+  {\rm Li}_{2}(1-z)   \Bigg(  - 2 + \frac{8}{3} \frac{1}{z} - 16 z + \frac{116}{3} z^2 \Bigg) +  {\rm Li}_{2}(-z)   \Bigg( 20 - 16 \frac{1}{z} + 36 z \Bigg)
\nonumber\\&+  \log(1+z) {\rm Li}_{2}(-z)   \Bigg(  - 72 - 32 \frac{1}{z} - 48 z - 8 z^2 \Bigg)
\nonumber\\&+  \log(1-z)   \Bigg(  - \frac{235}{3} - \frac{256}{9} \frac{1}{z} + \frac{436}{3} z - \frac{347}{9} z^2 \Bigg)
\nonumber\\&+  \log(1-z) {\rm Li}_{2}(1-z)   \Bigg( 56 + 32 z - 32 z^2 \Bigg) +  \log^2(1-z)   \Bigg( 8 + \frac{64}{3} \frac{1}{z} - 32 z + \frac{8}{3} z^2 \Bigg)
\nonumber\\&+ \log(z)   \Bigg( \frac{25}{2} + \frac{200}{9} \frac{1}{z} - 144 z + \frac{433}{18} z^2 \Bigg) 
\nonumber\\&+ \log(z) {\rm Li}_{2}(1-z)   \Bigg(  - 2 - 8 \frac{1}{z} - 40 z - 8 z^2 \Bigg)
\nonumber\\&+  \log(z) {\rm Li}_{2}(-z)   \Bigg( 72 + 64 \frac{1}{z} + 64 z + 4 z^2 \Bigg) +  \log(z) \log(1+z)   \Bigg( 20 - 16 \frac{1}{z} + 36 z \Bigg)
\nonumber\\&+  \log(z) \log^2(1+z)   \Bigg(  - 36 - 16 \frac{1}{z} - 24 z - 4 z^2 \Bigg)
\nonumber\\&+  \log(z) \log(1-z)   \Bigg(  - 30 - \frac{64}{3} \frac{1}{z} + 32 z + \frac{104}{3} z^2 \Bigg)
\nonumber\\&+  \log(z) \log^2(1-z)   \Bigg( 28 + 16 z - 16 z^2 \Bigg) +  \log^2(z)   \Bigg(  - 9 + \frac{16}{3} \frac{1}{z} - 37 z - 20 z^2 \Bigg)
\nonumber\\&+  \log^2(z) \log(1+z)   \Bigg( 54 + 24 \frac{1}{z} + 36 z + 6 z^2 \Bigg)
\nonumber\\&+  \log^2(z) \log(1-z)   \Bigg(  - 30 - 24 z + 8 z^2 \Bigg)
\nonumber\\&+  \log^3(z)   \Bigg( \frac{17}{3} - \frac{4}{3} z - \frac{8}{3} z^2 \Bigg)+  \delta(1-z)   \Bigg( \frac{61}{3} \Bigg)
\nonumber\\&+  \log \left(\frac{Q^2}{\mu_F^2}\right)\Bigg[   \Bigg(  - \frac{235}{6} - \frac{128}{9} \frac{1}{z} + \frac{218}{3} z - \frac{347}{18}  z^2 \Bigg)
\nonumber\\&+   {\rm Li}_{2}(1-z)   \Bigg( 28 + 16 z - 16 z^2 \Bigg)
+   \log(1-z)   \Bigg( 8 + \frac{64}{3} \frac{1}{z} - 32 z + \frac{8}{3} z^2\Bigg)
\nonumber\\&+   \log(z)   \Bigg(  - 15 - \frac{32}{3} \frac{1}{z} + 16 z + \frac{52}{3}  z^2 \Bigg)
+  \log(z) \log(1-z)   \Bigg( 28 + 16 z - 16 z^2 \Bigg)
\nonumber\\&+   \log^2(z)   \Bigg(  - 15 - 12 z + 4 z^2 \Bigg)
+  \delta(1-z)   \Bigg(  - 4 \Bigg)\Bigg]
\nonumber\\&+  \log^2 \left(\frac{Q^2}{\mu_F^2}\right)  \Bigg[ \Bigg( 2 + \frac{16}{3} \frac{1}{z} - 8 z + \frac{2}{3} z^2 \Bigg)
+  \log(z)   \Bigg( 7 + 4 z - 4 z^2 \Bigg)\Bigg]
\nonumber\\&+  \zeta_3   \Bigg(  - 18 - 56 \frac{1}{z} - 36 z + 4 z^2 \Bigg)
+  \zeta_3 \delta(1-z)   \Bigg(  - 16 \Bigg)
\nonumber\\&+  \zeta_2   \Bigg( 2 - \frac{88}{3} \frac{1}{z} + 50 z - \frac{8}{3} z^2 \Bigg)
+  \zeta_2 \log(1+z)   \Bigg(  - 36 - 16 \frac{1}{z} - 24 z - 4 z^2 \Bigg)
\nonumber\\&+  \zeta_2 \log(z)   \Bigg(  - 10 - 8 \frac{1}{z} - 12 z + 20 z^2 \Bigg)\Bigg\}
\nonumber\\&+\mathbf{C_A n_f }  \Bigg\{ \Bigg( \frac{41}{108} - \frac{226}{27} \frac{1}{z} + \frac{224}{27} \frac{1}{1-z} - \frac{196}{27} z + \frac{751}{108}  z^2 \Bigg)
\nonumber\\&+  S_{1,2}(1-z)   \Bigg(  - 50 + 40 \frac{1}{z} + 28 z - 2 z^2 \Bigg) +  S_{1,2}(-z)   \Bigg( 36 + 16 \frac{1}{z} + 24 z + 4 z^2 \Bigg)
\nonumber\\&+  {\rm Li}_{3}(1-z)   \Bigg( 16 - 16 \frac{1}{z} - 8 z \Bigg) +  {\rm Li}_{3}(-z)   \Bigg( 54 + 24 \frac{1}{z} + 36 z + 6 z^2 \Bigg)
\nonumber\\&+  {\rm Li}_{2}(1-z)   \Bigg(  - 8 + \frac{16}{3} \frac{1}{z} + \frac{8}{3} \frac{1}{1-z} - 12 z + \frac{8}{3} z^2\Bigg)
\nonumber\\&+  {\rm Li}_{2}(-z)   \Bigg(  - 20 - \frac{16}{3} \frac{1}{z} - 24 z - \frac{28}{3} z^2 \Bigg)
\nonumber\\&+  \log(1+z) {\rm Li}_{2}(-z)   \Bigg( 36 + 16 \frac{1}{z} + 24 z + 4 z^2 \Bigg)
\nonumber\\&+  \log(1-z)   \Bigg(  - \frac{176}{9} - \frac{142}{9} \frac{1}{z} + \frac{400}{9} \frac{1}{1-z} - \frac{110}{9} z + \frac{16}{9} z^2 \Bigg)
\nonumber\\&+  \log(1-z) {\rm Li}_{2}(1-z)   \Bigg(  - 16 + 16 \frac{1}{z} + 8 z \Bigg)
\nonumber\\&+  \log^2(1-z)   \Bigg(  - \frac{40}{3} + 8 \frac{1}{z} + \frac{32}{3} \frac{1}{1-z} + \frac{8}{3} z - 8  z^2 \Bigg)
\nonumber\\&+  \log(z)   \Bigg(  - \frac{187}{9} + \frac{2}{9} \frac{1}{z} - \frac{200}{9} \frac{1}{1-z} + \frac{53}{9} z - \frac{625}{18} z^2 \Bigg)
\nonumber\\&+  \log(z) {\rm Li}_{2}(-z)   \Bigg(  - 54 - 24 \frac{1}{z} - 36 z - 6 z^2 \Bigg)
\nonumber\\&+  \log(z) \log(1+z)   \Bigg(  - 20 - \frac{16}{3} \frac{1}{z} - 24 z - \frac{28}{3} z^2 \Bigg)
\nonumber\\&+  \log(z) \log^2(1+z)   \Bigg( 18 + 8 \frac{1}{z} + 12 z + 2 z^2 \Bigg)
\nonumber\\&+  \log(z) \log(1-z)   \Bigg(  - 16 - \frac{32}{3} \frac{1}{1-z} + 4 z + \frac{16}{3} z^2\Bigg)
\nonumber\\&+  \log^2(z)   \Bigg( 37 - \frac{8}{3} \frac{1}{z} + \frac{4}{3} \frac{1}{1-z} + 3 z + \frac{27}{2} z^2 \Bigg)
\nonumber\\&+  \log^2(z) \log(1+z)   \Bigg(  - 27 - 12 \frac{1}{z} - 18 z - 3 z^2 \Bigg) +  \delta(1-z)   \Bigg(  - \frac{2983}{162} \Bigg)
\nonumber\\&+  \log \left(\frac{Q^2}{\mu_F^2}\right)  \Bigg[ \Bigg(  - \frac{316}{9} + \frac{140}{9} \frac{1}{z} + \frac{200}{9} \frac{1}{1-z} + \frac{116}{9} z - \frac{140}{9} z^2 \Bigg)
\nonumber\\&+   \log(1-z)   \Bigg( \frac{64}{3} - \frac{32}{3} \frac{1}{z} - \frac{32}{3}  \frac{1}{1-z} - \frac{32}{3} z + \frac{32}{3} z^2 \Bigg)
\nonumber\\&+ \log(z)   \Bigg(  - 16 + \frac{16}{3} \frac{1}{z} + \frac{16}{3} \frac{1}{1-z} - \frac{16}{3} z^2 \Bigg)
+ \delta(1-z)   \Bigg( \frac{647}{27} \Bigg)\Bigg]
\nonumber\\&+  \log^2 \left(\frac{Q^2}{\mu_F^2}\right) \Bigg[  \Bigg( 16 - 8 \frac{1}{z} - 8 \frac{1}{1-z} - 8 z + 8 z^2 \Bigg)
+   \delta(1-z)   \Bigg(  - \frac{44}{9} \Bigg)\Bigg]
\nonumber\\&+  \zeta_3   \Bigg( 36 + 16 \frac{1}{z} + 24 z + 4 z^2 \Bigg)
+  \zeta_3 \delta(1-z)   \Bigg( \frac{64}{3} \Bigg)
\nonumber\\&+  \zeta_2   \Bigg(  - \frac{86}{3} - \frac{32}{3} \frac{1}{1-z} + \frac{52}{3} z - \frac{22}{3} z^2 \Bigg)
+  \zeta_2 \log(1+z)   \Bigg( 18 + 8 \frac{1}{z} + 12 z + 2 z^2 \Bigg)
\nonumber\\&+  \zeta_2 \delta(1-z)   \Bigg(  - \frac{94}{9} \Bigg)
+  \zeta_2 \log \left(\frac{Q^2}{\mu_F^2}\right) \delta(1-z)   \Bigg(  - \frac{16}{3} \Bigg)\Bigg\}
\nonumber\\&+ \mathbf{C_A^2}\Bigg\{   \Bigg( \frac{20137}{27} - \frac{21595}{27} \frac{1}{z} - \frac{1616}{27} \frac{1}{1-z} - \frac{6605}{27} z + \frac{9679}{27} z^2 \Bigg)
\nonumber\\&+ S_{1,2}(1-z)   \Bigg( 64 \frac{1}{z^2} - 112 \frac{1}{z} - 144 \frac{1}{1-z} + 64 \frac{1}{1+z} - 528 z + 112 z^2 \Bigg)
\nonumber\\&+ S_{1,2}(-z)   \Bigg(  - 480 - 192 \frac{1}{z^2} - 320 \frac{1}{z} - 32 \frac{1}{1+z} - 288 z - 64 z^2 \Bigg)
\nonumber\\&+ {\rm Li}_{3}\left(\frac{1-z}{1+z}\right)   \Bigg(  - 128 - 64 \frac{1}{z} + 64 \frac{1}{1+z} - 64 z - 64 z^2 \Bigg)
\nonumber\\&+ {\rm Li}_{3}\left(-\frac{1-z}{1+z}\right)   \Bigg( 128 + 64 \frac{1}{z} - 64 \frac{1}{1+z} + 64 z + 64  z^2 \Bigg)
\nonumber\\&+ {\rm Li}_{3}(1-z)   \Bigg( 744 + 32 \frac{1}{z^2} - 16 \frac{1}{z} - 8 \frac{1}{1-z} - 64  \frac{1}{1+z} + 536 z + 88 z^2 \Bigg)
\nonumber\\&+ {\rm Li}_{3}(-z)   \Bigg(  - 272 - 160 \frac{1}{z^2} + 320 \frac{1}{z} - 16 \frac{1}{1+z} + 304 z - 64 z^2 \Bigg)
\nonumber\\&+ {\rm Li}_{2}(1-z)   \Bigg(  - 376 - \frac{748}{3} \frac{1}{z} - \frac{44}{3} \frac{1}{1-z} - 400 z - \frac{704}{3} z^2 \Bigg)
\nonumber\\&+ {\rm Li}_{2}(-z)   \Bigg( 296 + 264 \frac{1}{z} + 120 z + 88 z^2 \Bigg)
\nonumber\\&+ \log(1+z) {\rm Li}_{2}(-z)   \Bigg(  - 480 - 192 \frac{1}{z^2} - 320 \frac{1}{z} - 32  \frac{1}{1+z} - 288 z - 64 z^2 \Bigg)
\nonumber\\&+ \log(1-z)   \Bigg(  - \frac{1492}{9} + \frac{4396}{9} \frac{1}{z} - \frac{2176}{9} \frac{1}{1-z} + \frac{4040}{9}  z - \frac{4756}{9} z^2 \Bigg)
\nonumber\\&+ \log(1-z) {\rm Li}_{2}(1-z)   \Bigg(  - 496 - 16 \frac{1}{z} - 544 z \Bigg)
\nonumber\\&+ \log(1-z) {\rm Li}_{2}(-z)   \Bigg( 128 + 64 \frac{1}{z} - 64 \frac{1}{1+z} + 64 z + 64  z^2 \Bigg)
\nonumber\\&+ \log^2(1-z)   \Bigg( \frac{1324}{3} - 572 \frac{1}{z} - \frac{176}{3} \frac{1}{1-z} - \frac{1148}{3} z + 572 z^2 \Bigg)
\nonumber\\&+ \log^3(1-z)   \Bigg(  - 256 + 128 \frac{1}{z} + 128 \frac{1}{1-z} + 128 z - 128  z^2 \Bigg)
\nonumber\\&+ \log(z)   \Bigg( \frac{946}{9} - \frac{1264}{3} \frac{1}{z} + \frac{1088}{9} \frac{1}{1-z} - \frac{2582}{9} z + \frac{5740}{9} z^2 \Bigg)
\nonumber\\&+ \log(z) {\rm Li}_{2}(1-z)   \Bigg( 8 + 48 \frac{1}{z} - 56 \frac{1}{1-z} + 48 \frac{1}{1+z} + 24 z + 8 z^2 \Bigg)
\nonumber\\&+ \log(z) {\rm Li}_{2}(-z)   \Bigg( 256 + 160 \frac{1}{z^2} - 64 \frac{1}{z} + 64 \frac{1}{1+z} - 64 z + 16 z^2 \Bigg)
\nonumber\\&+ \log(z) \log(1+z)   \Bigg( 296 + 264 \frac{1}{z} + 120 z + 88 z^2 \Bigg)
\nonumber\\&+ \log(z) \log^2(1+z)   \Bigg(  - 240 - 96 \frac{1}{z^2} - 160 \frac{1}{z} - 16  \frac{1}{1+z} - 144 z - 32 z^2 \Bigg)
\nonumber\\&+ \log(z) \log(1-z)   \Bigg(  - 832 + \frac{1232}{3} \frac{1}{z} + \frac{176}{3} \frac{1}{1-z} + 72 z - 968 z^2 \Bigg)
\nonumber\\&+ \log(z) \log(1-z) \log(1+z)   \Bigg( 128 + 64 \frac{1}{z} - 64 \frac{1}{1+z} + 64 z + 64 z^2 \Bigg)
\nonumber\\&+ \log(z) \log^2(1-z)   \Bigg( 240 - 248 \frac{1}{z} - 248 \frac{1}{1-z} - 504 z + 248 z^2 \Bigg)
\nonumber\\&+ \log^2(z)   \Bigg( 28 - \frac{220}{3} \frac{1}{z} - \frac{22}{3} \frac{1}{1-z} + 44 z + \frac{550}{3}  z^2 \Bigg)
\nonumber\\&+ \log^2(z) \log(1+z)   \Bigg( 120 + 80 \frac{1}{z^2} + 96 \frac{1}{z} + 56 \frac{1}{1+z} + 88 z - 16 z^2 \Bigg)
\nonumber\\&+ \log^2(z) \log(1-z)   \Bigg(  - 96 + 112 \frac{1}{z} + 128 \frac{1}{1-z} + 16  \frac{1}{1+z} + 304 z - 144 z^2 \Bigg)
\nonumber\\&+ \log^3(z)   \Bigg(  - \frac{16}{3} - 16 \frac{1}{z} - \frac{56}{3} \frac{1}{1-z} - \frac{16}{3} \frac{1}{1+z} - \frac{184}{3} z + 24 z^2 \Bigg)
\nonumber\\&+ \delta(1-z)   \Bigg( \frac{7801}{324} \Bigg)
\nonumber\\&+ \log \left(\frac{Q^2}{\mu_F^2}\right) \Bigg[  \Bigg(  - \frac{518}{9} + 150 \frac{1}{z} - \frac{1088}{9} \frac{1}{1-z} + \frac{1606}{9} z - 150 z^2 \Bigg)
\nonumber\\&+ {\rm Li}_{2}(1-z)   \Bigg(  - 256 - 256 z \Bigg)
+ {\rm Li}_{2}(-z)   \Bigg( 64 + 32 \frac{1}{z} - 32 \frac{1}{1+z} + 32 z + 32 z^2 \Bigg)
\nonumber\\&+ \log(1-z)   \Bigg( \frac{752}{3} - \frac{1408}{3} \frac{1}{z} + \frac{176}{3}  \frac{1}{1-z} - \frac{928}{3} z + \frac{1408}{3} z^2 \Bigg)
\nonumber\\&+ \log^2(1-z)   \Bigg(  - 384 + 192 \frac{1}{z} + 192  \frac{1}{1-z} + 192 z - 192 z^2 \Bigg)
\nonumber\\&+ \log(z)   \Bigg(  - 376 + \frac{440}{3} \frac{1}{z} - \frac{88}{3}  \frac{1}{1-z} - 440 z^2 \Bigg)
\nonumber\\&+ \log(z) \log(1+z)   \Bigg( 64 + 32 \frac{1}{z} - 32  \frac{1}{1+z} + 32 z + 32 z^2 \Bigg)
\nonumber\\&+ \log(z) \log(1-z)   \Bigg( 192 - 224 \frac{1}{z} - 224  \frac{1}{1-z} - 480 z + 224 z^2 \Bigg)
\nonumber\\&+ \log^2(z)   \Bigg( 32 \frac{1}{z} + 40 \frac{1}{1-z} + 8  \frac{1}{1+z} + 128 z - 48 z^2 \Bigg)
+ \delta(1-z)   \Bigg(  - \frac{1657}{27} \Bigg)\Bigg]
\nonumber\\&+ \log^2 \left(\frac{Q^2}{\mu_F^2}\right) \Bigg[  \Bigg( 8 - \frac{220}{3} \frac{1}{z} + 44 \frac{1}{1-z} - 52 z + \frac{220}{3}  z^2 \Bigg)
\nonumber\\&+ \log(1-z)   \Bigg(  - 128 + 64 \frac{1}{z} + 64 \frac{1}{1-z} + 64 z - 64 z^2 \Bigg)
\nonumber\\&+ \log(z)   \Bigg(  - 32 \frac{1}{z} - 32 \frac{1}{1-z} - 96 z + 32 z^2 \Bigg)
+ \delta(1-z)   \Bigg( \frac{121}{9} \Bigg)\Bigg]
\nonumber\\&+ \zeta_3   \Bigg(  - 768 - 96 \frac{1}{z^2} + 592 \frac{1}{z} + 312 \frac{1}{1-z} - 8  \frac{1}{1+z} + 576 z - 352 z^2 \Bigg)
\nonumber\\&+ \zeta_3 \delta(1-z)   \Bigg(  - \frac{88}{3} \Bigg)
+ \zeta_3 \log \left(\frac{Q^2}{\mu_F^2}\right) \delta(1-z)   \Bigg( 152 \Bigg)
\nonumber\\&+ \zeta_2   \Bigg(  - \frac{436}{3} + \frac{1628}{3} \frac{1}{z} + \frac{176}{3} \frac{1}{1-z} + \frac{884}{3} z - \frac{1100}{3} z^2 \Bigg)
\nonumber\\&+ \zeta_2 \log(1+z)   \Bigg(  - 240 - 96 \frac{1}{z^2} - 160 \frac{1}{z} - 16 \frac{1}{1+z}
          - 144 z - 32 z^2 \Bigg)
\nonumber\\&+ \zeta_2 \log(1-z)   \Bigg( 384 - 128 \frac{1}{z} - 160 \frac{1}{1-z} - 32 \frac{1}{1+z}
          - 128 z + 192 z^2 \Bigg)
\nonumber\\&+ \zeta_2 \log(z)   \Bigg(  - 40 + 272 \frac{1}{z} + 144 \frac{1}{1-z} + 24 \frac{1}{1+z}+ 520 z - 168 z^2 \Bigg)
\nonumber\\&+ \zeta_2 \delta(1-z)   \Bigg(  - \frac{224}{9} \Bigg)
\nonumber\\&+ \zeta_2 \log \left(\frac{Q^2}{\mu_F^2}\right)   \Bigg( 192 - 64 \frac{1}{z} - 80 \frac{1}{1-z} - 16 
         \frac{1}{1+z} - 64 z + 96 z^2 \Bigg)
\nonumber\\&+ \zeta_2 \log \left(\frac{Q^2}{\mu_F^2}\right) \delta(1-z)   \Bigg( \frac{88}{3} \Bigg)
+ \zeta_2 \log^2 \left(\frac{Q^2}{\mu_F^2}\right) \delta(1-z)   \Bigg(  - 32 \Bigg)
\nonumber\\&+ \zeta_2^2 \delta(1-z)   \Bigg(  - \frac{4}{5} \Bigg)\Bigg\}\,,
\nonumber\\
   {\bar{\Delta}}_{gq}^{h,(1)} &=  \Bigg(\frac{9}{2} - 6 \frac{1}{z} + 9 z - \frac{7}{2} z^2\Bigg)
+ \log(1-z)   \Bigg(  - 14 + 16 \frac{1}{z} + 4 z + 4 z^2 \Bigg)
\nonumber\\&+ \log(z)   \Bigg( 7 - 8 \frac{1}{z} - 2 z - 2 z^2 \Bigg)
+ \log \left(\frac{Q^2}{\mu_F^2}\right)   \Bigg(  - 7 + 8 \frac{1}{z} + 2 z + 2 z^2 \Bigg)\,,
\nonumber\\
  {\bar{\Delta}}_{gq}^{h,(2)} &= \mathbf{n_f}\Bigg\{   \Bigg(  - \frac{160}{27} + \frac{304}{27} \frac{1}{z} + \frac{380}{27} z \Bigg)
+  {\rm Li}_{2}(1-z)   \Bigg(  - \frac{32}{3} + \frac{32}{3} \frac{1}{z} + \frac{4}{3} z \Bigg)
\nonumber\\&+  \log(1-z)   \Bigg(  - \frac{200}{9} + \frac{200}{9} \frac{1}{z} + \frac{88}{9} z + \frac{16}{3} z^2 \Bigg)
\nonumber\\&+  \log^2(1-z)   \Bigg(  - \frac{8}{3} + \frac{8}{3} \frac{1}{z} + \frac{4}{3} z \Bigg)
+  \log(z)   \Bigg(  - \frac{4}{3} z - \frac{16}{3} z^2 \Bigg)
\nonumber\\&+  \log(z) \log(1-z)   \Bigg(  - 4 z \Bigg)
+  \log^2(z)   \Bigg( 2 z \Bigg)
\nonumber\\&+  \log \left(\frac{Q^2}{\mu_F^2}\right) \Bigg[  \Bigg(  - \frac{40}{3} + \frac{40}{3} \frac{1}{z} + \frac{4}{3} z \Bigg)
+   \log(1-z)   \Bigg( \frac{32}{3} - \frac{32}{3} \frac{1}{z} - \frac{16}{3} z \Bigg)\Bigg]
\nonumber\\&+  \log^2 \left(\frac{Q^2}{\mu_F^2}\right)   \Bigg( \frac{16}{3} - \frac{16}{3} \frac{1}{z} - \frac{8}{3} z \Bigg)\Bigg\}
\nonumber\\&+ \mathbf{C_F}  \Bigg\{ \Bigg( \frac{1507}{36} - \frac{4057}{27} \frac{1}{z} + \frac{2935}{18} z - \frac{4499}{108} z^2 \Bigg)
\nonumber\\&+  S_{1,2}(1-z)   \Bigg(  - 118 + 64 \frac{1}{z} + 92 z - 68 z^2 \Bigg)
\nonumber\\&+  {\rm Li}_{3}(1-z)   \Bigg( 30 - 80 \frac{1}{z} + 4 z + 36 z^2 \Bigg)
+  {\rm Li}_{3}(-z)   \Bigg( 144 - 64 \frac{1}{z} - 160 z + 32 z^2 \Bigg)
\nonumber\\&+  {\rm Li}_{2}(1-z)   \Bigg( 167 - \frac{392}{3} \frac{1}{z} - 52 z + \frac{248}{3} z^2 \Bigg)
\nonumber\\&+  {\rm Li}_{2}(-z)   \Bigg( 40 - \frac{160}{3} \frac{1}{z} + 48 z - \frac{136}{3} z^2 \Bigg)
\nonumber\\&+  \log(1-z)   \Bigg(  - \frac{346}{3} + \frac{1360}{9} \frac{1}{z} - \frac{593}{3} z + \frac{728}{9} z^2 \Bigg)
\nonumber\\&+  \log(1-z) {\rm Li}_{2}(1-z)   \Bigg(  - 70 + 96 \frac{1}{z} + 44 z - 52 z^2 \Bigg)
\nonumber\\&+  \log^2(1-z)   \Bigg( 133 - \frac{392}{3} \frac{1}{z} + 22 z - \frac{157}{3} z^2 \Bigg)
\nonumber\\&+  \log^3(1-z)   \Bigg(  - 23 + \frac{104}{3} \frac{1}{z} - 6 z + \frac{70}{3} z^2 \Bigg)
\nonumber\\&+  \log(z)   \Bigg(  - \frac{19}{2} - \frac{284}{9} \frac{1}{z} + \frac{667}{6} z - \frac{919}{9} z^2 \Bigg)
\nonumber\\&+  \log(z) {\rm Li}_{2}(1-z)   \Bigg( 50 - 48 \frac{1}{z} - 44 z \Bigg)
\nonumber\\&+  \log(z) {\rm Li}_{2}(-z)   \Bigg(  - 72 + 32 \frac{1}{z} + 80 z - 16 z^2 \Bigg)
\nonumber\\&+  \log(z) \log(1+z)   \Bigg( 40 - \frac{160}{3} \frac{1}{z} + 48 z - \frac{136}{3} z^2 \Bigg)
\nonumber\\&+  \log(z) \log(1-z)   \Bigg(  - 130 + 72 \frac{1}{z} + 32 z + 96 z^2 \Bigg)
\nonumber\\&+  \log(z) \log^2(1-z)   \Bigg( 59 - 48 \frac{1}{z} + 2 z - 66 z^2 \Bigg)
+  \log^2(z)   \Bigg( \frac{325}{4} - 92 z + \frac{61}{3} z^2 \Bigg)
\nonumber\\&+  \log^2(z) \log(1-z)   \Bigg(  - 36 + 16 \frac{1}{z} + 40 z^2 \Bigg)
+  \log^3(z)   \Bigg( \frac{23}{6} + \frac{7}{3} z - \frac{26}{3} z^2 \Bigg)
\nonumber\\&+  \log \left(\frac{Q^2}{\mu_F^2}\right)\Bigg[   \Bigg(  - 40 + \frac{212}{9} \frac{1}{z} - 26 z - \frac{77}{9} z^2 \Bigg)
+   {\rm Li}_{2}(1-z)   \Bigg( 32 \frac{1}{z} - 24 z^2 \Bigg)
\nonumber\\&+  \log(1-z)   \Bigg( 80 - 72 \frac{1}{z} + 44 z - 46 z^2 \Bigg)
\nonumber\\&+   \log^2(1-z)   \Bigg(  - 30 + 48 \frac{1}{z} - 12 z + 36 z^2
          \Bigg)
+   \log(z)   \Bigg(  - 11 - 12 z + \frac{170}{3} z^2 \Bigg)
\nonumber\\&+   \log(z) \log(1-z)   \Bigg( 44 - 32 \frac{1}{z} + 8 z - 64 
         z^2 \Bigg)
+   \log^2(z)   \Bigg(  - 12 + 16 z^2 \Bigg)\Bigg]
\nonumber\\&+  \log^2 \left(\frac{Q^2}{\mu_F^2}\right) \Bigg[  \Bigg( \frac{13}{2} + 4 z \Bigg)
+  \log(1-z)   \Bigg(  - 10 + 16 \frac{1}{z} - 4 z + 12 z^2 \Bigg)
\nonumber\\&+   \log(z)   \Bigg( 5 + 2 z - 12 z^2 \Bigg)\Bigg]
\nonumber\\&+  \zeta_3   \Bigg( 82 + 16 \frac{1}{z} - 164 z + 100 z^2 \Bigg)
+  \zeta_2   \Bigg(  - 214 + \frac{512}{3} \frac{1}{z} + 156 z - 88 z^2 \Bigg)
\nonumber\\&+  \zeta_2 \log(1-z)   \Bigg( 120 - 128 \frac{1}{z} - 48 z - 16 z^2 \Bigg)
+  \zeta_2 \log(z)   \Bigg(  - 52 + 48 \frac{1}{z} - 8 z + 48 z^2 \Bigg)
\nonumber\\&+  \zeta_2 \log \left(\frac{Q^2}{\mu_F^2}\right)   \Bigg( 60 - 64 \frac{1}{z} - 24 z - 8 z^2 \Bigg)\Bigg\}
\nonumber\\&+ \mathbf{C_A}\Bigg\{   \Bigg( \frac{42329}{54} - \frac{21895}{27} \frac{1}{z} - \frac{2474}{27} z - \frac{137}{18} z^2 \Bigg)
\nonumber\\&+  S_{1,2}(1-z)   \Bigg( 20 + 128 \frac{1}{z^2} - 144 \frac{1}{z} - 200 z + 16 z^2 \Bigg)
\nonumber\\&+  S_{1,2}(-z)   \Bigg(  - 448 - 384 \frac{1}{z^2} - 640 \frac{1}{z} - 128 z \Bigg)
\nonumber\\&+  {\rm Li}_{3}\left(\frac{1-z}{1+z}\right)   \Bigg(  - 56 - 64 \frac{1}{z} - 16 z + 16 z^2 \Bigg)
\nonumber\\&+  {\rm Li}_{3}\left(-\frac{1-z}{1+z}\right)   \Bigg( 56 + 64 \frac{1}{z} + 16 z - 16 z^2 \Bigg)
\nonumber\\&+  {\rm Li}_{3}(1-z)   \Bigg( 722 + 64 \frac{1}{z^2} + 112 \frac{1}{z} + 220 z - 24 z^2 \Bigg)
\nonumber\\&+  {\rm Li}_{3}(-z)   \Bigg(  - 324 - 320 \frac{1}{z^2} + 608 \frac{1}{z} + 136 z - 8 z^2 \Bigg)
\nonumber\\&+  {\rm Li}_{2}(1-z)   \Bigg(  - \frac{1822}{3} - \frac{28}{3} \frac{1}{z} - \frac{532}{3} z - \frac{88}{3} z^2 \Bigg)
\nonumber\\&+  {\rm Li}_{2}(-z)   \Bigg( 516 + 504 \frac{1}{z} + 68 z + 48 z^2 \Bigg)
\nonumber\\&+  \log(1+z) {\rm Li}_{2}(-z)   \Bigg(  - 448 - 384 \frac{1}{z^2} - 640 \frac{1}{z} - 128 z \Bigg)
\nonumber\\&+  \log(1-z)   \Bigg(  - \frac{4618}{9} + \frac{3952}{9} \frac{1}{z} + \frac{143}{9} z - \frac{10}{3} z^2 \Bigg)
\nonumber\\&+  \log(1-z) {\rm Li}_{2}(1-z)   \Bigg(  - 500 - 224 \frac{1}{z} - 272 z + 20 z^2 \Bigg)
\nonumber\\&+  \log(1-z) {\rm Li}_{2}(-z)   \Bigg( 56 + 64 \frac{1}{z} + 16 z - 16 z^2 \Bigg)
\nonumber\\&+  \log^2(1-z)   \Bigg( \frac{1112}{3} - \frac{1484}{3} \frac{1}{z} + \frac{452}{3} z + 2 z^2 \Bigg)
\nonumber\\&+  \log^3(1-z)   \Bigg(  - 89 + \frac{280}{3} \frac{1}{z} + 38 z + \frac{26}{3} z^2 \Bigg)
\nonumber\\&+  \log(z)   \Bigg( \frac{1174}{3} - \frac{1544}{3} \frac{1}{z} - 98 z - \frac{11}{3} z^2 \Bigg)
\nonumber\\&+  \log(z) {\rm Li}_{2}(1-z)   \Bigg(  - 48 + 192 \frac{1}{z} + 68 z - 8 z^2 \Bigg)
\nonumber\\&+  \log(z) {\rm Li}_{2}(-z)   \Bigg( 328 + 320 \frac{1}{z^2} - 64 \frac{1}{z} - 16 z + 16 z^2 \Bigg)
\nonumber\\&+  \log(z) \log(1+z)   \Bigg( 516 + 504 \frac{1}{z} + 68 z + 48 z^2 \Bigg)
\nonumber\\&+  \log(z) \log^2(1+z)   \Bigg(  - 224 - 192 \frac{1}{z^2} - 320 \frac{1}{z} - 64 z \Bigg)
\nonumber\\&+  \log(z) \log(1-z)   \Bigg(  - 838 + \frac{1208}{3} \frac{1}{z} - 416 z + \frac{70}{3} z^2 \Bigg)
\nonumber\\&+  \log(z) \log(1-z) \log(1+z)   \Bigg( 56 + 64 \frac{1}{z} + 16 z - 16 z^2 \Bigg)
\nonumber\\&+  \log(z) \log^2(1-z)   \Bigg(  - 118 - 264 \frac{1}{z} - 184 z - 12 z^2 \Bigg)
\nonumber\\&+  \log^2(z)   \Bigg(  - \frac{341}{2} - \frac{284}{3} \frac{1}{z} + 130 z - \frac{181}{3} z^2 \Bigg)
\nonumber\\&+  \log^2(z) \log(1+z)   \Bigg( 166 + 160 \frac{1}{z^2} + 240 \frac{1}{z} + 52 z + 12 
         z^2 \Bigg)
\nonumber\\&+  \log^2(z) \log(1-z)   \Bigg( 90 + 128 \frac{1}{z} + 100 z + 4 z^2 \Bigg)
+  \log^3(z)   \Bigg(  - \frac{95}{3} - \frac{80}{3} \frac{1}{z} - \frac{28}{3} z \Bigg)
\nonumber\\&+  \log \left(\frac{Q^2}{\mu_F^2}\right) \Bigg[  \Bigg(  - \frac{709}{3} + \frac{620}{3} \frac{1}{z} + \frac{64}{3} z + 17 z^2 \Bigg)
\nonumber\\&+   {\rm Li}_{2}(1-z)   \Bigg(  - 276 - 96 \frac{1}{z} - 120 z + 8 z^2 \Bigg)
+   {\rm Li}_{2}(-z)   \Bigg( 28 + 32 \frac{1}{z} + 8 z - 8 z^2 \Bigg)
\nonumber\\&+   \log(1-z)   \Bigg( \frac{1042}{3} - 456 \frac{1}{z} + \frac{460}{3} z + \frac{50}{3}
          z^2 \Bigg)
\nonumber\\&+   \log^2(1-z)   \Bigg(  - 138 + 144 \frac{1}{z} + 60 z + 12 
         z^2 \Bigg)
\nonumber\\&+  \log(z)   \Bigg(  - 458 + \frac{568}{3} \frac{1}{z} - 188 z - 8 z^2
          \Bigg)
\nonumber\\&+   \log(z) \log(1+z)   \Bigg( 28 + 32 \frac{1}{z} + 8 z - 8 
         z^2 \Bigg)
\nonumber\\&+  \log(z) \log(1-z)   \Bigg(  - 124 - 256 \frac{1}{z} - 184 z
          - 8 z^2 \Bigg)
+   \log^2(z)   \Bigg( 60 + 48 \frac{1}{z} + 44 z \Bigg)\Bigg]
\nonumber\\&+  \log^2 \left(\frac{Q^2}{\mu_F^2}\right) \Bigg[  \Bigg( \frac{203}{3} - \frac{280}{3} \frac{1}{z} + \frac{104}{3} z + \frac{17}{3} z^2 \Bigg)
\nonumber\\&+   \log(1-z)   \Bigg(  - 46 + 48 \frac{1}{z} + 20 z + 4 z^2 \Bigg)
+   \log(z)   \Bigg(  - 46 - 48 \frac{1}{z} - 40 z \Bigg)\Bigg]
\nonumber\\&+  \zeta_3   \Bigg(  - 434 - 192 \frac{1}{z^2} + 784 \frac{1}{z} + 240 z - 4 z^2 \Bigg)
\nonumber\\&+  \zeta_2   \Bigg( 138 + \frac{1636}{3} \frac{1}{z} - 164 z + \frac{230}{3} z^2 \Bigg)
\nonumber\\&+  \zeta_2 \log(1+z)   \Bigg(  - 224 - 192 \frac{1}{z^2} - 320 \frac{1}{z} - 64 z \Bigg)
+  \zeta_2 \log(1-z)   \Bigg( 48 + 16 z - 32 z^2 \Bigg)
\nonumber\\&+  \zeta_2 \log(z)   \Bigg( 240 + 416 \frac{1}{z} + 184 z + 16 z^2 \Bigg)
+  \zeta_2 \log \left(\frac{Q^2}{\mu_F^2}\right)   \Bigg( 24 + 8 z - 16 z^2 \Bigg)\Bigg\}\,,
\nonumber\\
{\bar{\Delta}}_{q{\bar q}}^{h,(0)} &= \delta(1-z) \,,
\nonumber\\
   {\bar{\Delta}}_{q{\bar q}}^{h,(1)} &= \mathbf{C_F}\Bigg\{   \Bigg( \frac{16}{3} \frac{1}{z} - \frac{16}{3} z^2 \Bigg)
+  \log(1-z)   \Bigg(  - 8 + 16 \frac{1}{1-z} - 8 z \Bigg)
\nonumber\\&+  \log(z)   \Bigg( 4 - 8 \frac{1}{1-z} + 4 z \Bigg)
+  \delta(1-z)   \Bigg(  - 20 \Bigg)
\nonumber\\&+  \log \left(\frac{Q^2}{\mu_F^2}\right) \Bigg[  \Bigg(  - 4 + 8 \frac{1}{1-z} - 4 z \Bigg)
+   \delta(1-z)   \Bigg( 6 \Bigg)\Bigg]
+  \zeta_2 \delta(1-z)   \Bigg( 8 \Bigg)\Bigg\}\,,
\nonumber\\
   {\bar{\Delta}}_{q{\bar q}}^{h,(2)} &= \mathbf{C_F}\Bigg\{   \Bigg( \frac{5417}{9} - \frac{12304}{27} \frac{1}{z} - \frac{1550}{9} z + \frac{703}{27} z^2 \Bigg)
\nonumber\\&+  S_{1,2}(1-z)   \Bigg( 144 + 128 \frac{1}{z^2} + 64 \frac{1}{z} + 64 z \Bigg)
\nonumber\\&+  S_{1,2}(-z)   \Bigg(  - 288 - 384 \frac{1}{z^2} - 576 \frac{1}{z} - 48 z \Bigg)
\nonumber\\&+  {\rm Li}_{3}(1-z)   \Bigg( 312 + 64 \frac{1}{z^2} + 128 \frac{1}{z} + 12 z \Bigg)
\nonumber\\&+  {\rm Li}_{3}(-z)   \Bigg(  - 144 - 320 \frac{1}{z^2} + 672 \frac{1}{z} + 40 z \Bigg)
\nonumber\\&+  {\rm Li}_{2}(1-z)   \Bigg(  - 248 + \frac{736}{3} \frac{1}{z} - 108 z + \frac{32}{3} z^2 \Bigg)
+  {\rm Li}_{2}(-z)   \Bigg( 488 + 448 \frac{1}{z} + 40 z \Bigg)
\nonumber\\&+  \log(1+z) {\rm Li}_{2}(-z)   \Bigg(  - 288 - 384 \frac{1}{z^2} - 576 \frac{1}{z} - 48 z \Bigg)
\nonumber\\&+  \log(1-z)   \Bigg(  - \frac{1160}{3} + \frac{3272}{9} \frac{1}{z} + \frac{128}{3} z - \frac{176}{9} z^2 \Bigg)
\nonumber\\&+  \log(1-z) {\rm Li}_{2}(1-z)   \Bigg(  - 224 - 256 \frac{1}{z} - 32 z \Bigg)
\nonumber\\&+  \log^2(1-z)   \Bigg( 136 - \frac{544}{3} \frac{1}{z} + 56 z - \frac{32}{3} z^2 \Bigg)
\nonumber\\&+  \log(z)   \Bigg( \frac{974}{3} - 284 \frac{1}{z} - \frac{68}{3} z + \frac{40}{9} z^2 \Bigg)
+  \log(z) {\rm Li}_{2}(1-z)   \Bigg( 32 + 224 \frac{1}{z} + 52 z \Bigg)
\nonumber\\&+  \log(z) {\rm Li}_{2}(-z)   \Bigg( 192 + 320 \frac{1}{z^2} - 96 \frac{1}{z} \Bigg)
+  \log(z) \log(1+z)   \Bigg( 488 + 448 \frac{1}{z} + 40 z \Bigg)
\nonumber\\&+  \log(z) \log^2(1+z)   \Bigg(  - 144 - 192 \frac{1}{z^2} - 288 \frac{1}{z} - 24 z \Bigg)
\nonumber\\&+  \log(z) \log(1-z)   \Bigg(  - 232 + 192 \frac{1}{z} - 96 z + 32 z^2 \Bigg)
\nonumber\\&+  \log(z) \log^2(1-z)   \Bigg(  - 112 - 128 \frac{1}{z} - 16 z \Bigg)
\nonumber\\&+  \log^2(z)   \Bigg(  - 277 - 48 \frac{1}{z} - 35 z - \frac{40}{3} z^2 \Bigg)
\nonumber\\&+  \log^2(z) \log(1+z)   \Bigg( 120 + 160 \frac{1}{z^2} + 240 \frac{1}{z} + 20 z \Bigg)
\nonumber\\&+  \log^2(z) \log(1-z)   \Bigg( 48 + 64 \frac{1}{z} \Bigg)
+  \log^3(z)   \Bigg(  - \frac{14}{3} - \frac{64}{3} \frac{1}{z} + \frac{34}{3} z \Bigg)
\nonumber\\&+  \log \left(\frac{Q^2}{\mu_F^2}\right) \Bigg[  \Bigg(  - \frac{580}{3} + \frac{1636}{9} \frac{1}{z} + \frac{64}{3} z - \frac{88}{9} z^2 \Bigg)
\nonumber\\&+   {\rm Li}_{2}(1-z)   \Bigg(  - 112 - 128 \frac{1}{z} - 16 z \Bigg)
\nonumber\\&+   \log(1-z)   \Bigg( 136 - \frac{544}{3} \frac{1}{z} + 56 z - \frac{32}{3} z^2
          \Bigg)
+   \log(z)   \Bigg(  - 116 + 96 \frac{1}{z} - 48 z + 16 z^2 \Bigg)
\nonumber\\&+  \log(z) \log(1-z)   \Bigg(  - 112 - 128 \frac{1}{z} - 16 z
          \Bigg)
+   \log^2(z)   \Bigg( 24 + 32 \frac{1}{z} \Bigg)\Bigg]
\nonumber\\&+  \log^2 \left(\frac{Q^2}{\mu_F^2}\right) \Bigg[  \Bigg( 34 - \frac{136}{3} \frac{1}{z} + 14 z - \frac{8}{3} z^2 \Bigg)
+   \log(z)   \Bigg(  - 28 - 32 \frac{1}{z} - 4 z \Bigg)\Bigg]
\nonumber\\&+  \zeta_3   \Bigg(  - 72 - 192 \frac{1}{z^2} + 576 \frac{1}{z} + 36 z \Bigg)
+  \zeta_2   \Bigg( 108 + \frac{1216}{3} \frac{1}{z} - 36 z + \frac{32}{3} z^2 \Bigg)
\nonumber\\&+  \zeta_2 \log(1+z)   \Bigg(  - 144 - 192 \frac{1}{z^2} - 288 \frac{1}{z} - 24 z \Bigg)
+  \zeta_2 \log(z)   \Bigg( 136 + 416 \frac{1}{z} + 36 z \Bigg)\Bigg\}
\nonumber\\&+ \mathbf{C_F n_f}\Bigg\{   \Bigg( \frac{1280}{27} - \frac{8}{9} \frac{1}{z} + \frac{224}{27} \frac{1}{1-z} - \frac{1252}{27} z - \frac{76}{9} z^2
          \Bigg)
\nonumber\\&+  {\rm Li}_{2}(1-z)   \Bigg( \frac{8}{3} - \frac{32}{3} \frac{1}{z} - \frac{8}{3} \frac{1}{1-z} + \frac{8}{3} z \Bigg)
+  {\rm Li}_{2}(-z)   \Bigg(  - 48 - \frac{64}{3} \frac{1}{z} - 32 z - \frac{16}{3} z^2 \Bigg)
\nonumber\\&+  \log(1-z)   \Bigg(  - \frac{16}{9} + \frac{64}{9} \frac{1}{z} - \frac{160}{9} \frac{1}{1-z} + \frac{176}{9} z
          - \frac{64}{9} z^2 \Bigg)
\nonumber\\&+  \log^2(1-z)   \Bigg(  - \frac{16}{3} + \frac{32}{3} \frac{1}{1-z} - \frac{16}{3} z \Bigg)
\nonumber\\&+  \log(z)   \Bigg( \frac{88}{3} - \frac{40}{3} \frac{1}{z} + \frac{40}{3} \frac{1}{1-z} + \frac{52}{3} z + \frac{164}{9} 
         z^2 \Bigg)
\nonumber\\&+  \log(z) \log(1+z)   \Bigg(  - 48 - \frac{64}{3} \frac{1}{z} - 32 z - \frac{16}{3} z^2 \Bigg)
\nonumber\\&+  \log(z) \log(1-z)   \Bigg( \frac{32}{3} - \frac{64}{3} \frac{1}{1-z} + \frac{32}{3} z \Bigg)
\nonumber\\&+  \log^2(z)   \Bigg( \frac{26}{3} + 8 \frac{1}{1-z} + \frac{14}{3} z + \frac{4}{3} z^2 \Bigg)
+  \delta(1-z)   \Bigg( \frac{461}{18} \Bigg)
\nonumber\\&+  \log \left(\frac{Q^2}{\mu_F^2}\right) \Bigg[  \Bigg(  - \frac{8}{9} + \frac{32}{9} \frac{1}{z} - \frac{80}{9} \frac{1}{1-z} + \frac{88}{9} 
         z - \frac{32}{9} z^2 \Bigg)
\nonumber\\&+   \log(1-z)   \Bigg(  - \frac{16}{3} + \frac{32}{3} \frac{1}{1-z} - \frac{16}{3} 
         z \Bigg)
\nonumber\\&+   \log(z)   \Bigg( \frac{16}{3} - \frac{32}{3} \frac{1}{1-z} + \frac{16}{3} z \Bigg)
+   \delta(1-z)   \Bigg(  - 14 \Bigg)\Bigg]
\nonumber\\&+  \log^2 \left(\frac{Q^2}{\mu_F^2}\right)\Bigg[   \Bigg(  - \frac{4}{3} + \frac{8}{3} \frac{1}{1-z} - \frac{4}{3} z \Bigg)
+   \delta(1-z)   \Bigg( 2 \Bigg)\Bigg]
+  \zeta_3 \delta(1-z)   \Bigg( 8 \Bigg)
\nonumber\\&+  \zeta_2   \Bigg(  - \frac{56}{3} - \frac{32}{3} \frac{1}{z} - \frac{32}{3} \frac{1}{1-z} - \frac{32}{3} z - \frac{8}{3} 
         z^2 \Bigg)
+ \zeta_2 \delta(1-z)   \Bigg(  - \frac{64}{9} \Bigg)\Bigg\}
\nonumber\\&+ \mathbf{C_F^2}\Bigg\{   \Bigg( 96 + 136 \frac{1}{z} - 236 z + 12 z^2 \Bigg)
\nonumber\\&+  S_{1,2}(1-z)   \Bigg( 696 - 256 \frac{1}{z} - 208 \frac{1}{1-z} + 696 z + 64 z^2 \Bigg)
\nonumber\\&+  S_{1,2}(-z)   \Bigg(  - 864 - 384 \frac{1}{z} - 576 z - 96 z^2 \Bigg)
\nonumber\\&+  {\rm Li}_{3}(1-z)   \Bigg(  - 80 + 96 \frac{1}{z} + 48 \frac{1}{1-z} - 48 z \Bigg)
\nonumber\\&+  {\rm Li}_{3}(-z)   \Bigg(  - 144 - 192 \frac{1}{z} - 224 z - 16 z^2 \Bigg)
\nonumber\\&+  {\rm Li}_{2}(1-z)   \Bigg( 232 + \frac{320}{3} \frac{1}{z} - 24 \frac{1}{1-z} - 392 z - \frac{88}{3} z^2
          \Bigg)
\nonumber\\&+  {\rm Li}_{2}(-z)   \Bigg( 688 + \frac{640}{3} \frac{1}{z} + 608 z + \frac{400}{3} z^2 \Bigg)
\nonumber\\&+  \log(1+z) {\rm Li}_{2}(-z)   \Bigg(  - 864 - 384 \frac{1}{z} - 576 z - 96 z^2 \Bigg)
\nonumber\\&+  \log(1-z)   \Bigg(  - \frac{32}{3} + 96 \frac{1}{z} - 320 \frac{1}{1-z} + \frac{500}{3} z + 64 
         z^2 \Bigg)
\nonumber\\&+  \log(1-z) {\rm Li}_{2}(1-z)   \Bigg( 56 - 128 \frac{1}{z} - 16 \frac{1}{1-z} + 56 z \Bigg)
\nonumber\\&+  \log^2(1-z)   \Bigg(  - 64 + \frac{128}{3} \frac{1}{z} + 64 z - \frac{128}{3} z^2 \Bigg)
\nonumber\\&+  \log^3(1-z)   \Bigg(  - 64 + 128 \frac{1}{1-z} - 64 z \Bigg)
\nonumber\\&+  \log(z)   \Bigg( \frac{1360}{3} - 64 \frac{1}{z} + 192 \frac{1}{1-z} + 12 z - 64 z^2 \Bigg)
\nonumber\\&+  \log(z) {\rm Li}_{2}(1-z)   \Bigg( 344 + 32 \frac{1}{z} - 96 \frac{1}{1-z} + 312 z + 32
          z^2 \Bigg)
\nonumber\\&+  \log(z) {\rm Li}_{2}(-z)   \Bigg( 432 + 256 \frac{1}{z} + 352 z + 48 z^2 \Bigg)
\nonumber\\&+  \log(z) \log(1+z)   \Bigg( 688 + \frac{640}{3} \frac{1}{z} + 608 z + \frac{400}{3} z^2 \Bigg)
\nonumber\\&+  \log(z) \log^2(1+z)   \Bigg(  - 432 - 192 \frac{1}{z} - 288 z - 48 z^2 \Bigg)
\nonumber\\&+  \log(z) \log(1-z)   \Bigg( 96 - \frac{128}{3} \frac{1}{z} - 48 \frac{1}{1-z} - 144 z + 
         128 z^2 \Bigg)
\nonumber\\&+  \log(z) \log^2(1-z)   \Bigg( 156 - 248 \frac{1}{1-z} + 156 z \Bigg)
\nonumber\\&+  \log^2(z)   \Bigg(  - 184 + 30 \frac{1}{1-z} - 384 z - \frac{500}{3} z^2 \Bigg)
\nonumber\\&+  \log^2(z) \log(1+z)   \Bigg( 360 + 160 \frac{1}{z} + 240 z + 40 z^2 \Bigg)
\nonumber\\&+  \log^2(z) \log(1-z)   \Bigg(  - 80 + 112 \frac{1}{1-z} - 80 z \Bigg)
\nonumber\\&+  \log^3(z)   \Bigg( \frac{110}{3} - \frac{40}{3} \frac{1}{1-z} + \frac{134}{3} z + \frac{8}{3} z^2 \Bigg)
+  \delta(1-z)   \Bigg( \frac{2293}{12} \Bigg)
\nonumber\\&+  \log \left(\frac{Q^2}{\mu_F^2}\right) \Bigg[  \Bigg( \frac{248}{3} - 160 \frac{1}{1-z} + \frac{232}{3} z \Bigg)
+   {\rm Li}_{2}(1-z)   \Bigg( 32 + 32 z \Bigg)
\nonumber\\&+   \log(1-z)   \Bigg(  - 112 + \frac{128}{3} \frac{1}{z} + 96 
         \frac{1}{1-z} + 16 z - \frac{128}{3} z^2 \Bigg)
\nonumber\\&+   \log^2(1-z)   \Bigg(  - 96 + 192 \frac{1}{1-z} - 96 z \Bigg)
\nonumber\\&+  \log(z)   \Bigg( 40 - 72 \frac{1}{1-z} - 40 z + \frac{128}{3} 
         z^2 \Bigg)
\nonumber\\&+   \log(z) \log(1-z)   \Bigg( 144 - 224 \frac{1}{1-z} + 
         144 z \Bigg)
\nonumber\\&+   \log^2(z)   \Bigg(  - 28 + 32 \frac{1}{1-z} - 28 z \Bigg)
+   \delta(1-z)   \Bigg(  - 117 \Bigg)\Bigg]
\nonumber\\&+  \log^2 \left(\frac{Q^2}{\mu_F^2}\right) \Bigg[  \Bigg(  - 40 + 48 \frac{1}{1-z} - 8 z \Bigg)
\nonumber\\&+   \log(1-z)   \Bigg(  - 32 + 64 \frac{1}{1-z} - 32 z \Bigg)
\nonumber\\&+   \log(z)   \Bigg( 24 - 32 \frac{1}{1-z} + 24 z \Bigg)
+   \delta(1-z)   \Bigg( 18 \Bigg)\Bigg]
\nonumber\\&+  \zeta_3   \Bigg(  - 128 - 96 \frac{1}{z} + 256 \frac{1}{1-z} - 224 z \Bigg)
+  \zeta_3 \delta(1-z)   \Bigg(  - 124 \Bigg)
\nonumber\\&+  \zeta_3 \log \left(\frac{Q^2}{\mu_F^2}\right) \delta(1-z)   \Bigg( 176 \Bigg)
+  \zeta_2   \Bigg( 408 + \frac{448}{3} \frac{1}{z} + 240 z + 24 z^2 \Bigg)
\nonumber\\&+  \zeta_2 \log(1+z)   \Bigg(  - 432 - 192 \frac{1}{z} - 288 z - 48 z^2 \Bigg)
\nonumber\\&+  \zeta_2 \log(1-z)   \Bigg( 64 - 128 \frac{1}{1-z} + 64 z \Bigg)
\nonumber\\&+  \zeta_2 \log(z)   \Bigg( 48 + 32 \frac{1}{z} + 128 \frac{1}{1-z} - 32 z + 16 z^2 \Bigg)
+  \zeta_2 \delta(1-z)   \Bigg(  - 70 \Bigg)
\nonumber\\&+  \zeta_2 \log \left(\frac{Q^2}{\mu_F^2}\right)   \Bigg( 32 - 64 \frac{1}{1-z} + 32 z \Bigg)
+  \zeta_2 \log \left(\frac{Q^2}{\mu_F^2}\right) \delta(1-z)   \Bigg( 24 \Bigg)
\nonumber\\&+  \zeta_2 \log^2 \left(\frac{Q^2}{\mu_F^2}\right) \delta(1-z)   \Bigg(  - 32 \Bigg)
+  \zeta_2^2 \delta(1-z)   \Bigg( \frac{8}{5} \Bigg)\Bigg\}
\nonumber\\&+ \mathbf{C_A C_F}\Bigg\{   \Bigg(  - \frac{4154}{27} - \frac{584}{27} \frac{1}{z} - \frac{1616}{27} \frac{1}{1-z} + \frac{7300}{27} z - 
         \frac{1162}{27} z^2 \Bigg)
\nonumber\\&+  S_{1,2}(1-z)   \Bigg(  - 432 + 256 \frac{1}{z} + 64 \frac{1}{1-z} - 240 z - 32 z^2
          \Bigg)
\nonumber\\&+  S_{1,2}(-z)   \Bigg( 432 + 192 \frac{1}{z} + 288 z + 48 z^2 \Bigg)
\nonumber\\&+  {\rm Li}_{3}(1-z)   \Bigg( 88 - 96 \frac{1}{z} - 56 \frac{1}{1-z} - 8 z \Bigg)
\nonumber\\&+  {\rm Li}_{3}(-z)   \Bigg( 72 + 32 \frac{1}{z} + 48 z + 8 z^2 \Bigg)
\nonumber\\&+  {\rm Li}_{2}(1-z)   \Bigg(  - \frac{464}{3} - \frac{272}{3} \frac{1}{z} + \frac{44}{3} \frac{1}{1-z} + \frac{340}{3} z
   + \frac{140}{3} z^2 \Bigg)
\nonumber\\&+  {\rm Li}_{2}(-z)   \Bigg(  - 408 - 160 \frac{1}{z} - 336 z - 88 z^2 \Bigg)
\nonumber\\&+  \log(1+z) {\rm Li}_{2}(-z)   \Bigg( 432 + 192 \frac{1}{z} + 288 z + 48 z^2 \Bigg)
\nonumber\\&+  \log(1-z)   \Bigg( \frac{2440}{9} - \frac{664}{3} \frac{1}{z} + \frac{1072}{9} \frac{1}{1-z} - \frac{2252}{9} z
  + \frac{256}{3} z^2 \Bigg)
\nonumber\\&+  \log(1-z) {\rm Li}_{2}(1-z)   \Bigg(  - 64 + 128 \frac{1}{z} + 16 \frac{1}{1-z} + 32 z
          \Bigg)
\nonumber\\&+  \log^2(1-z)   \Bigg( \frac{88}{3} - \frac{176}{3} \frac{1}{1-z} + \frac{88}{3} z \Bigg)
\nonumber\\&+  \log(z)   \Bigg(  - \frac{1136}{3} + \frac{232}{3} \frac{1}{z} - \frac{280}{3} \frac{1}{1-z} + \frac{296}{3} z
          - \frac{256}{3} z^2 \Bigg)
\nonumber\\&+  \log(z) {\rm Li}_{2}(1-z)   \Bigg(  - 168 - 32 \frac{1}{z} + 40 \frac{1}{1-z} - 168 z
          - 16 z^2 \Bigg)
\nonumber\\&+  \log(z) {\rm Li}_{2}(-z)   \Bigg(  - 216 - 96 \frac{1}{z} - 144 z - 24 z^2 \Bigg)
\nonumber\\&+  \log(z) \log(1+z)   \Bigg(  - 408 - 160 \frac{1}{z} - 336 z - 88 z^2 \Bigg)
\nonumber\\&+  \log(z) \log^2(1+z)   \Bigg( 216 + 96 \frac{1}{z} + 144 z + 24 z^2 \Bigg)
\nonumber\\&+  \log(z) \log(1-z)   \Bigg(  - \frac{512}{3} + \frac{352}{3} \frac{1}{1-z} - \frac{176}{3} z - 
         \frac{128}{3} z^2 \Bigg)
\nonumber\\&+  \log^2(z)   \Bigg( \frac{631}{3} - 44 \frac{1}{1-z} + \frac{667}{3} z + \frac{314}{3} z^2 \Bigg)
\nonumber\\&+  \log^2(z) \log(1+z)   \Bigg(  - 180 - 80 \frac{1}{z} - 120 z - 20 z^2 \Bigg)
\nonumber\\&+  \log^2(z) \log(1-z)   \Bigg(  - 8 + 16 \frac{1}{1-z} - 8 z \Bigg)
\nonumber\\&+  \log^3(z)   \Bigg(  - 10 - \frac{16}{3} \frac{1}{1-z} - 14 z - \frac{4}{3} z^2 \Bigg)
+  \delta(1-z)   \Bigg(  - \frac{5941}{36} \Bigg)
\nonumber\\&+  \log \left(\frac{Q^2}{\mu_F^2}\right) \Bigg[  \Bigg( \frac{212}{9} - \frac{176}{9} \frac{1}{z} + \frac{536}{9} \frac{1}{1-z} - \frac{748}{9} z + \frac{176}{9} z^2 \Bigg)
\nonumber\\&+   \log(1-z)   \Bigg( \frac{88}{3} - \frac{176}{3} \frac{1}{1-z} + \frac{88}{3} z
          \Bigg)
+   \log(z)   \Bigg(  - \frac{64}{3} + \frac{176}{3} \frac{1}{1-z} - \frac{64}{3} z
          \Bigg)
\nonumber\\&+   \log^2(z)   \Bigg(  - 4 + 8 \frac{1}{1-z} - 4 z \Bigg)
+   \delta(1-z)   \Bigg( 79 \Bigg)\Bigg]
\nonumber\\&+  \log^2 \left(\frac{Q^2}{\mu_F^2}\right)\Bigg[   \Bigg( \frac{22}{3} - \frac{44}{3} \frac{1}{1-z} + \frac{22}{3} z \Bigg)
+   \delta(1-z)   \Bigg(  - 11 \Bigg)\Bigg]
\nonumber\\&+  \zeta_3   \Bigg(  - 28 + 56 \frac{1}{1-z} - 28 z \Bigg)
+  \zeta_3 \delta(1-z)   \Bigg( 92 \Bigg)
\nonumber\\&+  \zeta_3 \log \left(\frac{Q^2}{\mu_F^2}\right) \delta(1-z)   \Bigg(  - 24 \Bigg)
\nonumber\\&+  \zeta_2   \Bigg(  - \frac{688}{3} - \frac{272}{3} \frac{1}{z} + \frac{176}{3} \frac{1}{1-z} - \frac{604}{3} z - 
         \frac{100}{3} z^2 \Bigg)
\nonumber\\&+  \zeta_2 \log(1+z)   \Bigg( 216 + 96 \frac{1}{z} + 144 z + 24 z^2 \Bigg)
+  \zeta_2 \log(1-z)   \Bigg( 16 - 32 \frac{1}{1-z} + 16 z \Bigg)
\nonumber\\&+  \zeta_2 \log(z)   \Bigg(  - 80 - 32 \frac{1}{z} + 16 \frac{1}{1-z} - 56 z - 8 z^2
          \Bigg)
+  \zeta_2 \delta(1-z)   \Bigg( \frac{328}{9} \Bigg)
\nonumber\\&+  \zeta_2 \log \left(\frac{Q^2}{\mu_F^2}\right)   \Bigg( 8 - 16 \frac{1}{1-z} + 8 z \Bigg)
+  \zeta_2^2 \delta(1-z)   \Bigg(  - \frac{12}{5} \Bigg)\Bigg\}\,,
\nonumber\\
{\bar{\Delta}}_{qq}^{h,(2)} &=\mathbf{C_F}\Bigg\{   \Bigg( \frac{5417}{9} - \frac{12304}{27} \frac{1}{z} - \frac{1550}{9} z + \frac{703}{27} z^2 \Bigg)
\nonumber\\&+  S_{1,2}(1-z)   \Bigg( 144 + 128 \frac{1}{z^2} + 64 \frac{1}{z} + 64 z \Bigg)
\nonumber\\&+  S_{1,2}(-z)   \Bigg(  - 288 - 384 \frac{1}{z^2} - 576 \frac{1}{z} - 48 z \Bigg)
\nonumber\\&+  {\rm Li}_{3}(1-z)   \Bigg( 312 + 64 \frac{1}{z^2} + 128 \frac{1}{z} + 12 z \Bigg)
\nonumber\\&+  {\rm Li}_{3}(-z)   \Bigg(  - 144 - 320 \frac{1}{z^2} + 672 \frac{1}{z} + 40 z \Bigg)
\nonumber\\&+  {\rm Li}_{2}(1-z)   \Bigg(  - 248 + \frac{736}{3} \frac{1}{z} - 108 z + \frac{32}{3} z^2 \Bigg)
+  {\rm Li}_{2}(-z)   \Bigg( 488 + 448 \frac{1}{z} + 40 z \Bigg)
\nonumber\\&+  \log(1+z) {\rm Li}_{2}(-z)   \Bigg(  - 288 - 384 \frac{1}{z^2} - 576 \frac{1}{z} - 48 z \Bigg)
\nonumber\\&+  \log(1-z)   \Bigg(  - \frac{1160}{3} + \frac{3272}{9} \frac{1}{z} + \frac{128}{3} z - \frac{176}{9} z^2 \Bigg)
\nonumber\\&+  \log(1-z) {\rm Li}_{2}(1-z)   \Bigg(  - 224 - 256 \frac{1}{z} - 32 z \Bigg)
\nonumber\\&+  \log^2(1-z)   \Bigg( 136 - \frac{544}{3} \frac{1}{z} + 56 z - \frac{32}{3} z^2 \Bigg)
\nonumber\\&+  \log(z)   \Bigg( \frac{974}{3} - 284 \frac{1}{z} - \frac{68}{3} z + \frac{40}{9} z^2 \Bigg)
+  \log(z) {\rm Li}_{2}(1-z)   \Bigg( 32 + 224 \frac{1}{z} + 52 z \Bigg)
\nonumber\\&+  \log(z) {\rm Li}_{2}(-z)   \Bigg( 192 + 320 \frac{1}{z^2} - 96 \frac{1}{z} \Bigg)
+  \log(z) \log(1+z)   \Bigg( 488 + 448 \frac{1}{z} + 40 z \Bigg)
\nonumber\\&+  \log(z) \log^2(1+z)   \Bigg(  - 144 - 192 \frac{1}{z^2} - 288 \frac{1}{z} - 24 z \Bigg)
\nonumber\\&+  \log(z) \log(1-z)   \Bigg(  - 232 + 192 \frac{1}{z} - 96 z + 32 z^2 \Bigg)
\nonumber\\&+  \log(z) \log^2(1-z)   \Bigg(  - 112 - 128 \frac{1}{z} - 16 z \Bigg)
\nonumber\\&+  \log^2(z)   \Bigg(  - 277 - 48 \frac{1}{z} - 35 z - \frac{40}{3} z^2 \Bigg)
\nonumber\\&+  \log^2(z) \log(1+z)   \Bigg( 120 + 160 \frac{1}{z^2} + 240 \frac{1}{z} + 20 z \Bigg)
\nonumber\\&+  \log^2(z) \log(1-z)   \Bigg( 48 + 64 \frac{1}{z} \Bigg)
+  \log^3(z)   \Bigg(  - \frac{14}{3} - \frac{64}{3} \frac{1}{z} + \frac{34}{3} z \Bigg)
\nonumber\\&+  \log \left(\frac{Q^2}{\mu_F^2}\right)\Bigg[   \Bigg(  - \frac{580}{3} + \frac{1636}{9} \frac{1}{z} + \frac{64}{3} z - \frac{88}{9} z^2 \Bigg)
\nonumber\\&+   {\rm Li}_{2}(1-z)   \Bigg(  - 112 - 128 \frac{1}{z} - 16 z \Bigg)
+   \log(1-z)   \Bigg( 136 - \frac{544}{3} \frac{1}{z} + 56 z - \frac{32}{3} z^2
          \Bigg)
\nonumber\\&+   \log(z)   \Bigg(  - 116 + 96 \frac{1}{z} - 48 z + 16 z^2 \Bigg)
\nonumber\\&+   \log(z) \log(1-z)   \Bigg(  - 112 - 128 \frac{1}{z} - 16 z
          \Bigg)
+   \log^2(z)   \Bigg( 24 + 32 \frac{1}{z} \Bigg)\Bigg]
\nonumber\\&+  \log^2 \left(\frac{Q^2}{\mu_F^2}\right)\Bigg[   \Bigg( 34 - \frac{136}{3} \frac{1}{z} + 14 z - \frac{8}{3} z^2 \Bigg)
+   \log(z)   \Bigg(  - 28 - 32 \frac{1}{z} - 4 z \Bigg)\Bigg]
\nonumber\\&+  \zeta_3   \Bigg(  - 72 - 192 \frac{1}{z^2} + 576 \frac{1}{z} + 36 z \Bigg)
+  \zeta_2   \Bigg( 108 + \frac{1216}{3} \frac{1}{z} - 36 z + \frac{32}{3} z^2 \Bigg)
\nonumber\\&+  \zeta_2 \log(1+z)   \Bigg(  - 144 - 192 \frac{1}{z^2} - 288 \frac{1}{z} - 24 z \Bigg)
+  \zeta_2 \log(z)   \Bigg( 136 + 416 \frac{1}{z} + 36 z \Bigg)\Bigg\}
\nonumber\\&+ \mathbf{C_F^2}\Bigg\{   \Bigg(  - 194 + 72 \frac{1}{z} + 148 z - 26 z^2 \Bigg)
\nonumber\\&+  S_{1,2}(1-z)   \Bigg(  - 208 - 192 \frac{1}{z} + 128 \frac{1}{1+z} + 160 z - 16 z^2
          \Bigg)
\nonumber\\&+  S_{1,2}(-z)   \Bigg( 256 \frac{1}{z} - 64 \frac{1}{1+z} \Bigg)
+  {\rm Li}_{3}\left(\frac{1-z}{1+z}\right)   \Bigg(  - 64 + 128 \frac{1}{1+z} + 64 z \Bigg)
\nonumber\\&+  {\rm Li}_{3}\left(-\frac{1-z}{1+z}\right)   \Bigg( 64 - 128 \frac{1}{1+z} - 64 z \Bigg)
\nonumber\\&+  {\rm Li}_{3}(1-z)   \Bigg( 208 - 64 \frac{1}{z} - 128 \frac{1}{1+z} - 160 z + 16 z^2 \Bigg)
\nonumber\\&+  {\rm Li}_{3}(-z)   \Bigg( 32 - 128 \frac{1}{z} - 32 \frac{1}{1+z} - 32 z \Bigg)
\nonumber\\&+  {\rm Li}_{2}(1-z)   \Bigg( 88 - 96 \frac{1}{z} + 96 z - 24 z^2 \Bigg)
+  {\rm Li}_{2}(-z)   \Bigg(  - 112 - 192 \frac{1}{z} + 80 z \Bigg)
\nonumber\\&+  \log(1+z) {\rm Li}_{2}(-z)   \Bigg( 256 \frac{1}{z} - 64 \frac{1}{1+z} \Bigg)
+  \log(1-z)   \Bigg( 64 - 64 z \Bigg)
\nonumber\\&+  \log(1-z) {\rm Li}_{2}(-z)   \Bigg( 64 - 128 \frac{1}{1+z} - 64 z \Bigg)
+  \log(z)   \Bigg(  - 32 + 68 z \Bigg)
\nonumber\\&+  \log(z) {\rm Li}_{2}(1-z)   \Bigg(  - 192 - 64 \frac{1}{z} + 96 \frac{1}{1+z} + 144 z
          - 16 z^2 \Bigg)
\nonumber\\&+  \log(z) {\rm Li}_{2}(-z)   \Bigg(  - 64 + 128 \frac{1}{1+z} + 64 z \Bigg)
\nonumber\\&+  \log(z) \log(1+z)   \Bigg(  - 112 - 192 \frac{1}{z} + 80 z \Bigg)
\nonumber\\&+  \log(z) \log^2(1+z)   \Bigg( 128 \frac{1}{z} - 32 \frac{1}{1+z} \Bigg)
+  \log(z) \log(1-z)   \Bigg( 32 + 32 z \Bigg)
\nonumber\\&+  \log(z) \log(1-z) \log(1+z)   \Bigg( 64 - 128 \frac{1}{1+z} - 64 z \Bigg)
+  \log^2(z)   \Bigg( 44 - 12 z^2 \Bigg)
\nonumber\\&+  \log^2(z) \log(1+z)   \Bigg(  - 48 - 64 \frac{1}{z} + 112 \frac{1}{1+z} + 48 z
          \Bigg)
\nonumber\\&+  \log^2(z) \log(1-z)   \Bigg(  - 16 + 32 \frac{1}{1+z} + 16 z \Bigg)
\nonumber\\&+  \log^3(z)   \Bigg(  - 20 - \frac{32}{3} \frac{1}{1+z} + 12 z - \frac{8}{3} z^2 \Bigg)
\nonumber\\&+  \log \left(\frac{Q^2}{\mu_F^2}\right) \Bigg[  \Bigg( 32 - 32 z \Bigg)
+   {\rm Li}_{2}(-z)   \Bigg( 32 - 64 \frac{1}{1+z} - 32 z \Bigg)
\nonumber\\&+   \log(z)   \Bigg( 16 + 16 z \Bigg)
+   \log(z) \log(1+z)   \Bigg( 32 - 64 \frac{1}{1+z} - 32 z
          \Bigg)
\nonumber\\&+   \log^2(z)   \Bigg(  - 8 + 16 \frac{1}{1+z} + 8 z \Bigg)\Bigg]
\nonumber\\&+  \zeta_3   \Bigg( 24 - 128 \frac{1}{z} - 16 \frac{1}{1+z} - 24 z \Bigg)
+  \zeta_2   \Bigg(  - 56 - 96 \frac{1}{z} + 40 z \Bigg)
\nonumber\\&+  \zeta_2 \log(1+z)   \Bigg( 128 \frac{1}{z} - 32 \frac{1}{1+z} \Bigg)
+  \zeta_2 \log(1-z)   \Bigg( 32 - 64 \frac{1}{1+z} - 32 z \Bigg)
\nonumber\\&+  \zeta_2 \log(z)   \Bigg(  - 16 - 64 \frac{1}{z} + 48 \frac{1}{1+z} + 16 z \Bigg)
\nonumber\\&+  \zeta_2 \log \left(\frac{Q^2}{\mu_F^2}\right)   \Bigg( 16 - 32 \frac{1}{1+z} - 16 z \Bigg)\Bigg\}
\nonumber\\&+ \mathbf{C_A C_F}\Bigg\{   \Bigg( 97 - 36 \frac{1}{z} - 74 z + 13 z^2 \Bigg)
\nonumber\\&+  S_{1,2}(1-z)   \Bigg( 104 + 96 \frac{1}{z} - 64 \frac{1}{1+z} - 80 z + 8 z^2 \Bigg)
\nonumber\\&+  S_{1,2}(-z)   \Bigg(  - 128 \frac{1}{z} + 32 \frac{1}{1+z} \Bigg)
+  {\rm Li}_{3}\left(\frac{1-z}{1+z}\right)   \Bigg( 32 - 64 \frac{1}{1+z} - 32 z \Bigg)
\nonumber\\&+  {\rm Li}_{3}\left(-\frac{1-z}{1+z}\right)   \Bigg(  - 32 + 64 \frac{1}{1+z} + 32 z \Bigg)
\nonumber\\&+  {\rm Li}_{3}(1-z)   \Bigg(  - 104 + 32 \frac{1}{z} + 64 \frac{1}{1+z} + 80 z - 8 z^2 \Bigg)
\nonumber\\&+  {\rm Li}_{3}(-z)   \Bigg(  - 16 + 64 \frac{1}{z} + 16 \frac{1}{1+z} + 16 z \Bigg)
\nonumber\\&+  {\rm Li}_{2}(1-z)   \Bigg(  - 44 + 48 \frac{1}{z} - 48 z + 12 z^2 \Bigg)
+  {\rm Li}_{2}(-z)   \Bigg( 56 + 96 \frac{1}{z} - 40 z \Bigg)
\nonumber\\&+  \log(1+z) {\rm Li}_{2}(-z)   \Bigg(  - 128 \frac{1}{z} + 32 \frac{1}{1+z} \Bigg)
+  \log(1-z)   \Bigg(  - 32 + 32 z \Bigg)
\nonumber\\&+  \log(1-z) {\rm Li}_{2}(-z)   \Bigg(  - 32 + 64 \frac{1}{1+z} + 32 z \Bigg)
+  \log(z)   \Bigg( 16 - 34 z \Bigg)
\nonumber\\&+  \log(z) {\rm Li}_{2}(1-z)   \Bigg( 96 + 32 \frac{1}{z} - 48 \frac{1}{1+z} - 72 z + 8 
         z^2 \Bigg)
\nonumber\\&+  \log(z) {\rm Li}_{2}(-z)   \Bigg( 32 - 64 \frac{1}{1+z} - 32 z \Bigg)
+  \log(z) \log(1+z)   \Bigg( 56 + 96 \frac{1}{z} - 40 z \Bigg)
\nonumber\\&+  \log(z) \log^2(1+z)   \Bigg(  - 64 \frac{1}{z} + 16 \frac{1}{1+z} \Bigg)
+  \log(z) \log(1-z)   \Bigg(  - 16 - 16 z \Bigg)
\nonumber\\&+  \log(z) \log(1-z) \log(1+z)   \Bigg(  - 32 + 64 \frac{1}{1+z} + 32 z \Bigg)
+  \log^2(z)   \Bigg(  - 22 + 6 z^2 \Bigg)
\nonumber\\&+  \log^2(z) \log(1+z)   \Bigg( 24 + 32 \frac{1}{z} - 56 \frac{1}{1+z} - 24 z \Bigg)
\nonumber\\&+  \log^2(z) \log(1-z)   \Bigg( 8 - 16 \frac{1}{1+z} - 8 z \Bigg)
+  \log^3(z)   \Bigg( 10 + \frac{16}{3} \frac{1}{1+z} - 6 z + \frac{4}{3} z^2 \Bigg)
\nonumber\\&+  \log \left(\frac{Q^2}{\mu_F^2}\right)\Bigg[   \Bigg(  - 16 + 16 z \Bigg)
+   {\rm Li}_{2}(-z)   \Bigg(  - 16 + 32 \frac{1}{1+z} + 16 z \Bigg)
\nonumber\\&+   \log(z)   \Bigg(  - 8 - 8 z \Bigg)
+   \log(z) \log(1+z)   \Bigg(  - 16 + 32 \frac{1}{1+z} + 
         16 z \Bigg)
\nonumber\\&+   \log^2(z)   \Bigg( 4 - 8 \frac{1}{1+z} - 4 z \Bigg)\Bigg]
\nonumber\\&+  \zeta_3   \Bigg(  - 12 + 64 \frac{1}{z} + 8 \frac{1}{1+z} + 12 z \Bigg)
+  \zeta_2   \Bigg( 28 + 48 \frac{1}{z} - 20 z \Bigg)
\nonumber\\&+  \zeta_2 \log(1+z)   \Bigg(  - 64 \frac{1}{z} + 16 \frac{1}{1+z} \Bigg)
+  \zeta_2 \log(1-z)   \Bigg(  - 16 + 32 \frac{1}{1+z} + 16 z \Bigg)
\nonumber\\&+  \zeta_2 \log(z)   \Bigg( 8 + 32 \frac{1}{z} - 24 \frac{1}{1+z} - 8 z \Bigg)
+  \zeta_2 \log \left(\frac{Q^2}{\mu_F^2}\right)   \Bigg(  - 8 + 16 \frac{1}{1+z} + 8 z \Bigg)\Bigg\}\,,
\nonumber\\
   {\bar{\Delta}}_{q_1q_2}^{h,(2)} &= \mathbf{C_F}\Bigg\{   \Bigg( \frac{5417}{9} - \frac{12304}{27} \frac{1}{z} - \frac{1550}{9} z + \frac{703}{27} z^2 \Bigg)
\nonumber\\&+  S_{1,2}(1-z)   \Bigg( 144 + 128 \frac{1}{z^2} + 64 \frac{1}{z} + 64 z \Bigg)
\nonumber\\&+  S_{1,2}(-z)   \Bigg(  - 288 - 384 \frac{1}{z^2} - 576 \frac{1}{z} - 48 z \Bigg)
\nonumber\\&+  {\rm Li}_{3}(1-z)   \Bigg( 312 + 64 \frac{1}{z^2} + 128 \frac{1}{z} + 12 z \Bigg)
\nonumber\\&+  {\rm Li}_{3}(-z)   \Bigg(  - 144 - 320 \frac{1}{z^2} + 672 \frac{1}{z} + 40 z \Bigg)
\nonumber\\&+  {\rm Li}_{2}(1-z)   \Bigg(  - 248 + \frac{736}{3} \frac{1}{z} - 108 z + \frac{32}{3} z^2 \Bigg)
+  {\rm Li}_{2}(-z)   \Bigg( 488 + 448 \frac{1}{z} + 40 z \Bigg)
\nonumber\\&+  \log(1+z) {\rm Li}_{2}(-z)   \Bigg(  - 288 - 384 \frac{1}{z^2} - 576 \frac{1}{z} - 48 z \Bigg)
\nonumber\\&+  \log(1-z)   \Bigg(  - \frac{1160}{3} + \frac{3272}{9} \frac{1}{z} + \frac{128}{3} z - \frac{176}{9} z^2 \Bigg)
\nonumber\\&+  \log(1-z) {\rm Li}_{2}(1-z)   \Bigg(  - 224 - 256 \frac{1}{z} - 32 z \Bigg)
\nonumber\\&+  \log^2(1-z)   \Bigg( 136 - \frac{544}{3} \frac{1}{z} + 56 z - \frac{32}{3} z^2 \Bigg)
+  \log(z)   \Bigg( \frac{974}{3} - 284 \frac{1}{z} - \frac{68}{3} z + \frac{40}{9} z^2 \Bigg)
\nonumber\\&+  \log(z) {\rm Li}_{2}(1-z)   \Bigg( 32 + 224 \frac{1}{z} + 52 z \Bigg)
+  \log(z) {\rm Li}_{2}(-z)   \Bigg( 192 + 320 \frac{1}{z^2} - 96 \frac{1}{z} \Bigg)
\nonumber\\&+  \log(z) \log(1+z)   \Bigg( 488 + 448 \frac{1}{z} + 40 z \Bigg)
\nonumber\\&+  \log(z) \log^2(1+z)   \Bigg(  - 144 - 192 \frac{1}{z^2} - 288 \frac{1}{z} - 24 z \Bigg)
\nonumber\\&+  \log(z) \log(1-z)   \Bigg(  - 232 + 192 \frac{1}{z} - 96 z + 32 z^2 \Bigg)
\nonumber\\&+  \log(z) \log^2(1-z)   \Bigg(  - 112 - 128 \frac{1}{z} - 16 z \Bigg)
+  \log^2(z)   \Bigg(  - 277 - 48 \frac{1}{z} - 35 z - \frac{40}{3} z^2 \Bigg)
\nonumber\\&+  \log^2(z) \log(1+z)   \Bigg( 120 + 160 \frac{1}{z^2} + 240 \frac{1}{z} + 20 z \Bigg)
\nonumber\\&+  \log^2(z) \log(1-z)   \Bigg( 48 + 64 \frac{1}{z} \Bigg)
+  \log^3(z)   \Bigg(  - \frac{14}{3} - \frac{64}{3} \frac{1}{z} + \frac{34}{3} z \Bigg)
\nonumber\\&+  \log \left(\frac{Q^2}{\mu_F^2}\right) \Bigg[  \Bigg(  - \frac{580}{3} + \frac{1636}{9} \frac{1}{z} + \frac{64}{3} z - \frac{88}{9} z^2 \Bigg)
\nonumber\\&+  {\rm Li}_{2}(1-z)   \Bigg(  - 112 - 128 \frac{1}{z} - 16 z \Bigg)
+   \log(1-z)   \Bigg( 136 - \frac{544}{3} \frac{1}{z} + 56 z - \frac{32}{3} z^2
          \Bigg)
\nonumber\\&+   \log(z)   \Bigg(  - 116 + 96 \frac{1}{z} - 48 z + 16 z^2 \Bigg)
\nonumber\\&+   \log(z) \log(1-z)   \Bigg(  - 112 - 128 \frac{1}{z} - 16 z
          \Bigg)
+   \log^2(z)   \Bigg( 24 + 32 \frac{1}{z} \Bigg)\Bigg]
\nonumber\\&+  \log^2 \left(\frac{Q^2}{\mu_F^2}\right) \Bigg[  \Bigg( 34 - \frac{136}{3} \frac{1}{z} + 14 z - \frac{8}{3} z^2 \Bigg)
+   \log(z)   \Bigg(  - 28 - 32 \frac{1}{z} - 4 z \Bigg)\Bigg]
\nonumber\\&+  \zeta_3   \Bigg(  - 72 - 192 \frac{1}{z^2} + 576 \frac{1}{z} + 36 z \Bigg)
+  \zeta_2   \Bigg( 108 + \frac{1216}{3} \frac{1}{z} - 36 z + \frac{32}{3} z^2 \Bigg)
\nonumber\\&+  \zeta_2 \log(1+z)   \Bigg(  - 144 - 192 \frac{1}{z^2} - 288 \frac{1}{z} - 24 z \Bigg)
\nonumber\\&+  \zeta_2 \log(z)   \Bigg( 136 + 416 \frac{1}{z} + 36 z \Bigg)\Bigg\}\,,
\nonumber\\
{\bar{\Delta}}_{q_1\bar{q}_2}^{h,(2)} &={\bar{\Delta}}_{q_1q_2}^{h,(2)} .
\end{align}

In the above expressions, the $S_{n,p}(z)$ and ${\rm Li}_n(z)$ are defined through
\begin{align}
\label{eq:SnpLi}
&S_{n,p}(z) = \frac{(-1)^{n+p-1}}{(n-1)!p!} \int\limits_{0}^{1} \frac{dy}{y} \log^{n-1}(y) \log^p(1-y z)\,,
\intertext{and}
&{\rm Li}_n(z) \equiv S_{n-1,1}(z)\,.
\end{align}
The constants $\zeta_2=\frac{\pi^2}{6}$ and $\zeta_3=1.20205690\ldots$ .

\section{Identities}
\label{app:ident}

The identities which have been employed to get the results manifestly
real and to perform mass factorisation in an effective manner are
listed below: 

\begin{align}
\label{eq:29}
{\rm Li}_{3}\left( \frac{1}{1+z} \right) &= - {\rm Li}_{3}\left( -\frac{1}{z}
  \right)  - {\rm Li}_{3}\left( 1+\frac{1}{z} \right) + \frac{1}{3} \log^{3} \left( \frac{z}{1+z} \right)
   \nonumber\\
&- \frac{1}{2} \log \left( -\frac{1}{1+z} \right) \log^{2}\left(
  \frac{z}{1+z} \right) - \frac{1}{6} \pi^2 \log \left( \frac{z}{1+z}
  \right) + \zeta_{3}\,,
\nonumber\\
{\rm Li}_3\left( 1+\frac{1}{z} \right) &= -S_{1,2}\left( -\frac{1}{z} \right) +
  \log\left( 1+\frac{1}{z} \right) {\rm Li}_2\left( 1+\frac{1}{z} \right) +
  \frac{1}{2} \log\left( -\frac{1}{z} \right)
                                         \log^{2}\left(1+\frac{1}{z}
                                         \right) 
\nonumber\\
&+ \zeta_{3}\,, 
\nonumber\\
{\rm Li}_3\left( \frac{1-z}{2} \right) &= - {\rm Li}_3\left( -\frac{1-z}{1+z}
  \right) - {\rm Li}_3\left( \frac{2}{1+z} \right) +
  \frac{1}{3} \log^{3}\left( \frac{1+z}{2}\right) 
\nonumber\\
&- \frac{1}{2}
                                         \log\left(
                                         -\frac{1-z}{2}\right) \log^{2}\left( \frac{1+z}{2}\right) -
  \frac{1}{6} \pi^2 \log\left( \frac{1+z}{2}\right) + \zeta_{3}\,, 
\nonumber\\ 
{\rm Li}_3\left( \frac{2}{1+z} \right) &= {\rm Li}_3\left( \frac{1+z}{2}\right)
                                   - \frac{1}{6} \log^{3}\left(-\frac{2}{1+z}
                                   \right) -
                                   \frac{1}{6} \pi^2 \log\left( -\frac{2}{1+z}
                                   \right)\,,
\nonumber\\
 {\rm Li}_3\left( z \right) =  - ~\,&S_{1,2}\left( 1-z \right) +
                             \log\left( z \right) {\rm Li}_2\left( z
                             \right)  + \frac{1}{2} \log\left( 1-z
                             \right) \log^{2}\left( z \right) + \zeta_{3}\,,
\nonumber\\
{\rm Li}_2\left( 1+\frac{1}{z} \right) &=  - {\rm Li}_2\left(
                                         -\frac{1}{z} \right) -
                                         \log\left( 1+\frac{1}{z}
                                         \right) \log\left(
                                         -\frac{1}{z} \right) + \zeta_{2}\,,
\nonumber\\
{\rm Li}_2\left( -\frac{1}{z} \right) ~\;&= - {\rm Li}_2\left( -z \right)
                                        - \frac{1}{2} \log^{2}\left( z
                                        \right) - \zeta_{2}\,,
\nonumber\\
{\rm Li}_2\left( z \right) = - ~&{\rm Li}_2\left( 1-z \right) -
                             \log\left( z \right) \log\left( 1-z 
                             \right) + \zeta_{2}\,,
\nonumber\\
S_{1,2}\left( -\frac{1}{z} \right) ~&= -S_{1,2}\left( -z \right) + {\rm
                                     Li}_3\left( -\frac{1}{z} \right)
                                     + \log\left( -z \right) {\rm
                                     Li}_2\left( -\frac{1}{z} \right)
                                     + \frac{1}{6} \log^{3}\left( -z
                                     \right) - \frac{\pi^{2}}{2}
                                     \log\left( -z \right) 
\nonumber\\
&+ \zeta_{3}
             + i \pi \Bigg( \frac{\pi^2}{6} - {\rm Li}_2\left( -\frac{1}{z}
                                     \right) -
                                     \frac{1}{2} \log^{2}\left( -z
  \right) \Bigg) \,.
\end{align}

\bibliography{main} 
\bibliographystyle{utphysM}
  
\end{document}